\RequirePackage{fix-cm}
\documentclass[9pt,twoside]{extarticle}
\usepackage[OT1]{fontenc}
\usepackage[paperwidth=210mm,paperheight=276mm,textwidth=178mm,textheight=244mm,
  inner=16.87mm,outer=15.13mm,top=18mm,footskip=4.434mm]{geometry}

\usepackage{amsmath,amssymb,graphicx,booktabs,array,longtable,tabularx}
\usepackage[table]{xcolor}
\usepackage{caption,needspace,etoolbox,titlesec,multicol,balance}
\usepackage[authoryear,round]{natbib}
\usepackage[unicode,colorlinks=true,allcolors=black,bookmarksnumbered=false]{hyperref}
\hypersetup{pdftitle={Breaking News Out of the Filter Bubble: Generative AI Search Diversifies Collective Attention and Raises Shared Information Consumption},
 pdfauthor={Heeseung Andrew Lee; Dokyun (DK) Lee; Gwanhoo Lee; Dongwon Lee},
 pdfsubject={Generative AI search and news consumption},pdfkeywords={generative AI, news consumption, diversity, field experiment}}
\makeatletter
\renewcommand\normalsize{\@setfontsize\normalsize{9bp}{11.5pt}\abovedisplayskip=7bp plus2bp minus2bp\belowdisplayskip=7bp plus2bp minus2bp}
\renewcommand\small{\@setfontsize\small{8bp}{11pt}}
\renewcommand\footnotesize{\@setfontsize\footnotesize{8bp}{11pt}}
\makeatother
\normalsize
\titleformat{\section}{\raggedright\hyphenpenalty=10000\exhyphenpenalty=10000\fontsize{15.5bp}{17pt}\selectfont\bfseries}{}{0pt}{}
\titleformat{\subsection}{\raggedright\hyphenpenalty=10000\exhyphenpenalty=10000\fontsize{13bp}{17pt}\selectfont\bfseries}{}{0pt}{}
\titleformat{\subsubsection}{\raggedright\hyphenpenalty=10000\exhyphenpenalty=10000\fontsize{10.5bp}{13pt}\selectfont\bfseries}{}{0pt}{}
\titlespacing*{\section}{0pt}{14bp plus2bp minus2bp}{5bp}
\titlespacing*{\subsection}{0pt}{10bp plus2bp minus2bp}{4bp}
\DeclareCaptionFont{tablefont}{\fontsize{9bp}{11.5pt}\selectfont}
\newcommand{\addtabletext}[1]{\vspace{3bp}{\fontsize{8bp}{9.6bp}\selectfont #1}\par}

\begin{document}
\twocolumn[{
\vspace*{14.833bp}
{\noindent\fontsize{7bp}{9pt}\selectfont\bfseries Working Paper\par}
\vspace{4.71bp}\hrule height 1pt\vspace{12bp}
{\noindent\raggedright\fontsize{20bp}{21pt}\selectfont\bfseries Breaking News Out of the Filter Bubble: Generative AI Search Diversifies Collective Attention and Raises Shared Information Consumption\par}
\vspace{5.977bp}
{\noindent\raggedright\fontsize{11bp}{15pt}\selectfont\bfseries
Heeseung Andrew Lee\textsuperscript{1,*}\quad Dokyun (DK) Lee\textsuperscript{2,3,*}\quad Gwanhoo Lee\textsuperscript{4}\quad Dongwon Lee\textsuperscript{5}\par}
\vspace{7.711bp}\hrule height 1pt\vspace{8.446bp}
{\fontsize{8bp}{11pt}\selectfont\noindent
\textsuperscript{1}Naveen Jindal School of Management, University of Texas at Dallas, Richardson, TX 75080, USA\\
\textsuperscript{2}Questrom School of Business, Boston University, Boston, MA 02215, USA\\
\textsuperscript{3}Faculty of Computing \& Data Sciences, Boston University, Boston, MA 02215, USA\\
\textsuperscript{4}Kogod School of Business, American University, Washington, DC 20016, USA\\
\textsuperscript{5}School of Business and Management, Hong Kong University of Science and Technology, Hong Kong SAR\\
\textsuperscript{*}Correspondence: \href{mailto:heeseung.lee@utdallas.edu}{heeseung.lee@utdallas.edu} and \href{mailto:dokyun@bu.edu}{dokyun@bu.edu}.\par}
\vspace{16.439bp}
{\noindent\fontsize{13bp}{17pt}\selectfont\bfseries Abstract\par}\vspace{4bp}
{\noindent\fontsize{9.5bp}{11.5pt}\selectfont\bfseries Generative AI search and AI overviews are transforming access to information and news, renewing concerns that readers will encounter a narrower range of topics and have less in common. We examine these concerns via a randomized field experiment with 37,561 readers at The Washington Post. Both groups searched the same archive, but treatment readers also received AI answers with article citations above conventional results. Measuring consumption across displayed answers and opened articles, we find that AI search expands the reach of widely read topics and increases overlap in readers' topic consumption. At the same time, consumption becomes less concentrated and shifts toward less-popular topics, both within readers and across the audience. AI answers account for most of the increase in shared information, delivering it without requiring article clicks and broadening exposure beyond the articles readers open. Cited articles also contribute to the shift toward less-popular topics. Readers shift from conventional-result clicks and browsing toward cited articles and follow-up searches. More frequent searching offsets lower article consumption per search, producing a small increase in article consumption per reader. Total information consumption per minute also rises. Generative AI search can thus diversify collective attention while strengthening the information readers have in common.\par}
\vspace{8bp}
{\noindent\fontsize{9bp}{11.5pt}\selectfont Keywords\quad generative AI, AI answer, news consumption, filter bubble, diversity, field experiment, retrieval-augmented generation\par}
\vspace{13bp}
}]
\section*{Introduction}
Generative AI answers (henceforth AI answers), which serve synthesized summaries from multiple retrieved articles using techniques like retrieval-augmented generation (RAG), are transforming how people search for and consume online content and news.
When a user submits a query, RAG retrieves relevant articles, which a language model synthesizes into an answer \citep{lewis2020rag}.
Semantic (i.e., conventional) search returns a ranked list of articles, whereas RAG search can add a synthesized answer with links to its sources on top of those results.
Google's AI Overviews and Perplexity already provide such answers \citep{egan2026report,perplexity2026}, and news publishers are following \citep{wapo2024pr}.
As generative AI search emerges as a gateway to news, it may shape not only how society finds information but also how public attention is distributed across issues.

The rise of RAG search with AI answers brings renewed attention to Pariser's filter-bubble concern that curation algorithms may narrow users' information exposure and separate them into distinct informational environments \citep{pariser2011}. Such segregation may limit exposure to competing perspectives and weaken shared information, the topics that many readers encounter in common and that public deliberation and consensus formation depend on \citep{flaxman2016,sunstein2017}. In news, the stakes
extend beyond individual exposure because media shape the public agenda by directing attention across issues \citep{mccombs1972}.\footnote{Although filter-bubble research focuses largely on political ideology \citep{flaxman2016,gonzalez2023,cinelli2021}, we examine the diversity of consumption across all news topics, because ideological measures apply only to political content.}

This study examines three aspects of news consumption: 1) shared information consumption, 2) topic concentration, and 3) topic popularity \citep{hosanagar2014,lee2019}. Shared information consumption measures how much shared information each reader consumes and thus whether RAG search shrinks the common ground. Topic concentration measures whether consumption spans many topics or is confined to a few, the narrowing Pariser describes \citep{pariser2011}. Topic popularity measures whether consumption favors topics receiving the most total consumption across readers, that is, whether attention tilts toward already-popular topics and away from the rest of the public agenda \citep{mccombs1972}. Because the concern applies both to what each reader sees and to where the audience's attention goes as a whole, we measure concentration and popularity at the individual and aggregate levels.

We estimate these effects in a 38-day randomized field experiment with The Washington Post involving 37,561 readers, conducted before the public launch of Ask The Post AI \citep{wapo2024pr}.
At their first on-site search, readers were randomly assigned to one of two groups.
The Control group was shown the usual conventional ranked semantic-search results.
The Treatment group received an AI answer citing up to five source articles above the same semantic-search results shown alone in Control (Supplementary material, Fig.
S1).
Both groups drew from the same article archive.
We used BERTopic to represent each article as a distribution over news topics and calculated the three outcomes in each condition (Materials and Methods).
For each outcome, we compare the groups using two measures of consumption.
Total information consumption includes the AI answer displayed on the results page and the articles readers open.
Article consumption includes only opened articles and captures changes in article selection.
Under Control, the two measures coincide because no AI answer is delivered.

We find that RAG search changes all three aspects of news consumption.
First, shared information consumption per reader rises by 69\%, with the AI answer accounting for most of the increase.
RAG search therefore adds to the common informational ground across readers.
Second, topic concentration declines at both the individual and aggregate levels when total information is considered.
Third, both total information and article consumption shift toward less-popular topics at both levels.
These results show why the three aspects must be distinguished: readers consume more information in common even as their attention spreads across a broader and less-popular set of topics.

The engagement pattern helps explain how RAG search can expand common informational ground and topical breadth simultaneously.
Total information consumption per reader rises by 82\%, reflecting both 29\% more searches and 41\% more information per search.
The AI answer increases the information delivered within each search.
Activity becomes more oriented toward consuming information through search and AI answers, with less browsing through the homepage or other section fronts.
Total information consumption per minute rises by 62\%.
These results suggest that RAG search efficiently delivers a common informational core while continued searching extends consumption across a wider range of less-popular topics.

The mechanism analyses support this account. In Treatment, the AI answer appears above conventional results. Compared with Control, conventional-result clicks fall by 21 percentage points and browsing by 5 points, while cited-source clicks rise by 14 points and follow-up searches without an article click rise by 11 points. About 21\% of searches end the session in both groups. Across all article opens after the first search, cited-source articles contribute a broader, less-popular set of topics to aggregate consumption than other articles opened by Treatment readers. AI answers also broaden readers' total information consumption beyond the articles they open and have lower mean topic popularity than articles opened by Control readers. The shift toward less-popular topics could reflect Treatment readers asking about less-popular topics (demand) or cited-source lists offering less-popular articles than conventional results (supply). Our comparisons show neither pattern.

Contrary to the filter-bubble concern, RAG search broadens individual topic exposure while strengthening common informational ground.
The AI answer efficiently delivers information shared across readers, and both the answer and the cited articles readers open contribute to the shift in total information consumption toward less-popular topics.
Thus, generative AI search diversifies collective attention without increasing reader fragmentation.

\section*{Results}

\subsection*{RAG search increases shared information consumption}

\begin{figure*}[t!]
\centering
\includegraphics[width=\textwidth]{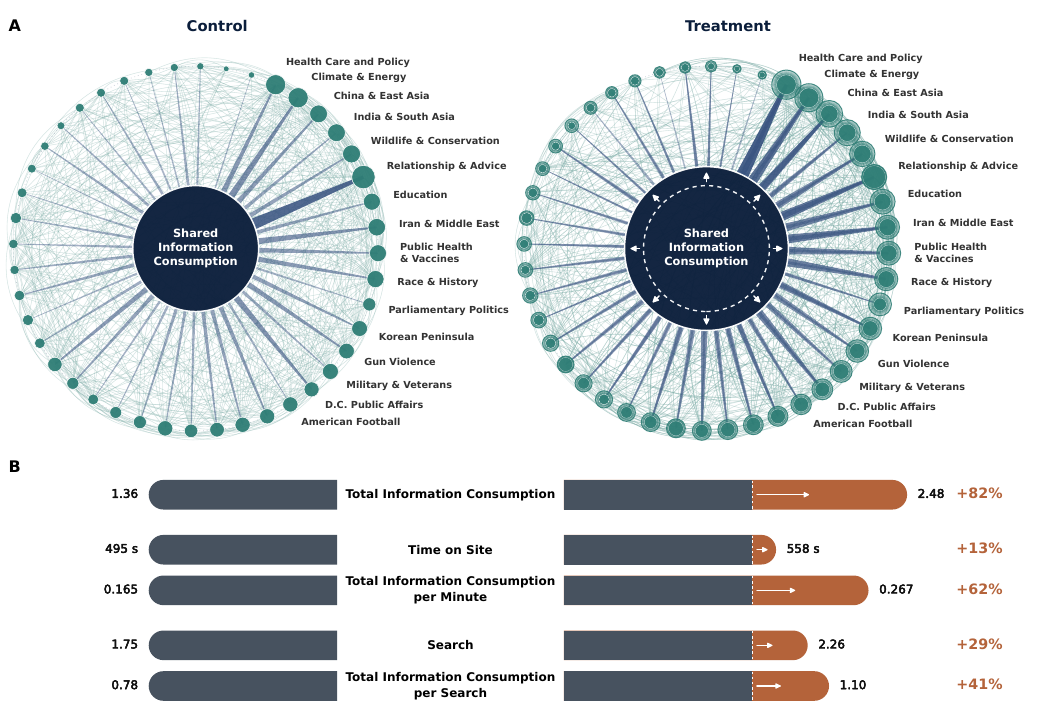}
\caption{(A) RAG search increases consumption of both shared and non-core information. Node area shows the amount of information consumed per reader, and edges connect topics consumed following the same search (Supplementary material, section S27). (B) RAG search increases searches per reader and total information consumption per search (search efficiency). Total information consumption per reader increases more than time on site per reader, so total information consumption per minute (consumption efficiency) also increases.}
\label{fig:one}
\end{figure*}

We define shared information as the most widely read news topics in the Control group, selecting the top quartile based on the number of distinct readers who read about each topic (Materials and Methods).
Treatment readers consume 69\% more shared information on average than Control readers (\textit{P} \ensuremath{<} 0.001, Fig.~\ref{fig:one}A and Supplementary material, Table S1).
The proportion of readers consuming shared information rises from 37\% in Control to 55\% in Treatment (Supplementary material, section S3).
The increase in shared information consumption holds under alternative definitions of shared information and after political topics are excluded (Supplementary material, section S4).
Across all topics, pairs of Treatment readers also have more similar topic consumption and a greater amount of information in common than pairs of Control readers (both \textit{P} \ensuremath{<} 0.001, Supplementary material, section S6).

Readers also consume more non-core information, defined as information on topics in the remaining three quartiles outside the shared core (Fig.~\ref{fig:one}A and Supplementary material, section S2).
AI answers account for most of the increase in both shared and non-core information consumption by summarizing multiple cited-source articles within a single search (Supplementary material, section S5 and Materials and Methods).

Treatment readers consume 82\% more information on average than Control readers, reflecting 29\% more searches per reader and 41\% more information per search (Fig.~\ref{fig:one}B and Supplementary material, section S7).
Time on site per reader is only 13\% higher, so consumption efficiency, measured as total information consumed per minute, is 62\% higher. The increase in total information consumption per reader also holds when measured by word counts and time spent on article and search pages (Supplementary material, section S8).
We next examine how this consumption is distributed across topics.

\subsection*{Consumption disperses toward less popular topics}

\begin{figure*}[t!]
\centering
\includegraphics[width=\textwidth]{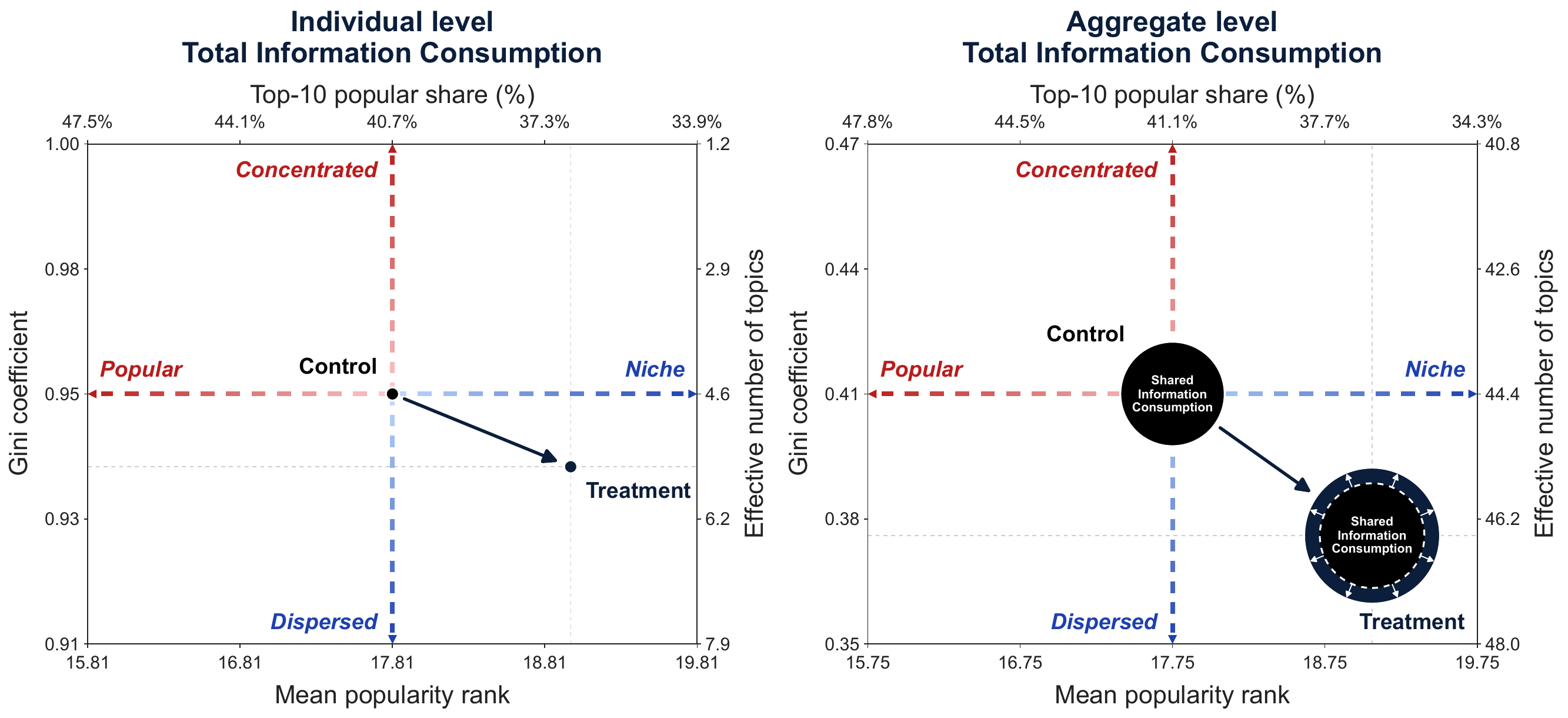}
\caption{At both the individual and aggregate levels, RAG search disperses total information consumption toward less-popular topics. In the aggregate panel, dot positions show the concentration and popularity of each group's overall consumption. Dot area is proportional to shared information consumption averaged over all readers in each group, which is higher in Treatment than in Control.}
\label{fig:two}
\end{figure*}

With RAG search, total information consumption becomes less concentrated at both the individual and aggregate levels (Fig.~\ref{fig:two} and Table~\ref{tab:one}).
The individual level describes each reader's consumption, while the aggregate level describes each group's overall consumption.
We measure topic concentration using the Gini coefficient and the effective number of topics (Materials and Methods).
The Gini coefficient measures how unevenly consumption is distributed across topics, with lower values indicating less concentration.
Its values are lower in Treatment than in Control at both levels (both \textit{P} \ensuremath{<} 0.001).
The effective number of topics measures how widely consumption is spread across topics, with higher values indicating less concentration.
At the individual level, the mean effective number of topics increases from 4.58 in Control to 5.55 in Treatment (\textit{P} \ensuremath{<} 0.001).
At the aggregate level, the effective number of topics increases from 44.4 in Control to 46.4 in Treatment (\textit{P} \ensuremath{<} 0.001).
The decrease in individual concentration persists after equalizing consumption volume per reader (Supplementary material, section S9).

Total information consumption also shifts toward less-popular topics at both the individual and aggregate levels (Fig.~\ref{fig:two} and Table~\ref{tab:one}).
We rank topics from most to least popular by total article consumption in Control, so higher mean popularity ranks indicate greater weight on less-popular topics (Materials and Methods).
Mean popularity rank is higher in Treatment than in Control at both levels (both \textit{P} \ensuremath{<} 0.001, Table~\ref{tab:one}).
The mean individual share of consumption on the ten most popular topics (Top-10 popular share) falls by 4.0 percentage points (40.7\% in Control to 36.7\% in Treatment), while the aggregate share falls by 4.4 points (41.1\% to 36.7\%; both \textit{P} \ensuremath{<} 0.001).
These concentration and popularity results persist when we exclude political topics, use an alternative topic model, or measure answer content differently (Supplementary material, sections S10 and S11, and Table S16).

\begin{table*}[t!]
\caption{Topic concentration and popularity of total information consumption}
\label{tab:one}
\centering
\fontsize{7pt}{10pt}\selectfont
\setlength{\tabcolsep}{4pt}
\begin{tabular*}{\textwidth}{@{\extracolsep{\fill}}llccccc@{}}
\toprule
Outcome & Measure & Control & Treatment & Difference & 95\% CI & \textit{P} \\
\midrule
\multicolumn{7}{l}{\textbf{Individual level}} \\
\addlinespace[2pt]
Concentration & Gini coefficient & 0.953 & 0.940 & \ensuremath{-}0.013 & [\ensuremath{-}0.015, \ensuremath{-}0.012] & \ensuremath{<} 0.001 \\
 & Effective number of topics & 4.58 & 5.55 & +0.97 & [0.86, 1.08] & \ensuremath{<} 0.001 \\
Popularity & Mean popularity rank & 17.81 & 18.98 & +1.16 & [0.88, 1.45] & \ensuremath{<} 0.001 \\
 & Top-10 popular share (\%) & 40.7 & 36.7 & \ensuremath{-}4.0 & [\ensuremath{-}5.0, \ensuremath{-}3.0] & \ensuremath{<} 0.001 \\
\addlinespace[5pt]
\multicolumn{7}{l}{\textbf{Aggregate level}} \\
\addlinespace[2pt]
Concentration & Gini coefficient & 0.414 & 0.380 & \ensuremath{-}0.034 & [\ensuremath{-}0.043, \ensuremath{-}0.024] & \ensuremath{<} 0.001 \\
 & Effective number of topics & 44.37 & 46.42 & +2.05 & [1.46, 2.58] & \ensuremath{<} 0.001 \\
Popularity & Mean popularity rank & 17.75 & 19.06 & +1.31 & [1.03, 1.58] & \ensuremath{<} 0.001 \\
 & Top-10 popular share (\%) & 41.1 & 36.7 & \ensuremath{-}4.4 & [\ensuremath{-}5.4, \ensuremath{-}3.4] & \ensuremath{<} 0.001 \\
\bottomrule
\end{tabular*}\par
\raggedright
\addtabletext{\textit{Notes.} Individual estimates include only readers who consume at least two articles or AI answers in total. Brackets show 95\% confidence intervals for each difference. \textit{P} values are adjusted for the eight comparisons in this table using the Holm method (Materials and Methods). Differences are calculated before rounding.}
\end{table*}

RAG search increases shared information consumption while broadening total information consumption toward less-popular topics (Fig.~\ref{fig:two}).
These findings run counter to the filter-bubble concern that RAG search both narrows the range of topics individual readers consume and leaves readers with less information in common.
We next examine how AI answers and the articles readers open contribute to these changes.

\subsection*{AI answers and cited articles change how readers consume}
AI answers deliver shared information without an article click.
Counting articles alone, shared information reaches 36.9\% of Control readers and 35.1\% of Treatment readers; including answers raises the Treatment share by 20.3 percentage points, to 55.4\% (Supplementary material, section S3).

Individual concentration is statistically insignificant between Treatment and Control when only articles are counted (Supplementary material, section S13), but this misses the AI answer consumption and therefore provides an incomplete understanding of consumption.
Among Treatment readers who opened articles, the mean effective number of topics rises from 3.81 for articles alone to 5.48 when AI answers are included (\textit{P} \ensuremath{<} 0.001).
For these same readers, including AI answers reduces the share of consumption on the ten most popular topics from 37.85\% to 37.12\%, indicating a shift toward less-popular topics (Supplementary material, section S14). Aggregate AI-answer consumption in Treatment is less concentrated and favors less-popular topics than article consumption in Control (Supplementary material, Table S19).

Compared with Control, article consumption in Treatment favors less-popular topics at both levels and is less concentrated in aggregate (Supplementary material, section S13). Within Treatment, aggregate consumption of cited-source articles is less concentrated and favors less-popular topics than consumption of other articles, including those opened through conventional results or browsing (Supplementary material, section S15). We find no evidence that Treatment readers search for less-popular topics than Control readers (demand), or that cited-source lists offer less-popular articles than conventional results (supply) (Supplementary material, section S16).

\begin{figure*}[t!]
\centering
\includegraphics[width=\textwidth]{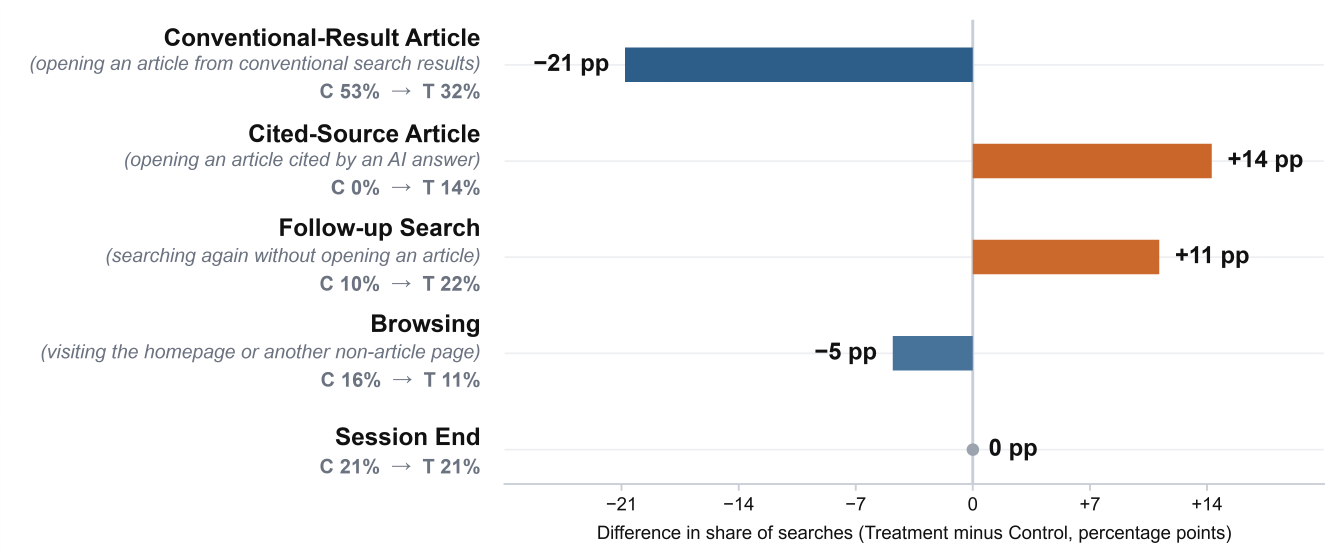}
\caption{RAG search changes the next action after a search (Materials and Methods). With an AI answer and cited sources on the results page, cited-source article opens and follow-up searches become more common, while conventional-result article opens and browsing elsewhere on the site decline. C and T denote Control and Treatment. Shares and differences are rounded separately, so displayed differences may not sum to zero.}
\label{fig:three}
\end{figure*}

With AI answers and cited sources above conventional results, the shares of searches followed by conventional-result clicks and browsing elsewhere on the site are 21 and 5 percentage points lower in Treatment than in Control, respectively.
The shares followed by cited-source clicks and follow-up searches without an article click are 14 and 11 percentage points higher, respectively (Fig.~\ref{fig:three} and Supplementary material, section S17).
About 21\% of searches end the session in both groups (Materials and Methods).

Treatment readers search 29\% more often and consume 41\% more total information per search than Control readers (Fig.~\ref{fig:one}B).
More frequent searching more than offsets 18\% lower article consumption per search, yielding 6\% higher article consumption per reader (Supplementary material, section S18 and Table S25).
Each additional search can deliver another AI answer without an article click and offer another opportunity to open cited-source articles.
These patterns help explain how RAG search delivers shared information efficiently while broadening total information consumption toward less-popular topics (Supplementary material, Table S7).

\section*{Discussion}

Adding AI answers and citations to search within the same news archive increased shared information consumption, its reach, and readers' topic overlap.
Total consumption became less concentrated and shifted toward less-popular topics at both individual and aggregate levels (Table~\ref{tab:one} and Supplementary material, section S6).
The concentration result matters because less-popular consumption can remain confined to a few topics.
These findings extend research showing that broader individual consumption can coexist with greater overlap \citep{hosanagar2014} and that individual and aggregate diversity can move in different directions \citep{lee2019}.

AI answers accounted for most of the increase in shared information consumption and broadened consumption beyond opened articles (Supplementary material, Tables S5 and S18).
Both answers and cited-source articles contributed to the shift toward less-popular topics (Supplementary material, section S14 and Table S21).
Treatment readers more often followed searches with another query and less often resumed site browsing (Fig.~\ref{fig:three}).
Total information consumption increased per search and per minute, and shared information consumption increased per minute (Fig.~\ref{fig:one}B and Supplementary material, Table S7).
More searching offset lower article consumption per search, slightly increasing article consumption per reader (Supplementary material, Table S25).

Research on conventional search distinguishes articles shown from articles opened \citep{robertson2023}.
AI answers change what a click represents by providing information before that choice.
Readers may click for detail or verification, or to fill gaps in an answer; no click may reflect satisfaction or disengagement.
Publishers may therefore need to reassess how clicks and time on site measure engagement and predict customer lifetime value.
Future studies should link answer use, article reading, and continued searching to satisfaction, return visits, and subscription renewal, and assess economic value after accounting for changes in revenue and the cost of providing answers.

Providing AI answers extends publishers' editorial responsibility to source selection, summarization, and attribution.
Broader topic coverage does not ensure accurate or unbiased summaries, which may omit qualifications, favor particular accounts, or obscure disagreement.
Publishers should assess fidelity to the reporting and how answer content and citation placement direct attention among source articles.
Our measures infer answer topics from cited articles; analyzing answer text and subsequent reading would distinguish which claims and topics are retained, emphasized, or omitted, which sources are prominent in citations, and which receive readership.
Experiments holding queries and source articles fixed could vary answer content and citation placement independently to test effects on article choice, continued searching, and comprehension, separating components introduced together in our treatment.

These effects may depend on who operates the search interface.
Here, additional searches occurred within one publisher's archive.
Third-party services such as Google AI Overviews and Perplexity draw on multiple publishers and may retain follow-up searches.
More searching may therefore increase use of the service without increasing publishers' readership, while broader topic consumption may coexist with concentrated source selection and referrals.
Future research should distinguish topic diversity from publisher diversity and examine how source selection, answer composition, and citations distribute attribution and readership, with longer-term effects on subscriptions, revenue, and incentives to produce original reporting.

Our results show that generative AI search can diversify collective attention while increasing shared information consumption.
\section*{Materials and Methods}
\subsection*{Setting and design}

We analyzed a randomized field experiment conducted by The Washington Post before the public launch of its RAG search feature.
Readers were assigned by browser cookie at their first on-site search between October 1 and November 7, 2024.
Control readers saw conventional search results.
For Treatment readers, an AI answer appeared above those results and cited up to five articles.
Both groups used the same article corpus and retrieval infrastructure.
Our analytic sample contains 37,561 readers, with 26,639 assigned to Control and 10,922 to Treatment (Supplementary material, section S19).

The publisher reports that assignment was independent of reader characteristics (Supplementary material, section S19).
The proportion assigned to Treatment varied by the day readers entered the experiment.
Differences in measured pre-search characteristics are small, although they are statistically significant when the characteristics are tested jointly (Supplementary material, section S20 and Table S28).
Adjusting for these characteristics and the week of each reader's first search yields estimates close to the original group differences (Supplementary material, section S21).
Our causal interpretation assumes that readers assigned to Control and Treatment on the same day are otherwise comparable after accounting for the measured pre-search characteristics.

\subsection*{Topic model}

We use BERTopic \citep{grootendorst2022} to represent each of 370,654 archived articles as a distribution over news topics. To keep routine service and publishing-related content outside the scope of the analysis, we exclude four of the 62 topics, covering news-production commentary, routine weather forecasts, non-article boilerplate, and sports-broadcast listings. We rescale each article's shares across the remaining 58 topics so that they sum to one and apply the same model to both Control and Treatment groups. Implementation details \citep{minilm_model,mcinnes2018umap,mcinnes2017hdbscan}, model-selection comparisons, and topic coverage are provided in Supplementary material, sections S22 and S23. Results from refitting the model with a different random seed are reported in section S11.

\subsection*{Measuring consumption}

We measure consumption from each reader's first search onward.
Article consumption sums the estimated topic shares from article opens, and total information consumption also includes those from displayed AI answers.
For each AI answer, we average its cited articles' topic distributions with equal weights, then exclude the same four topics and rescale the remaining shares to sum to one (Supplementary material, section S24).
Article opens and AI answer displays without topic estimates remain in the engagement records but do not contribute to topic-based consumption (Supplementary material, section S23).
These measures record information provided, without showing how fully it was read or understood.
We also compare word counts in opened articles and AI answers and time spent on article and search pages (Supplementary material, section S8).

For each group, search efficiency is total information consumption divided by the number of searches.
Consumption efficiency is total information consumption divided by recorded time on site in minutes (Supplementary material, section S25).

\subsection*{Shared information consumption}

Audience research distinguishes the reach of news outlets from overlap between their audiences \citep{fletcher2017}.
We apply this distinction to topics.
We rank topics by the number of distinct Control readers who open at least one article mainly about each topic.
The main topic of an article or AI answer is the topic with the largest estimated share.
The ranking includes reading throughout the study, October 1 to November 7, 2024, including before each reader's first search.
The top quartile, 14 of 58 topics, defines the shared core for both groups.
The quartile cutoff is an analytic choice, and the results persist with alternative cutoffs and a ranking based only on pre-search Control reading (Supplementary material, section S4).
Shared information consumption sums the estimated shares assigned to these 14 topics across opened articles and displayed AI answers.
Non-core information consumption sums the shares assigned to the remaining 44 topics.
We also measure the proportion of readers who open at least one article or receive at least one AI answer mainly about a topic in the shared core.
Both per-reader consumption and this proportion include all readers in each group's analytic sample (Supplementary material, section S26).

We also compare readers across all 58 topics to assess how similar their consumption is and how much information they have in common.
For each reader, we sum the estimated topic shares from opened articles and displayed AI answers.
We use cosine similarity to measure how similarly two readers distribute their consumption across topics.
We measure the amount of information they have in common by adding up the amount of consumption that overlaps on each topic.
For each group, we average both measures over all reader pairs, assigning zero to pairs in which either reader has no measured topic consumption.
These averages therefore reflect both the proportion of readers with measured topic consumption and the overlap between readers (Supplementary material, sections S6 and S26).

\subsection*{Topic concentration and popularity}

We measure topic concentration using the effective number of topics and the normalized Gini coefficient.
The effective number of topics is the exponential of Shannon entropy \citep{hill1973,jost2006}.
For example, consuming four topics equally gives an effective number of four, while uneven consumption across those same topics gives a lower value.
The normalized Gini coefficient ranges from zero for equal consumption across all topics to one for consumption on a single topic \citep{fleder2009,brynjolfsson2011}.
To measure topic popularity, we rank topics by their total article consumption in the Control group during the experimental period, including reading before each reader's first search.
Rank 1 denotes the most popular topic, and higher numerical ranks denote less-popular topics.
We use the same ranking for both groups.
Mean popularity rank averages these ranks, weighted by each topic's share of a reader's or group's consumption.
Top-10 popular share is the share of consumption on the ten most popular topics.
The ranking and a check using only pre-search Control reading are described in Supplementary material, section S28.

At the individual level, we calculate concentration and popularity for each reader and average these measures within each group. The individual analysis includes readers with at least two article opens or AI answer displays in total (Supplementary material, section S9). At the aggregate level, we combine all measured topic consumption within each group before calculating concentration and popularity. Results based on article consumption alone are reported in Supplementary material, section S13.

\subsection*{Search behavior}

We classify the next recorded action after each search as a cited-source article open, a conventional-result article open, browsing, a follow-up search, or session end (Fig.~\ref{fig:three}).
An article open is classified as cited-source if an AI answer in the same session cited that article at or before it was opened.
Browsing includes visits to the homepage and other non-article pages.
We calculate each share using all searches in the corresponding group (Supplementary material, section S27).
We also compare aggregate concentration and popularity between cited-source and other article consumption within Treatment.
This comparison includes all article opens after each reader's first search (Supplementary material, section S15).

\subsection*{Inference}

We report 95\% confidence intervals for each difference, calculated by resampling readers within each group (Supplementary material, section S30). We calculate \textit{P} values using permutation tests that shuffle Treatment and Control labels among readers who first searched on the same day, preserving that day's group sizes. To account for multiple comparisons, we use the Holm method \citep{holm1979} to adjust \textit{P} values separately for the eleven shared-information and engagement comparisons and the eight concentration and popularity comparisons in Table~\ref{tab:one}. Resampling counts and the treatment of topic rankings are described in Supplementary material, sections S8 and S29--S32.

The authors used Claude Fable 5 (Anthropic) and GPT-6 (OpenAI, via Codex) to assist with developing and reviewing the BERTopic topic modeling and the replication analyses, and with editing the text.
The authors verified all analyses.

\subsection*{Ethics}

We conducted a secondary analysis of de-identified reader-level records linked by anonymous cookie-based identifiers.
The records contain no direct identifiers, and we cannot readily identify individual readers.
We were not involved in the design or operation of the publisher's experiment.
The Institutional Review Board of The University of Texas at Dallas reviewed the study and determined it exempt under a limited IRB review (IRB-26-577).

\section*{Author contributions}
D.K.L., H.A.L., G.H.L., and D.W.L. designed research; H.A.L., D.K.L., G.H.L., and D.W.L. performed research; G.H.L. obtained data; H.A.L. analyzed data; and H.A.L., D.K.L., G.H.L., and D.W.L. wrote the paper.

\section*{Competing interests}
The Washington Post designed and fielded the randomized test in its product and provided the resulting de-identified records under a data use agreement. The analyses, their interpretation, and the manuscript are the authors' own, and the publisher's pre-publication review is limited under the agreement to identifying proprietary information. No author is employed by or holds a financial interest in The Washington Post, and the authors declare no other competing interests.

\section*{Data, materials, and software availability}
The Washington Post provided the de-identified reader-level clickstream data under a Data Use and Research Collaboration Agreement with WP Company LLC executed in 2024.
Under this agreement, the reader-level records cannot be shared publicly.
We plan to deposit the analysis code in a public repository.
The code will reproduce the reported figures and tables from the analytic dataset.
De-identified summary data underlying the reported figures and tables can be provided alongside the code, subject to agreement with The Washington Post.

\section*{Acknowledgments}
We are grateful to The Washington Post team for fielding the experiment and for their support of this research collaboration. This work received no external funding.

\balance
\bibliographystyle{plainnat}
\bibliography{references}

@article{lewis2020rag,
  author = {Lewis, P. and others},
  title = {{Retrieval-augmented generation for knowledge-intensive NLP tasks}},
  journal = {Adv. Neural Inf. Process. Syst.},
  volume = {33},
  pages = {9459--9474},
  year = {2020},
  note = {\href{https://papers.neurips.cc/paper/2020/hash/6b493230205f780e1bc26945df7481e5-Abstract.html}{Source}}
}

@book{egan2026report,
  author = {Egan, J. and others},
  title = {{Reuters Institute Digital News Report 2026}},
  publisher = {Reuters Institute for the Study of Journalism, University of Oxford},
  year = {2026},
  note = {\href{https://reutersinstitute.politics.ox.ac.uk/digital-news-report/2026}{Source} Accessed 8 September 2026.}
}

@misc{perplexity2026,
  author = {{Perplexity}},
  title = {{Practical tips for using Perplexity}},
  year = {2026},
  note = {\href{https://www.perplexity.ai/help-center/en/articles/10352971-practical-tips-for-using-perplexity}{Source} Accessed 9 September 2026.}
}

@misc{wapo2024pr,
  author = {{The Washington Post}},
  title = {{The Washington Post launches ``Ask The Post AI,'' a new search experience}},
  year = {2024},
  month = nov,
  day = {7},
  note = {Published 7 November 2024. \href{https://www.washingtonpost.com/pr/2024/11/07/washington-post-launches-ask-post-ai-new-search-experience/}{Source} Accessed 8 September 2026.}
}

@book{pariser2011,
  author = {Pariser, E.},
  title = {{The Filter Bubble: What the Internet Is Hiding from You}},
  publisher = {Penguin Press},
  address = {New York},
  year = {2011},
  note = {\href{https://www.penguinrandomhouse.com/books/309214/the-filter-bubble-by-eli-pariser/9781101515129/}{Source}}
}

@article{flaxman2016,
  author = {Flaxman, S. and Goel, S. and Rao, J. M.},
  title = {{Filter bubbles, echo chambers, and online news consumption}},
  journal = {Public Opin. Q.},
  volume = {80},
  pages = {298--320},
  year = {2016},
  note = {\href{https://doi.org/10.1093/poq/nfw006}{doi:10.1093/poq/nfw006}}
}

@book{sunstein2017,
  author = {Sunstein, C. R.},
  title = {{\#Republic: Divided Democracy in the Age of Social Media}},
  publisher = {Princeton University Press},
  year = {2017},
  note = {\href{https://books.google.com/books/about/Republic.html?id=gAZpDQAAQBAJ}{Source}}
}

@article{mccombs1972,
  author = {McCombs, M. E. and Shaw, D. L.},
  title = {{The agenda-setting function of mass media}},
  journal = {Public Opin. Q.},
  volume = {36},
  pages = {176--187},
  year = {1972},
  note = {\href{https://doi.org/10.1086/267990}{doi:10.1086/267990}}
}

@article{gonzalez2023,
  author = {Gonz{\'a}lez-Bail{\'o}n, S. and others},
  title = {{Asymmetric ideological segregation in exposure to political news on Facebook}},
  journal = {Science},
  volume = {381},
  pages = {392--398},
  year = {2023},
  note = {\href{https://doi.org/10.1126/science.ade7138}{doi:10.1126/science.ade7138}}
}

@article{cinelli2021,
  author = {Cinelli, M. and De Francisci Morales, G. and Galeazzi, A. and Quattrociocchi, W. and Starnini, M.},
  title = {{The echo chamber effect on social media}},
  journal = {Proc. Natl. Acad. Sci. U.S.A.},
  volume = {118},
  pages = {e2023301118},
  year = {2021},
  note = {\href{https://doi.org/10.1073/pnas.2023301118}{doi:10.1073/pnas.2023301118}}
}

@article{hosanagar2014,
  author = {Hosanagar, K. and Fleder, D. and Lee, D. and Buja, A.},
  title = {{Will the global village fracture into tribes? Recommender systems and their effects on consumer fragmentation}},
  journal = {Manage. Sci.},
  volume = {60},
  pages = {805--823},
  year = {2014},
  note = {\href{https://doi.org/10.1287/mnsc.2013.1808}{doi:10.1287/mnsc.2013.1808}}
}

@article{lee2019,
  author = {Lee, D. and Hosanagar, K.},
  title = {{How do recommender systems affect sales diversity? A cross-category investigation via randomized field experiment}},
  journal = {Inf. Syst. Res.},
  volume = {30},
  pages = {239--259},
  year = {2019},
  note = {\href{https://doi.org/10.1287/isre.2018.0800}{doi:10.1287/isre.2018.0800}}
}

@article{robertson2023,
  author = {Robertson, R. E. and Green, J. and Ruck, D. J. and Ognyanova, K. and Wilson, C. and Lazer, D.},
  title = {{Users choose to engage with more partisan news than they are exposed to on Google Search}},
  journal = {Nature},
  volume = {618},
  pages = {342--348},
  year = {2023},
  note = {\href{https://doi.org/10.1038/s41586-023-06078-5}{doi:10.1038/s41586-023-06078-5}}
}

@misc{grootendorst2022,
  author = {Grootendorst, M.},
  title = {{BERTopic: Neural topic modeling with a class-based TF-IDF procedure}},
  howpublished = {arXiv:2203.05794},
  year = {2022},
  note = {\href{https://doi.org/10.48550/arXiv.2203.05794}{doi:10.48550/arXiv.2203.05794}}
}

@misc{minilm_model,
  author = {{Sentence Transformers}},
  title = {{all-MiniLM-L6-v2}},
  howpublished = {Hugging Face},
  year = {n.d.},
  note = {\href{https://huggingface.co/sentence-transformers/all-MiniLM-L6-v2}{Source} Accessed 8 September 2026.}
}

@misc{mcinnes2018umap,
  author = {McInnes, L. and Healy, J. and Melville, J.},
  title = {{UMAP: Uniform manifold approximation and projection for dimension reduction}},
  howpublished = {arXiv:1802.03426},
  year = {2018},
  note = {\href{https://doi.org/10.48550/arXiv.1802.03426}{doi:10.48550/arXiv.1802.03426}}
}

@article{mcinnes2017hdbscan,
  author = {McInnes, L. and Healy, J. and Astels, S.},
  title = {{hdbscan: Hierarchical density based clustering}},
  journal = {J. Open Source Softw.},
  volume = {2},
  pages = {205},
  year = {2017},
  note = {\href{https://doi.org/10.21105/joss.00205}{doi:10.21105/joss.00205}}
}

@article{fleder2009,
  author = {Fleder, D. and Hosanagar, K.},
  title = {{Blockbuster culture's next rise or fall: The impact of recommender systems on sales diversity}},
  journal = {Manage. Sci.},
  volume = {55},
  pages = {697--712},
  year = {2009},
  note = {\href{https://doi.org/10.1287/mnsc.1080.0974}{doi:10.1287/mnsc.1080.0974}}
}

@article{brynjolfsson2011,
  author = {Brynjolfsson, E. and Hu, Y. and Simester, D.},
  title = {{Goodbye Pareto principle, hello long tail: The effect of search costs on the concentration of product sales}},
  journal = {Manage. Sci.},
  volume = {57},
  pages = {1373--1386},
  year = {2011},
  note = {\href{https://doi.org/10.1287/mnsc.1110.1371}{doi:10.1287/mnsc.1110.1371}}
}

@article{hill1973,
  author = {Hill, M. O.},
  title = {{Diversity and evenness: A unifying notation and its consequences}},
  journal = {Ecology},
  volume = {54},
  pages = {427--432},
  year = {1973},
  note = {\href{https://doi.org/10.2307/1934352}{doi:10.2307/1934352}}
}

@article{jost2006,
  author = {Jost, L.},
  title = {{Entropy and diversity}},
  journal = {Oikos},
  volume = {113},
  pages = {363--375},
  year = {2006},
  note = {\href{https://doi.org/10.1111/j.2006.0030-1299.14714.x}{doi:10.1111/j.2006.0030-1299.14714.x}}
}

@article{holm1979,
  author = {Holm, S.},
  title = {{A simple sequentially rejective multiple test procedure}},
  journal = {Scand. J. Stat.},
  volume = {6},
  pages = {65--70},
  year = {1979},
  note = {\href{https://www.jstor.org/stable/4615733}{Source}}
}

@article{fletcher2017,
  author = {Fletcher, Richard and Nielsen, Rasmus Kleis},
  title = {{Are News Audiences Increasingly Fragmented? A Cross-National Comparative Analysis of Cross-Platform News Audience Fragmentation and Duplication}},
  journal = {Journal of Communication},
  volume = {67},
  number = {4},
  pages = {476--498},
  year = {2017},
  doi = {10.1111/jcom.12315}
}
\clearpage\onecolumn
% Complete Supplementary material converted from R123_SI_v28_260929.docx.
% Requires array, booktabs, longtable, caption, graphicx, hyperref, needspace, colortbl/xcolor, natbib.
\clearpage
\section*{Supplementary material}
\phantomsection\label{sec:supplementary}
\addcontentsline{toc}{section}{Supplementary material}
\setcounter{table}{0}\renewcommand{\thetable}{S\arabic{table}}
\setcounter{figure}{0}\renewcommand{\thefigure}{S\arabic{figure}}
\renewcommand{\theHtable}{supplement.\arabic{table}}
\renewcommand{\theHfigure}{supplement.\arabic{figure}}
\subsection*{Contents}
\begingroup\fontsize{9bp}{11.5pt}\selectfont
\par\medskip\noindent\textbf{Interfaces and Results}\par\smallskip
\noindent\hangindent=1.5em\hangafter=1 \hyperref[sec:S1]{S1. What did readers in each group see?}\nobreak\dotfill\pageref{sec:S1}\par\smallskip
\noindent\hangindent=1.5em\hangafter=1 \hyperref[sec:S2]{S2. Does RAG search increase shared and non-core consumption?}\nobreak\dotfill\pageref{sec:S2}\par\smallskip
\noindent\hangindent=1.5em\hangafter=1 \hyperref[sec:S3]{S3. Does shared information reach more readers?}\nobreak\dotfill\pageref{sec:S3}\par\smallskip
\noindent\hangindent=1.5em\hangafter=1 \hyperref[sec:S4]{S4. Does the increase depend on how shared information is defined, or on political topics?}\nobreak\dotfill\pageref{sec:S4}\par\smallskip
\noindent\hangindent=1.5em\hangafter=1 \hyperref[sec:S5]{S5. How much of the increase comes from AI answers?}\nobreak\dotfill\pageref{sec:S5}\par\smallskip
\noindent\hangindent=1.5em\hangafter=1 \hyperref[sec:S6]{S6. Do pairs of readers share more information?}\nobreak\dotfill\pageref{sec:S6}\par\smallskip
\noindent\hangindent=1.5em\hangafter=1 \hyperref[sec:S7]{S7. Does consumption per search and per minute increase?}\nobreak\dotfill\pageref{sec:S7}\par\smallskip
\noindent\hangindent=1.5em\hangafter=1 \hyperref[sec:S8]{S8. Does higher consumption also mean more text and more time spent with news?}\nobreak\dotfill\pageref{sec:S8}\par\smallskip
\noindent\hangindent=1.5em\hangafter=1 \hyperref[sec:S9]{S9. Do the individual-level results depend on consumption volume or who is included?}\nobreak\dotfill\pageref{sec:S9}\par\smallskip
\noindent\hangindent=1.5em\hangafter=1 \hyperref[sec:S10]{S10. Do the concentration and popularity results hold without political topics?}\nobreak\dotfill\pageref{sec:S10}\par\smallskip
\noindent\hangindent=1.5em\hangafter=1 \hyperref[sec:S11]{S11. Do the results hold under a different topic model?}\nobreak\dotfill\pageref{sec:S11}\par\smallskip
\noindent\hangindent=1.5em\hangafter=1 \hyperref[sec:S12]{S12. Do the results depend on how answers are measured or weighted?}\nobreak\dotfill\pageref{sec:S12}\par\smallskip
\noindent\hangindent=1.5em\hangafter=1 \hyperref[sec:S13]{S13. What changes when only opened articles are counted?}\nobreak\dotfill\pageref{sec:S13}\par\smallskip
\noindent\hangindent=1.5em\hangafter=1 \hyperref[sec:S14]{S14. How do AI answers broaden consumption and add less-popular topics?}\nobreak\dotfill\pageref{sec:S14}\par\smallskip
\noindent\hangindent=1.5em\hangafter=1 \hyperref[sec:S15]{S15. How does cited-source article consumption differ from other article consumption?}\nobreak\dotfill\pageref{sec:S15}\par\smallskip
\noindent\hangindent=1.5em\hangafter=1 \hyperref[sec:S16]{S16. Do readers ask about, or does search offer, less-popular topics?}\nobreak\dotfill\pageref{sec:S16}\par\smallskip
\noindent\hangindent=1.5em\hangafter=1 \hyperref[sec:S17]{S17. What do readers do next after a search?}\nobreak\dotfill\pageref{sec:S17}\par\smallskip
\noindent\hangindent=1.5em\hangafter=1 \hyperref[sec:S18]{S18. How do more searches add up to more consumption?}\nobreak\dotfill\pageref{sec:S18}\par\smallskip
\par\medskip\noindent\textbf{Experimental Design, Measurement, and Inference}\par\smallskip
\noindent\hangindent=1.5em\hangafter=1 \hyperref[sec:S19]{S19. How were readers assigned, and who is in the analytic sample?}\nobreak\dotfill\pageref{sec:S19}\par\smallskip
\noindent\hangindent=1.5em\hangafter=1 \hyperref[sec:S20]{S20. Were the two groups alike before the first search?}\nobreak\dotfill\pageref{sec:S20}\par\smallskip
\noindent\hangindent=1.5em\hangafter=1 \hyperref[sec:S21]{S21. Do the results hold after adjusting for reader characteristics and entry timing?}\nobreak\dotfill\pageref{sec:S21}\par\smallskip
\noindent\hangindent=1.5em\hangafter=1 \hyperref[sec:S22]{S22. How was the topic model chosen and curated?}\nobreak\dotfill\pageref{sec:S22}\par\smallskip
\noindent\hangindent=1.5em\hangafter=1 \hyperref[sec:S23]{S23. How much of the observed consumption has topic information?}\nobreak\dotfill\pageref{sec:S23}\par\smallskip
\noindent\hangindent=1.5em\hangafter=1 \hyperref[sec:S24]{S24. How are article consumption and answer topics measured?}\nobreak\dotfill\pageref{sec:S24}\par\smallskip
\noindent\hangindent=1.5em\hangafter=1 \hyperref[sec:S25]{S25. How are time on site and efficiency measured?}\nobreak\dotfill\pageref{sec:S25}\par\smallskip
\noindent\hangindent=1.5em\hangafter=1 \hyperref[sec:S26]{S26. How are shared information and reader-pair comparisons defined?}\nobreak\dotfill\pageref{sec:S26}\par\smallskip
\noindent\hangindent=1.5em\hangafter=1 \hyperref[sec:S27]{S27. How are search windows and cited-source article opens defined?}\nobreak\dotfill\pageref{sec:S27}\par\smallskip
\noindent\hangindent=1.5em\hangafter=1 \hyperref[sec:S28]{S28. How are concentration and popularity measured?}\nobreak\dotfill\pageref{sec:S28}\par\smallskip
\noindent\hangindent=1.5em\hangafter=1 \hyperref[sec:S29]{S29. How are the shared-information tests adjusted for multiple comparisons?}\nobreak\dotfill\pageref{sec:S29}\par\smallskip
\noindent\hangindent=1.5em\hangafter=1 \hyperref[sec:S30]{S30. How are intervals and permutation tests computed?}\nobreak\dotfill\pageref{sec:S30}\par\smallskip
\noindent\hangindent=1.5em\hangafter=1 \hyperref[sec:S31]{S31. How are the concentration and popularity tests adjusted?}\nobreak\dotfill\pageref{sec:S31}\par\smallskip
\noindent\hangindent=1.5em\hangafter=1 \hyperref[sec:S32]{S32. How are the supplementary comparisons tested?}\nobreak\dotfill\pageref{sec:S32}\par\smallskip
\noindent\hangindent=1.5em\hangafter=1 \hyperref[sec:S33]{S33. What are the 58 topics and their labels?}\nobreak\dotfill\pageref{sec:S33}\par\smallskip
\endgroup
\clearpage

\Needspace{6\baselineskip}
\subsection*{S1. What did readers in each group see?}
\phantomsection\label{sec:S1}

Readers in both groups searched the same news archive and saw conventional search results. The Treatment interface differed only by adding an automatically displayed AI answer and links to up to five cited articles above those results (Fig. S1).

\begin{figure}[htbp]
\centering
\includegraphics[width=\linewidth]{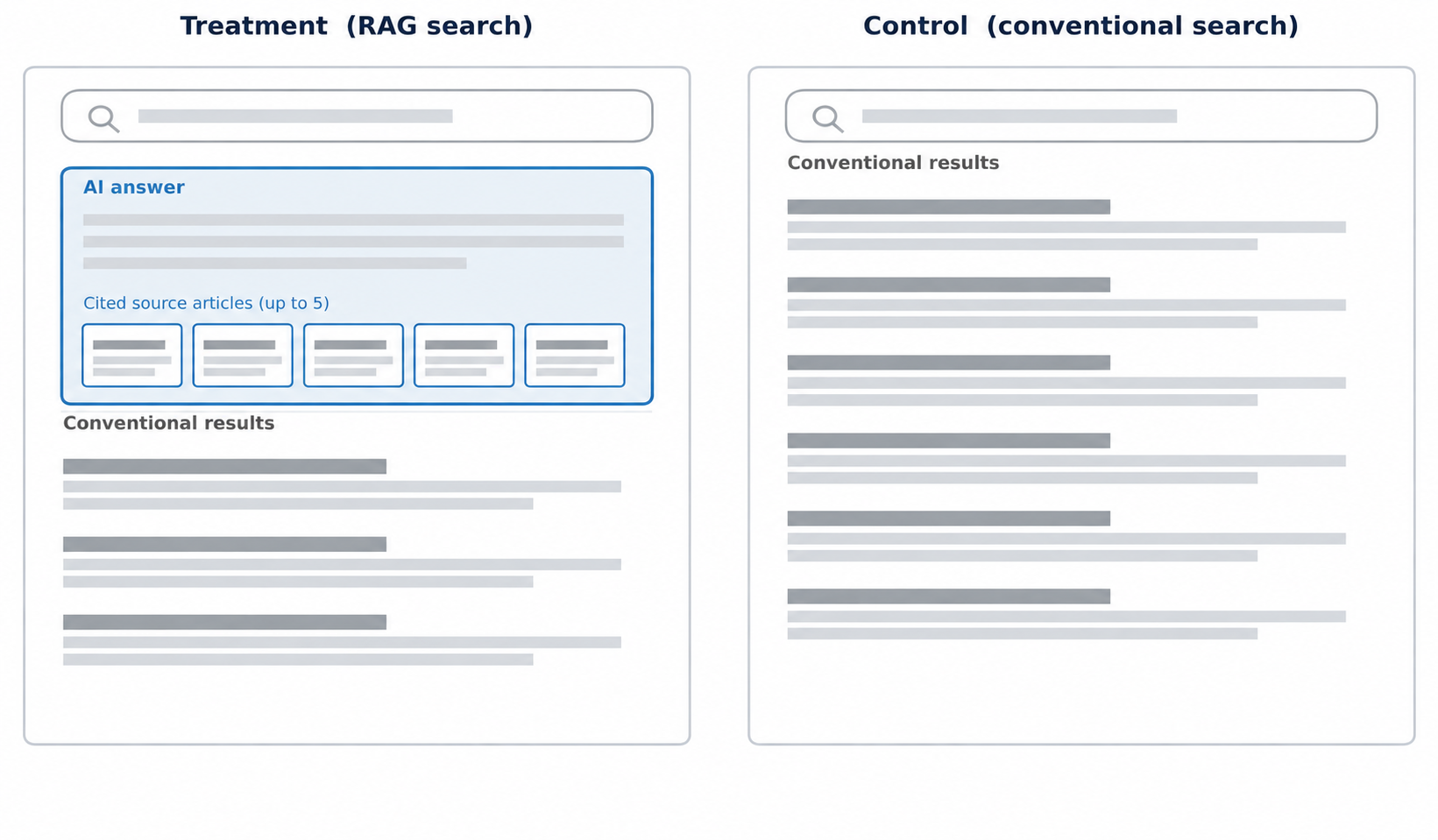}
\caption{Schematic of the experimental search interfaces. Treatment adds an AI answer and its cited-source articles above the conventional results shown alone in Control. This schematic illustrates the layout without reproducing the copyrighted Washington Post article content displayed on actual results pages.}
\label{fig:S1}
\end{figure}

\section*{Support for the Results}

\Needspace{6\baselineskip}
\subsection*{S2. Does RAG search increase shared and non-core consumption?}
\phantomsection\label{sec:S2}

RAG search increases both shared and non-core information consumption (Table S1). Shared information covers the 14 topics read by the largest shares of Control article readers. Non-core information covers the remaining topics (section S26). Consumption adds the portions of articles and answers devoted to each set. For example, an item that is 60\% about shared topics contributes 0.6 to shared consumption. Shared consumption rises from 0.676 to 1.144 per assigned reader, and non-core consumption from 0.687 to 1.339 (both Holm-adjusted P \ensuremath{<} 0.001).

\Needspace{12\baselineskip}
\setcounter{table}{0}
\captionof{table}{Shared and non-core information consumption}\label{tab:S1}

% SI physical table block 1; authoritative Table S1.
\begingroup
\fontsize{9bp}{11.5pt}\selectfont
\setlength{\tabcolsep}{3pt}
\renewcommand{\arraystretch}{1.13}
\setlength{\LTleft}{0pt}\setlength{\LTright}{0pt}
\setlength{\LTpre}{4pt}\setlength{\LTpost}{5pt}
\renewcommand{\theHtable}{supplement.block1.\arabic{table}}
\begin{longtable}{@{}>{\raggedright\arraybackslash}p{\dimexpr0.3230201\linewidth-2\tabcolsep\relax}>{\raggedright\arraybackslash}p{\dimexpr0.1077086\linewidth-2\tabcolsep\relax}>{\raggedright\arraybackslash}p{\dimexpr0.1307286\linewidth-2\tabcolsep\relax}>{\raggedright\arraybackslash}p{\dimexpr0.1000000\linewidth-2\tabcolsep\relax}>{\raggedright\arraybackslash}p{\dimexpr0.1616684\linewidth-2\tabcolsep\relax}>{\raggedright\arraybackslash}p{\dimexpr0.1768743\linewidth-2\tabcolsep\relax}@{}}
\toprule
\textbf{\textbf{Quantity}} & \textbf{\textbf{Control}} & \textbf{\textbf{Treatment}} & \textbf{\textbf{Difference}} & \textbf{\textbf{95\% CI}} & \textbf{\textbf{Holm-adjusted P}} \\
\midrule
\endfirsthead
\multicolumn{6}{@{}l@{}}{\textbf{Table S1 (continued)}}\\
\toprule
\textbf{\textbf{Quantity}} & \textbf{\textbf{Control}} & \textbf{\textbf{Treatment}} & \textbf{\textbf{Difference}} & \textbf{\textbf{95\% CI}} & \textbf{\textbf{Holm-adjusted P}} \\
\midrule
\endhead
\midrule
\multicolumn{6}{@{}r@{}}{Continued on next page}\\
\endfoot
\bottomrule
\endlastfoot
Shared information consumption & 0.676 & 1.144 & +0.469 & [0.435, 0.504] & \ensuremath{<} 0.001 \\
Non-core information consumption & 0.687 & 1.339 & +0.652 & [0.611, 0.693] & \ensuremath{<} 0.001 \\
Total information consumption & 1.363 & 2.483 & +1.120 & [1.058, 1.184] & \ensuremath{<} 0.001 \\
\end{longtable}
\endgroup

{\fontsize{8bp}{11pt}\selectfont\interlinepenalty=10000\noindent \textbf{Note.} Means include all 37,561 readers in the analytic sample, including those with no measured topic consumption. The 95\% confidence intervals (CIs) describe uncertainty for each estimate separately. \textit{P} values use Holm adjustment across 11 comparisons to account for multiple testing (sections S29 and S30). Section S24 explains how articles and answers contribute to topic consumption.\par}
\medskip

\Needspace{6\baselineskip}
\subsection*{S3. Does shared information reach more readers?}
\phantomsection\label{sec:S3}

Including AI answers raises the share of readers reached by shared information from 36.9\% to 55.4\% (Table S2). Counting articles alone gives 36.9\% in Control and 35.1\% in Treatment. The increase in reach therefore comes from including answers, rather than from more readers opening articles on shared topics.

\Needspace{12\baselineskip}
\setcounter{table}{1}
\captionof{table}{Readers reached by shared information}\label{tab:S2}

% SI physical table block 2; authoritative Table S2.
\begingroup
\fontsize{9bp}{11.5pt}\selectfont
\setlength{\tabcolsep}{3pt}
\renewcommand{\arraystretch}{1.13}
\setlength{\LTleft}{0pt}\setlength{\LTright}{0pt}
\setlength{\LTpre}{4pt}\setlength{\LTpost}{5pt}
\renewcommand{\theHtable}{supplement.block2.\arabic{table}}
\begin{longtable}{@{}>{\raggedright\arraybackslash}p{\dimexpr0.3230201\linewidth-2\tabcolsep\relax}>{\raggedright\arraybackslash}p{\dimexpr0.1077086\linewidth-2\tabcolsep\relax}>{\raggedright\arraybackslash}p{\dimexpr0.1307286\linewidth-2\tabcolsep\relax}>{\raggedright\arraybackslash}p{\dimexpr0.1000000\linewidth-2\tabcolsep\relax}>{\raggedright\arraybackslash}p{\dimexpr0.2154171\linewidth-2\tabcolsep\relax}>{\raggedright\arraybackslash}p{\dimexpr0.1231257\linewidth-2\tabcolsep\relax}@{}}
\toprule
\textbf{\textbf{Quantity}} & \textbf{\textbf{Control}} & \textbf{\textbf{Treatment}} & \textbf{\textbf{Difference}} & \textbf{\textbf{95\% CI}} & \textbf{\textbf{P}} \\
\midrule
\endfirsthead
\multicolumn{6}{@{}l@{}}{\textbf{Table S2 (continued)}}\\
\toprule
\textbf{\textbf{Quantity}} & \textbf{\textbf{Control}} & \textbf{\textbf{Treatment}} & \textbf{\textbf{Difference}} & \textbf{\textbf{95\% CI}} & \textbf{\textbf{P}} \\
\midrule
\endhead
\midrule
\multicolumn{6}{@{}r@{}}{Continued on next page}\\
\endfoot
\bottomrule
\endlastfoot
Readers consuming shared information (articles and answers) & 36.9\% & 55.4\% & +18.5 pp & [17.4, 19.6] & \ensuremath{<} 0.001 \\
Readers consuming shared information (articles only) & 36.9\% & 35.1\% & \ensuremath{-}1.8 pp & [\ensuremath{-}2.9, \ensuremath{-}0.8] & 0.0015 \\
\end{longtable}
\endgroup

{\fontsize{8bp}{11pt}\selectfont\interlinepenalty=10000\noindent \textbf{Note.} Reach is the share consuming an article or answer mainly about a shared topic (section S26), among all 26,639 Control and 10,922 Treatment readers. Differences and CIs are percentage points. The articles-and-answers P value uses the 11-comparison Holm adjustment. The article-only P value is unadjusted and uses 10,000 permutations within first-search day. CIs resample readers (sections S29 and S30).\par}
\medskip

\Needspace{6\baselineskip}
\subsection*{S4. Does the increase depend on how shared information is defined, or on political topics?}
\phantomsection\label{sec:S4}

Shared information consumption and reach increase when shared information is defined using the top 10, 14 or 18 topics, or using only Control reading before the first search (all P \ensuremath{<} 0.001, Table S3).

The top-quartile cutoff is an analytic choice. The top 10 and top 18 alternatives test whether the result depends on placing the boundary at 14 topics. Ranking topics using only pre-search Control reading tests a definition formed before exposure to the search interface.

\Needspace{12\baselineskip}
\setcounter{table}{2}
\captionof{table}{Shared information under alternative definitions}\label{tab:S3}

\Needspace{10\baselineskip}
\noindent\textbf{\textbf{A. }\textbf{Shared i}\textbf{nformation consumption per assigned reader}}\par
\smallskip

% SI physical table block 3; authoritative Table S3.
\begingroup
\fontsize{9bp}{11.5pt}\selectfont
\setlength{\tabcolsep}{3pt}
\renewcommand{\arraystretch}{1.13}
\setlength{\LTleft}{0pt}\setlength{\LTright}{0pt}
\setlength{\LTpre}{4pt}\setlength{\LTpost}{5pt}
\renewcommand{\theHtable}{supplement.block3.\arabic{table}}
\begin{longtable}{@{}>{\raggedright\arraybackslash}p{\dimexpr0.3076923\linewidth-2\tabcolsep\relax}>{\raggedright\arraybackslash}p{\dimexpr0.1538462\linewidth-2\tabcolsep\relax}>{\raggedright\arraybackslash}p{\dimexpr0.1538462\linewidth-2\tabcolsep\relax}>{\raggedright\arraybackslash}p{\dimexpr0.3846154\linewidth-2\tabcolsep\relax}@{}}
\toprule
\textbf{\textbf{Definition}} & \textbf{\textbf{Control}} & \textbf{\textbf{Treatment}} & \textbf{\textbf{Difference [95\% CI]}} \\
\midrule
\endfirsthead
\multicolumn{4}{@{}l@{}}{\textbf{Table S3 (continued)}}\\
\toprule
\textbf{\textbf{Definition}} & \textbf{\textbf{Control}} & \textbf{\textbf{Treatment}} & \textbf{\textbf{Difference [95\% CI]}} \\
\midrule
\endhead
\midrule
\multicolumn{4}{@{}r@{}}{Continued on next page}\\
\endfoot
\bottomrule
\endlastfoot
Top 10 & 0.555 & 0.935 & +0.380 [0.351, 0.411] \\
Top 14 & 0.676 & 1.144 & +0.469 [0.434, 0.504] \\
Top 18 & 0.816 & 1.387 & +0.571 [0.530, 0.611] \\
Before first search & 0.675 & 1.142 & +0.468 [0.431, 0.504] \\
\end{longtable}
\endgroup

\Needspace{10\baselineskip}
\noindent\textbf{\textbf{B. Readers reached by shared information}}\par
\smallskip

% SI physical table block 4; authoritative Table S3.
\begingroup
\fontsize{9bp}{11.5pt}\selectfont
\setlength{\tabcolsep}{3pt}
\renewcommand{\arraystretch}{1.13}
\setlength{\LTleft}{0pt}\setlength{\LTright}{0pt}
\setlength{\LTpre}{4pt}\setlength{\LTpost}{5pt}
\renewcommand{\theHtable}{supplement.block4.\arabic{table}}
\begin{longtable}{@{}>{\raggedright\arraybackslash}p{\dimexpr0.3076923\linewidth-2\tabcolsep\relax}>{\raggedright\arraybackslash}p{\dimexpr0.1538462\linewidth-2\tabcolsep\relax}>{\raggedright\arraybackslash}p{\dimexpr0.1538462\linewidth-2\tabcolsep\relax}>{\raggedright\arraybackslash}p{\dimexpr0.3846154\linewidth-2\tabcolsep\relax}@{}}
\toprule
\textbf{\textbf{Definition}} & \textbf{\textbf{Control}} & \textbf{\textbf{Treatment}} & \textbf{\textbf{Difference [95\% CI]}} \\
\midrule
\endfirsthead
\multicolumn{4}{@{}l@{}}{\textbf{Table S3 (continued)}}\\
\toprule
\textbf{\textbf{Definition}} & \textbf{\textbf{Control}} & \textbf{\textbf{Treatment}} & \textbf{\textbf{Difference [95\% CI]}} \\
\midrule
\endhead
\midrule
\multicolumn{4}{@{}r@{}}{Continued on next page}\\
\endfoot
\bottomrule
\endlastfoot
Top 10 & 31.4\% & 48.3\% & +16.9 pp [15.8, 18.0] \\
Top 14 & 36.9\% & 55.4\% & +18.5 pp [17.4, 19.6] \\
Top 18 & 41.9\% & 62.5\% & +20.7 pp [19.6, 21.8] \\
Before first search & 35.7\% & 54.8\% & +19.1 pp [17.9, 20.2] \\
\end{longtable}
\endgroup

{\fontsize{8bp}{11pt}\selectfont\interlinepenalty=10000\noindent \textbf{Note.} Panel A reports consumption per assigned reader. Panel B reports the percentage of assigned readers reached, with differences and CIs in percentage points. All differences have unadjusted P \ensuremath{<} 0.001. CIs resample readers, and permutation tests reassign labels within first-search day (sections S29 and S30).\par}
\medskip

Shared information consumption also increases after political topics are excluded (Table S4). This holds whether we keep the remaining shared topics or define shared information again as the top quartile of the topics left.

The narrow exclusion removes U.S. National Politics. The broad exclusion adds Abortion \& Reproductive Rights, Immigration \& Borders, Supreme Court \& Judicial Appointments, D.C. Public Affairs, Virginia Politics, Maryland Politics and Parliamentary Politics. The broadest also removes Gun Violence and Israel \& Palestine. These names match Table S35. We subtract only consumption on the excluded topics, leaving other consumption unchanged (section S26).

\Needspace{12\baselineskip}
\setcounter{table}{3}
\captionof{table}{Differences after excluding political topics}\label{tab:S4}

% SI physical table block 5; authoritative Table S4.
\begingroup
\fontsize{9bp}{11.5pt}\selectfont
\setlength{\tabcolsep}{3pt}
\renewcommand{\arraystretch}{1.13}
\setlength{\LTleft}{0pt}\setlength{\LTright}{0pt}
\setlength{\LTpre}{4pt}\setlength{\LTpost}{5pt}
\renewcommand{\theHtable}{supplement.block5.\arabic{table}}
\begin{longtable}{@{}>{\raggedright\arraybackslash}p{\dimexpr0.2000000\linewidth-2\tabcolsep\relax}>{\raggedright\arraybackslash}p{\dimexpr0.2538543\linewidth-2\tabcolsep\relax}>{\raggedright\arraybackslash}p{\dimexpr0.2538543\linewidth-2\tabcolsep\relax}>{\raggedright\arraybackslash}p{\dimexpr0.2922914\linewidth-2\tabcolsep\relax}@{}}
\toprule
\textbf{\textbf{Exclusion and shared information set}} & \textbf{\textbf{Shared information consumption difference}\newline{}\textbf{[95\% CI]}} & \textbf{\textbf{Non-core information consumption difference}\newline{}\textbf{[95\% CI]}} & \textbf{\textbf{Cosine similarity difference}\newline{}\textbf{[95\% CI]}} \\
\midrule
\endfirsthead
\multicolumn{4}{@{}l@{}}{\textbf{Table S4 (continued)}}\\
\toprule
\textbf{\textbf{Exclusion and shared information set}} & \textbf{\textbf{Shared information consumption difference}\newline{}\textbf{[95\% CI]}} & \textbf{\textbf{Non-core information consumption difference}\newline{}\textbf{[95\% CI]}} & \textbf{\textbf{Cosine similarity difference}\newline{}\textbf{[95\% CI]}} \\
\midrule
\endhead
\midrule
\multicolumn{4}{@{}r@{}}{Continued on next page}\\
\endfoot
\bottomrule
\endlastfoot
narrow / original set & +0.434 [0.402, 0.466] & +0.652 [0.612, 0.693] & +0.0419 [0.0405, 0.0434] \\
narrow / new set & +0.465 [0.432, 0.499] & +0.620 [0.583, 0.659] & +0.0419 [0.0404, 0.0434] \\
broad / original set & +0.314 [0.288, 0.340] & +0.589 [0.552, 0.628] & +0.0433 [0.0419, 0.0447] \\
broad / new set & +0.394 [0.365, 0.423] & +0.509 [0.474, 0.545] & +0.0433 [0.0418, 0.0447] \\
broadest / original set & +0.250 [0.227, 0.274] & +0.567 [0.531, 0.605] & +0.0437 [0.0423, 0.0451] \\
broadest / new set & +0.353 [0.326, 0.381] & +0.464 [0.432, 0.498] & +0.0437 [0.0424, 0.0451] \\
\end{longtable}
\endgroup

{\fontsize{8bp}{11pt}\selectfont\interlinepenalty=10000\noindent \textbf{Note.} The narrow, broad and broadest exclusions remove 1, 8 and 10 topics. Differences are Treatment minus Control, with reader-bootstrap CIs (section S30). All reported differences have unadjusted P \ensuremath{<} 0.001. The cosine comparison covers all reader pairs and topics remaining after each exclusion.\par}
\medskip

\Needspace{6\baselineskip}
\subsection*{S5. How much of the increase comes from AI answers?}
\phantomsection\label{sec:S5}

AI answers account for 98.0\% of the increase in shared information consumption and 89.2\% of the increase in non-core information consumption (Table S5).

\Needspace{12\baselineskip}
\setcounter{table}{4}
\captionof{table}{Article and answer contributions to information consumption}\label{tab:S5}

% SI physical table block 6; authoritative Table S5.
\begingroup
\fontsize{9bp}{11.5pt}\selectfont
\setlength{\tabcolsep}{3pt}
\renewcommand{\arraystretch}{1.13}
\setlength{\LTleft}{0pt}\setlength{\LTright}{0pt}
\setlength{\LTpre}{4pt}\setlength{\LTpost}{5pt}
\renewcommand{\theHtable}{supplement.block6.\arabic{table}}
\begin{longtable}{@{}>{\raggedright\arraybackslash}p{\dimexpr0.5307286\linewidth-2\tabcolsep\relax}>{\raggedright\arraybackslash}p{\dimexpr0.1538543\linewidth-2\tabcolsep\relax}>{\raggedright\arraybackslash}p{\dimexpr0.1538543\linewidth-2\tabcolsep\relax}>{\raggedright\arraybackslash}p{\dimexpr0.1615628\linewidth-2\tabcolsep\relax}@{}}
\toprule
\textbf{\textbf{Quantity}} & \textbf{\textbf{Control}} & \textbf{\textbf{Treatment}} & \textbf{\textbf{Difference}} \\
\midrule
\endfirsthead
\multicolumn{4}{@{}l@{}}{\textbf{Table S5 (continued)}}\\
\toprule
\textbf{\textbf{Quantity}} & \textbf{\textbf{Control}} & \textbf{\textbf{Treatment}} & \textbf{\textbf{Difference}} \\
\midrule
\endhead
\midrule
\multicolumn{4}{@{}r@{}}{Continued on next page}\\
\endfoot
\bottomrule
\endlastfoot
Shared information from articles & 0.676 & 0.685 & +0.010 \\
Shared information from AI answers & 0.000 & 0.459 & +0.459 \\
Non-core information from articles & 0.687 & 0.757 & +0.070 \\
Non-core information from AI answers & 0.000 & 0.582 & +0.582 \\
Total information from articles & 1.363 & 1.443 & +0.080 \\
Total information from AI answers & 0.000 & 1.041 & +1.041 \\
\end{longtable}
\endgroup

{\fontsize{8bp}{11pt}\selectfont\interlinepenalty=10000\noindent \textbf{Note.} Values are consumption per assigned reader. Answer contributions divide Treatment answer consumption by the corresponding total increase in Table S1. This separates the observed totals without isolating the causal effect of answers from the rest of the interface.\par}
\medskip

\Needspace{6\baselineskip}
\subsection*{S6. Do pairs of readers share more information?}
\phantomsection\label{sec:S6}

RAG search also increases overlap between readers across all topics, including topics outside the shared set (Table S6). Cosine similarity measures how similarly readers distribute consumption across topics. Shared information per reader pair measures the amount they have in common (section S26). Both increase (Holm-adjusted P \ensuremath{<} 0.001).

\Needspace{12\baselineskip}
\setcounter{table}{5}
\captionof{table}{Comparisons across all assigned reader pairs}\label{tab:S6}

% SI physical table block 7; authoritative Table S6.
\begingroup
\fontsize{9bp}{11.5pt}\selectfont
\setlength{\tabcolsep}{3pt}
\renewcommand{\arraystretch}{1.13}
\setlength{\LTleft}{0pt}\setlength{\LTright}{0pt}
\setlength{\LTpre}{4pt}\setlength{\LTpost}{5pt}
\renewcommand{\theHtable}{supplement.block7.\arabic{table}}
\begin{longtable}{@{}>{\raggedright\arraybackslash}p{\dimexpr0.3077086\linewidth-2\tabcolsep\relax}>{\raggedright\arraybackslash}p{\dimexpr0.1077086\linewidth-2\tabcolsep\relax}>{\raggedright\arraybackslash}p{\dimexpr0.1307286\linewidth-2\tabcolsep\relax}>{\raggedright\arraybackslash}p{\dimexpr0.1077086\linewidth-2\tabcolsep\relax}>{\raggedright\arraybackslash}p{\dimexpr0.1597677\linewidth-2\tabcolsep\relax}>{\raggedright\arraybackslash}p{\dimexpr0.1863780\linewidth-2\tabcolsep\relax}@{}}
\toprule
\textbf{\textbf{Quantity}} & \textbf{\textbf{Control}} & \textbf{\textbf{Treatment}} & \textbf{\textbf{Difference}} & \textbf{\textbf{95\% CI}} & \textbf{\textbf{Holm-adjusted P}} \\
\midrule
\endfirsthead
\multicolumn{6}{@{}l@{}}{\textbf{Table S6 (continued)}}\\
\toprule
\textbf{\textbf{Quantity}} & \textbf{\textbf{Control}} & \textbf{\textbf{Treatment}} & \textbf{\textbf{Difference}} & \textbf{\textbf{95\% CI}} & \textbf{\textbf{Holm-adjusted P}} \\
\midrule
\endhead
\midrule
\multicolumn{6}{@{}r@{}}{Continued on next page}\\
\endfoot
\bottomrule
\endlastfoot
Pairwise cosine similarity & 0.0256 & 0.0681 & +0.0425 & [0.0409, 0.0440] & \ensuremath{<} 0.001 \\
Shared information per reader pair,\newline{}all topics & 0.0454 & 0.1536 & +0.1082 & [0.1035, 0.1135] & \ensuremath{<} 0.001 \\
Shared information per reader pair,\newline{}shared topics & 0.0299 & 0.0897 & +0.0598 &  &  \\
Shared information per reader pair,\newline{}non-core topics & 0.0155 & 0.0639 & +0.0484 &  &  \\
\end{longtable}
\endgroup

{\fontsize{8bp}{11pt}\selectfont\interlinepenalty=10000\noindent \textbf{Note.} Both measures include articles and AI answers and average all assigned reader pairs. CIs resample readers, and P values use the 11-comparison Holm adjustment (sections S29 and S30). The last two rows divide the shared amount by topic set and have no separate tests.\par}
\medskip

Pairs involving a reader with no measured topic consumption contribute zero. The increases therefore reflect participation in measured consumption as well as overlap between readers.

\Needspace{6\baselineskip}
\subsection*{S7. Does consumption per search and per minute increase?}
\phantomsection\label{sec:S7}

Total information consumption grows faster than searches or observed time on site. It rises by 41\% per search and 62\% per minute (Table S7, panel A). Shared information consumption per minute rises by 50.3\% (95\% CI, 45.0\% to 55.7\%, panel B).

\Needspace{12\baselineskip}
\setcounter{table}{6}
\captionof{table}{Information consumption and efficiency}\label{tab:S7}

\Needspace{10\baselineskip}
\noindent\textbf{\textbf{A. Total information consumption and engagement}}\par
\smallskip

% SI physical table block 8; authoritative Table S7.
\begingroup
\fontsize{9bp}{11.5pt}\selectfont
\setlength{\tabcolsep}{3pt}
\renewcommand{\arraystretch}{1.13}
\setlength{\LTleft}{0pt}\setlength{\LTright}{0pt}
\setlength{\LTpre}{4pt}\setlength{\LTpost}{5pt}
\renewcommand{\theHtable}{supplement.block8.\arabic{table}}
\begin{longtable}{@{}>{\raggedright\arraybackslash}p{\dimexpr0.3460401\linewidth-2\tabcolsep\relax}>{\raggedright\arraybackslash}p{\dimexpr0.1046463\linewidth-2\tabcolsep\relax}>{\raggedright\arraybackslash}p{\dimexpr0.1077086\linewidth-2\tabcolsep\relax}>{\raggedright\arraybackslash}p{\dimexpr0.3184794\linewidth-2\tabcolsep\relax}>{\raggedright\arraybackslash}p{\dimexpr0.1231257\linewidth-2\tabcolsep\relax}@{}}
\toprule
\textbf{\textbf{Quantity}} & \textbf{\textbf{Control}} & \textbf{\textbf{Treatment}} & \textbf{\textbf{Difference [95\% CI]}} & \textbf{\textbf{Change}} \\
\midrule
\endfirsthead
\multicolumn{5}{@{}l@{}}{\textbf{Table S7 (continued)}}\\
\toprule
\textbf{\textbf{Quantity}} & \textbf{\textbf{Control}} & \textbf{\textbf{Treatment}} & \textbf{\textbf{Difference [95\% CI]}} & \textbf{\textbf{Change}} \\
\midrule
\endhead
\midrule
\multicolumn{5}{@{}r@{}}{Continued on next page}\\
\endfoot
\bottomrule
\endlastfoot
Total information consumption per reader & 1.363 & 2.483 & +1.120 [1.058, 1.184] & +82.2\% \\
Searches per reader & 1.746 & 2.260 & +0.514 [0.464, 0.566] & +29.4\% \\
Total information consumption per search (group totals) & 0.781 & 1.099 & +0.318 [0.289, 0.348] & +40.8\% \\
Time on site per reader, seconds & 495.3 & 558.0 & +62.7 [42.8, 82.7] & +12.7\% \\
Total information consumption per minute & 0.1651 & 0.2670 & +0.1019 [0.0949, 0.1092] & +61.7\% \\
\end{longtable}
\endgroup

{\fontsize{8bp}{11pt}\selectfont\interlinepenalty=10000\noindent \textbf{Note.} Panel A includes all assigned readers. CIs resample readers (section S30), and tests use the Holm adjustment in section S29. Per-search and per-minute measures are ratios of group totals, using the observed time defined in section S25.\par}
\medskip

\Needspace{10\baselineskip}
\noindent\textbf{\textbf{B. Shared information consumption per minute}}\par
\smallskip

% SI physical table block 9; authoritative Table S7.
\begingroup
\fontsize{9bp}{11.5pt}\selectfont
\setlength{\tabcolsep}{3pt}
\renewcommand{\arraystretch}{1.13}
\setlength{\LTleft}{0pt}\setlength{\LTright}{0pt}
\setlength{\LTpre}{4pt}\setlength{\LTpost}{5pt}
\renewcommand{\theHtable}{supplement.block9.\arabic{table}}
\begin{longtable}{@{}>{\raggedright\arraybackslash}p{\dimexpr0.4615628\linewidth-2\tabcolsep\relax}>{\raggedright\arraybackslash}p{\dimexpr0.1538543\linewidth-2\tabcolsep\relax}>{\raggedright\arraybackslash}p{\dimexpr0.1538543\linewidth-2\tabcolsep\relax}>{\raggedright\arraybackslash}p{\dimexpr0.2307286\linewidth-2\tabcolsep\relax}@{}}
\toprule
\textbf{\textbf{Measure}} & \textbf{\textbf{Control}} & \textbf{\textbf{Treatment}} & \textbf{\textbf{Difference [95\% CI]}} \\
\midrule
\endfirsthead
\multicolumn{4}{@{}l@{}}{\textbf{Table S7 (continued)}}\\
\toprule
\textbf{\textbf{Measure}} & \textbf{\textbf{Control}} & \textbf{\textbf{Treatment}} & \textbf{\textbf{Difference [95\% CI]}} \\
\midrule
\endhead
\midrule
\multicolumn{4}{@{}r@{}}{Continued on next page}\\
\endfoot
\bottomrule
\endlastfoot
Shared information consumption per minute & 0.0818 & 0.1230 & +0.0412 [0.0372, 0.0451] \\
\end{longtable}
\endgroup

{\fontsize{8bp}{11pt}\selectfont\interlinepenalty=10000\noindent \textbf{Note.} Panel B divides each group's shared information consumption by its total observed time on site. CIs resample readers (section S30).\par}
\medskip

\Needspace{6\baselineskip}
\subsection*{S8. Does higher consumption also mean more text and more time spent with news?}
\phantomsection\label{sec:S8}

An AI answer and an article can differ in length and in the attention they receive. We therefore examine whether higher consumption counts are accompanied by more text and more time spent with news.

These comparisons include all 26,639 Control and 10,922 Treatment readers from their first search onward. We count words in the full text of opened articles and AI answers and measure time spent on article and search pages.

\Needspace{3\baselineskip}
\paragraph*{Text volume}

Total words in opened articles and AI answers per reader increase from 1,710.3 in Control to 1,953.3 in Treatment, a 14.2\% increase (Table S8). Article text alone increases by 6.8\%, and AI answers add 126.1 words per Treatment reader. Total words remain 12.4\% higher when each answer is counted only once per reader.

\Needspace{12\baselineskip}
\setcounter{table}{7}
\captionof{table}{Word counts per reader}\label{tab:S8}

% SI physical table block 10; authoritative Table S8.
\begingroup
\fontsize{9bp}{11.5pt}\selectfont
\setlength{\tabcolsep}{3pt}
\renewcommand{\arraystretch}{1.13}
\setlength{\LTleft}{0pt}\setlength{\LTright}{0pt}
\setlength{\LTpre}{4pt}\setlength{\LTpost}{5pt}
\renewcommand{\theHtable}{supplement.block10.\arabic{table}}
\begin{longtable}{@{}>{\raggedright\arraybackslash}p{\dimexpr0.3460401\linewidth-2\tabcolsep\relax}>{\raggedright\arraybackslash}p{\dimexpr0.1046463\linewidth-2\tabcolsep\relax}>{\raggedright\arraybackslash}p{\dimexpr0.1077086\linewidth-2\tabcolsep\relax}>{\raggedright\arraybackslash}p{\dimexpr0.3184794\linewidth-2\tabcolsep\relax}>{\raggedright\arraybackslash}p{\dimexpr0.1231257\linewidth-2\tabcolsep\relax}@{}}
\toprule
\textbf{\textbf{Measure}} & \textbf{\textbf{Control}} & \textbf{\textbf{Treatment}} & \textbf{\textbf{Difference}\textbf{\newline{}[95\% CI]}} & \textbf{\textbf{Change}} \\
\midrule
\endfirsthead
\multicolumn{5}{@{}l@{}}{\textbf{Table S8 (continued)}}\\
\toprule
\textbf{\textbf{Measure}} & \textbf{\textbf{Control}} & \textbf{\textbf{Treatment}} & \textbf{\textbf{Difference}\textbf{\newline{}[95\% CI]}} & \textbf{\textbf{Change}} \\
\midrule
\endhead
\midrule
\multicolumn{5}{@{}r@{}}{Continued on next page}\\
\endfoot
\bottomrule
\endlastfoot
Words in opened articles & 1,710.3 & 1,827.2 & +116.9\newline{}[40.0, 196.5] & +6.8\% \\
Words in AI answers & 0.0 & 126.1 & +126.1\newline{}[123.4, 128.8] & \textemdash{} \\
Total words in articles and AI answers & 1,710.3 & 1,953.3 & +243.0\newline{}[165.4, 323.2] & +14.2\% \\
\end{longtable}
\endgroup

{\fontsize{8bp}{11pt}\selectfont\interlinepenalty=10000\noindent \textbf{Note.} Article text is available for 81.5\% of opens in Control and 82.0\% in Treatment. Counts include repeated opens and answer displays. They describe text in articles opened and answers shown, rather than words verified as read. Differences are Treatment minus Control. Brackets show pointwise 95\% confidence intervals from 5,000 reader-bootstrap resamples throughout this section.\par}
\medskip

\Needspace{3\baselineskip}
\paragraph*{Time spent}

Total time on article and search pages per reader increases from 409.0 seconds in Control to 489.7 seconds in Treatment, a 19.7\% increase (Table S9). Time on articles increases by 6.4\%, and time on search pages increases by 58.3\%.

\Needspace{12\baselineskip}
\setcounter{table}{8}
\captionof{table}{Time spent per reader in seconds}\label{tab:S9}

% SI physical table block 11; authoritative Table S9.
\begingroup
\fontsize{9bp}{11.5pt}\selectfont
\setlength{\tabcolsep}{3pt}
\renewcommand{\arraystretch}{1.13}
\setlength{\LTleft}{0pt}\setlength{\LTright}{0pt}
\setlength{\LTpre}{4pt}\setlength{\LTpost}{5pt}
\renewcommand{\theHtable}{supplement.block11.\arabic{table}}
\begin{longtable}{@{}>{\raggedright\arraybackslash}p{\dimexpr0.3460401\linewidth-2\tabcolsep\relax}>{\raggedright\arraybackslash}p{\dimexpr0.1046463\linewidth-2\tabcolsep\relax}>{\raggedright\arraybackslash}p{\dimexpr0.1077086\linewidth-2\tabcolsep\relax}>{\raggedright\arraybackslash}p{\dimexpr0.3184794\linewidth-2\tabcolsep\relax}>{\raggedright\arraybackslash}p{\dimexpr0.1231257\linewidth-2\tabcolsep\relax}@{}}
\toprule
\textbf{\textbf{Measure}} & \textbf{\textbf{Control}} & \textbf{\textbf{Treatment}} & \textbf{\textbf{Difference}\textbf{\newline{}[95\% CI]}} & \textbf{\textbf{Change}} \\
\midrule
\endfirsthead
\multicolumn{5}{@{}l@{}}{\textbf{Table S9 (continued)}}\\
\toprule
\textbf{\textbf{Measure}} & \textbf{\textbf{Control}} & \textbf{\textbf{Treatment}} & \textbf{\textbf{Difference}\textbf{\newline{}[95\% CI]}} & \textbf{\textbf{Change}} \\
\midrule
\endhead
\midrule
\multicolumn{5}{@{}r@{}}{Continued on next page}\\
\endfoot
\bottomrule
\endlastfoot
Time on article pages & 303.8 & 323.1 & +19.3\newline{}[4.1, 34.1] & +6.4\% \\
Time on search pages & 105.2 & 166.6 & +61.4\newline{}[54.5, 68.6] & +58.3\% \\
Total time on article and search pages & 409.0 & 489.7 & +80.7\newline{}[63.2, 97.9] & +19.7\% \\
\end{longtable}
\endgroup

{\fontsize{8bp}{11pt}\selectfont\interlinepenalty=10000\noindent \textbf{Note.} This total excludes time on other site pages, which is included in the time-on-site measure reported in Table S7 and Fig. 1B.\par}
\medskip

Time per search also increases from 60.3 seconds in Control to 73.7 seconds in Treatment (+22.3\%). Treatment readers therefore spend longer on each search, in addition to searching more often. This pattern is consistent with readers spending time reading the AI answers.

\Needspace{3\baselineskip}
\paragraph*{Time relative to text volume}

We next examine page time per word to assess time spent relative to text length. For articles, we divide time on article pages by the words in those articles. For search pages with an AI answer, we divide time on those pages by the words in the answers.

Article page time per word is 0.150 seconds in Control and 0.149 seconds in Treatment (Table S10). The estimated change is \ensuremath{-}0.3\%, with a 95\% confidence interval from \ensuremath{-}4.5\% to 3.9\%. Readers spend more time on articles as the amount of article text increases.

\Needspace{12\baselineskip}
\setcounter{table}{9}
\captionof{table}{Page time per word}\label{tab:S10}

% SI physical table block 12; authoritative Table S10.
\begingroup
\fontsize{9bp}{11.5pt}\selectfont
\setlength{\tabcolsep}{3pt}
\renewcommand{\arraystretch}{1.13}
\setlength{\LTleft}{0pt}\setlength{\LTright}{0pt}
\setlength{\LTpre}{4pt}\setlength{\LTpost}{5pt}
\renewcommand{\theHtable}{supplement.block12.\arabic{table}}
\begin{longtable}{@{}>{\raggedright\arraybackslash}p{\dimexpr0.3230769\linewidth-2\tabcolsep\relax}>{\raggedright\arraybackslash}p{\dimexpr0.3384615\linewidth-2\tabcolsep\relax}>{\raggedright\arraybackslash}p{\dimexpr0.3384615\linewidth-2\tabcolsep\relax}@{}}
\toprule
\textbf{\textbf{Page}} & \textbf{\textbf{Group}} & \textbf{\textbf{Seconds per word}\textbf{\newline{}[95\% CI]}} \\
\midrule
\endfirsthead
\multicolumn{3}{@{}l@{}}{\textbf{Table S10 (continued)}}\\
\toprule
\textbf{\textbf{Page}} & \textbf{\textbf{Group}} & \textbf{\textbf{Seconds per word}\textbf{\newline{}[95\% CI]}} \\
\midrule
\endhead
\midrule
\multicolumn{3}{@{}r@{}}{Continued on next page}\\
\endfoot
\bottomrule
\endlastfoot
Article pages & Control & 0.150\newline{}[0.147, 0.153] \\
Article pages & Treatment & 0.149\newline{}[0.144, 0.155] \\
Search pages with an AI answer & Treatment & 0.861\newline{}[0.827, 0.898] \\
\end{longtable}
\endgroup

{\fontsize{8bp}{11pt}\selectfont\interlinepenalty=10000\noindent \textbf{Note.} Ratios divide group totals. Article ratios use only articles with word counts and the time on those same pages. Search-page time includes viewing sources and conventional results as well as the answer. These measures do not directly assess comprehension or information quality.\par}
\medskip

On search pages displaying an AI answer, page time per answer word is 0.861 seconds. The ratio describes engagement with the search page as a whole.

The increase in total information consumption also holds when measured by text volume and time spent. The higher consumption count is therefore accompanied by more text in opened articles and AI answers and more time spent with news. Article page time per word remains similar across groups, suggesting that readers maintain their engagement with articles as the amount of article text increases.

\Needspace{6\baselineskip}
\subsection*{S9. Do the individual-level results depend on consumption volume or who is included?}
\phantomsection\label{sec:S9}

Individual consumption remains less concentrated in Treatment after equalizing consumption volume per reader. The shift toward less-popular topics also persists when readers with only one article or AI answer are included (Table S11).

The main individual analysis includes readers who consume at least two articles or AI answers. Counting answers allows more Treatment readers to meet this requirement, so we examine both the amount consumed and the readers included.

Readers who consume more articles or answers may encounter more topics simply because they consume more. We therefore compare concentration using the same number of articles or answers per reader. Consumption remains less concentrated in Treatment (panel A).

\Needspace{12\baselineskip}
\setcounter{table}{10}
\captionof{table}{Checks on individual concentration and popularity}\label{tab:S11}

\Needspace{10\baselineskip}
\noindent\textbf{\textbf{A. Individual concentration after equalizing consumption volume}}\par
\smallskip

% SI physical table block 13; authoritative Table S11.
\begingroup
\fontsize{9bp}{11.5pt}\selectfont
\setlength{\tabcolsep}{3pt}
\renewcommand{\arraystretch}{1.13}
\setlength{\LTleft}{0pt}\setlength{\LTright}{0pt}
\setlength{\LTpre}{4pt}\setlength{\LTpost}{5pt}
\renewcommand{\theHtable}{supplement.block13.\arabic{table}}
\begin{longtable}{@{}>{\raggedright\arraybackslash}p{\dimexpr0.2582893\linewidth-2\tabcolsep\relax}>{\raggedright\arraybackslash}p{\dimexpr0.1417107\linewidth-2\tabcolsep\relax}>{\raggedright\arraybackslash}p{\dimexpr0.1166843\linewidth-2\tabcolsep\relax}>{\raggedright\arraybackslash}p{\dimexpr0.1166843\linewidth-2\tabcolsep\relax}>{\raggedright\arraybackslash}p{\dimexpr0.3666315\linewidth-2\tabcolsep\relax}@{}}
\toprule
\textbf{\textbf{Measure}} & \textbf{\textbf{Consumption counted}} & \textbf{\textbf{Control}} & \textbf{\textbf{Treatment}} & \textbf{\textbf{Difference [95\% CI]}} \\
\midrule
\endfirsthead
\multicolumn{5}{@{}l@{}}{\textbf{Table S11 (continued)}}\\
\toprule
\textbf{\textbf{Measure}} & \textbf{\textbf{Consumption counted}} & \textbf{\textbf{Control}} & \textbf{\textbf{Treatment}} & \textbf{\textbf{Difference [95\% CI]}} \\
\midrule
\endhead
\midrule
\multicolumn{5}{@{}r@{}}{Continued on next page}\\
\endfoot
\bottomrule
\endlastfoot
Gini coefficient & All articles and answers & 0.9525 & 0.9396 & \ensuremath{-}0.0129 [\ensuremath{-}0.0144, \ensuremath{-}0.0114] \\
Gini coefficient & Two articles or answers & 0.9646 & 0.9519 & \ensuremath{-}0.0128 [\ensuremath{-}0.0138, \ensuremath{-}0.0118] \\
Effective number of topics & All articles and answers & 4.63 & 5.58 & +0.96 [0.85, 1.07] \\
Effective number of topics & Two articles or answers & 3.71 & 4.67 & +0.96 [0.89, 1.04] \\
\end{longtable}
\endgroup

{\fontsize{8bp}{11pt}\selectfont\interlinepenalty=10000\noindent \textbf{Note.} Both calculations use the same 7,783 Control and 6,526 Treatment readers with topic estimates for at least two articles or answers. The restricted calculation uses two randomly selected articles or answers per reader (section S32).\par}
\medskip

Including readers who consume only one article or answer leaves the shift toward less-popular topics intact (panel B). We examine popularity for this broader sample because concentration based on a single article or answer describes that content, rather than how a reader's consumption spans multiple articles or answers.

\Needspace{10\baselineskip}
\noindent\textbf{\textbf{B. Popularity when readers with only one article or answer are also included}}\par
\smallskip

% SI physical table block 14; authoritative Table S11.
\begingroup
\fontsize{9bp}{11.5pt}\selectfont
\setlength{\tabcolsep}{3pt}
\renewcommand{\arraystretch}{1.13}
\setlength{\LTleft}{0pt}\setlength{\LTright}{0pt}
\setlength{\LTpre}{4pt}\setlength{\LTpost}{5pt}
\renewcommand{\theHtable}{supplement.block14.\arabic{table}}
\begin{longtable}{@{}>{\raggedright\arraybackslash}p{\dimexpr0.2294615\linewidth-2\tabcolsep\relax}>{\raggedright\arraybackslash}p{\dimexpr0.1803590\linewidth-2\tabcolsep\relax}>{\raggedright\arraybackslash}p{\dimexpr0.1065470\linewidth-2\tabcolsep\relax}>{\raggedright\arraybackslash}p{\dimexpr0.1065470\linewidth-2\tabcolsep\relax}>{\raggedright\arraybackslash}p{\dimexpr0.2705385\linewidth-2\tabcolsep\relax}>{\raggedright\arraybackslash}p{\dimexpr0.1065470\linewidth-2\tabcolsep\relax}@{}}
\toprule
\textbf{\textbf{Measure}} & \textbf{\textbf{Readers included}} & \textbf{\textbf{Control}} & \textbf{\textbf{Treatment}} & \textbf{\textbf{Difference [95\% CI]}} & \textbf{\textbf{P}} \\
\midrule
\endfirsthead
\multicolumn{6}{@{}l@{}}{\textbf{Table S11 (continued)}}\\
\toprule
\textbf{\textbf{Measure}} & \textbf{\textbf{Readers included}} & \textbf{\textbf{Control}} & \textbf{\textbf{Treatment}} & \textbf{\textbf{Difference [95\% CI]}} & \textbf{\textbf{P}} \\
\midrule
\endhead
\midrule
\multicolumn{6}{@{}r@{}}{Continued on next page}\\
\endfoot
\bottomrule
\endlastfoot
Mean popularity rank & At least two articles or answers & 17.81 & 18.98 & +1.16 [0.88, 1.45] & \ensuremath{<} 0.001 \\
Mean popularity rank & At least one article or answer & 18.04 & 19.50 & +1.46 [1.23, 1.70] & \ensuremath{<} 0.001 \\
Top-10 popular share (\%) & At least two articles or answers & 40.7 & 36.7 & \ensuremath{-}4.0 [\ensuremath{-}5.0, \ensuremath{-}3.0] & \ensuremath{<} 0.001 \\
Top-10 popular share (\%) & At least one article or answer & 40.2 & 35.4 & \ensuremath{-}4.8 [\ensuremath{-}5.5, \ensuremath{-}4.0] & \ensuremath{<} 0.001 \\
\end{longtable}
\endgroup

{\fontsize{8bp}{11pt}\selectfont\interlinepenalty=10000\noindent \textbf{Note.} The main sample includes 7,992 Control and 6,676 Treatment readers; the broader sample includes 16,094 and 9,747, respectively. All included readers have some measured topic consumption. Main-sample P values use the eight-comparison Holm adjustment; broader-sample P values are unadjusted. Confidence intervals follow section S30.\par}
\medskip

Adjusting for pre-search characteristics and entry timing also yields similar individual estimates (section S21). These checks do not eliminate possible differences between readers selected by their consumption after assignment.

\Needspace{6\baselineskip}
\subsection*{S10. Do the concentration and popularity results hold without political topics?}
\phantomsection\label{sec:S10}

Concentration decreases and consumption shifts toward less-popular topics after political topics are excluded, although the popularity differences become smaller (Table S12).

We apply the three definitions of political topics described in section S4.

\Needspace{12\baselineskip}
\setcounter{table}{11}
\captionof{table}{Results after excluding political topics}\label{tab:S12}

% SI physical table block 15; authoritative Table S12.
\begingroup
\fontsize{9bp}{11.5pt}\selectfont
\setlength{\tabcolsep}{3pt}
\renewcommand{\arraystretch}{1.13}
\setlength{\LTleft}{0pt}\setlength{\LTright}{0pt}
\setlength{\LTpre}{4pt}\setlength{\LTpost}{5pt}
\renewcommand{\theHtable}{supplement.block15.\arabic{table}}
\begin{longtable}{@{}>{\raggedright\arraybackslash}p{\dimexpr0.2461538\linewidth-2\tabcolsep\relax}>{\raggedright\arraybackslash}p{\dimexpr0.2507479\linewidth-2\tabcolsep\relax}>{\raggedright\arraybackslash}p{\dimexpr0.2507479\linewidth-2\tabcolsep\relax}>{\raggedright\arraybackslash}p{\dimexpr0.2523504\linewidth-2\tabcolsep\relax}@{}}
\toprule
\textbf{\textbf{Measure}} & \textbf{\textbf{Narrow exclusion}} & \textbf{\textbf{Broad exclusion}} & \textbf{\textbf{Broadest exclusion}} \\
\midrule
\endfirsthead
\multicolumn{4}{@{}l@{}}{\textbf{Table S12 (continued)}}\\
\toprule
\textbf{\textbf{Measure}} & \textbf{\textbf{Narrow exclusion}} & \textbf{\textbf{Broad exclusion}} & \textbf{\textbf{Broadest exclusion}} \\
\midrule
\endhead
\midrule
\multicolumn{4}{@{}r@{}}{Continued on next page}\\
\endfoot
\bottomrule
\endlastfoot
\addlinespace[3pt]
\multicolumn{4}{@{}p{\linewidth}@{}}{\textit{Individual level}} \\*
Gini coefficient & \ensuremath{-}0.0135 [\ensuremath{-}0.0149, \ensuremath{-}0.0120]\newline{}P \ensuremath{<} 0.001 & \ensuremath{-}0.0136 [\ensuremath{-}0.0150, \ensuremath{-}0.0121]\newline{}P \ensuremath{<} 0.001 & \ensuremath{-}0.0141 [\ensuremath{-}0.0156, \ensuremath{-}0.0127]\newline{}P \ensuremath{<} 0.001 \\
Effective number of topics & +0.98 [0.88, 1.09]\newline{}P \ensuremath{<} 0.001 & +0.88 [0.79, 0.98]\newline{}P \ensuremath{<} 0.001 & +0.89 [0.80, 0.98]\newline{}P \ensuremath{<} 0.001 \\
Mean popularity rank & +0.99 [0.72, 1.27]\newline{}P \ensuremath{<} 0.001 & +0.67 [0.40, 0.92]\newline{}P \ensuremath{<} 0.001 & +0.56 [0.32, 0.81]\newline{}P \ensuremath{<} 0.001 \\
Top-10 popular share (\%) & \ensuremath{-}2.8 [\ensuremath{-}3.8, \ensuremath{-}1.8]\newline{}P \ensuremath{<} 0.001 & \ensuremath{-}2.2 [\ensuremath{-}3.3, \ensuremath{-}1.2]\newline{}P \ensuremath{<} 0.001 & \ensuremath{-}1.8 [\ensuremath{-}2.9, \ensuremath{-}0.8]\newline{}P = 0.002 \\
\addlinespace[3pt]
\multicolumn{4}{@{}p{\linewidth}@{}}{\textit{Aggregate level}} \\*
Gini coefficient & \ensuremath{-}0.0305 [\ensuremath{-}0.0402, \ensuremath{-}0.0208]\newline{}P \ensuremath{<} 0.001 & \ensuremath{-}0.0265 [\ensuremath{-}0.0372, \ensuremath{-}0.0156]\newline{}P \ensuremath{<} 0.001 & \ensuremath{-}0.0288 [\ensuremath{-}0.0401, \ensuremath{-}0.0171]\newline{}P \ensuremath{<} 0.001 \\
Effective number of topics & +1.77 [1.18, 2.30]\newline{}P \ensuremath{<} 0.001 & +1.27 [0.72, 1.79]\newline{}P \ensuremath{<} 0.001 & +1.33 [0.80, 1.82]\newline{}P \ensuremath{<} 0.001 \\
Mean popularity rank & +1.20 [0.92, 1.48]\newline{}P \ensuremath{<} 0.001 & +0.96 [0.69, 1.23]\newline{}P \ensuremath{<} 0.001 & +1.01 [0.73, 1.28]\newline{}P \ensuremath{<} 0.001 \\
Top-10 popular share (\%) & \ensuremath{-}3.9 [\ensuremath{-}4.9, \ensuremath{-}2.9]\newline{}P \ensuremath{<} 0.001 & \ensuremath{-}3.7 [\ensuremath{-}4.8, \ensuremath{-}2.6]\newline{}P \ensuremath{<} 0.001 & \ensuremath{-}3.5 [\ensuremath{-}4.7, \ensuremath{-}2.3]\newline{}P \ensuremath{<} 0.001 \\
\end{longtable}
\endgroup

{\fontsize{8bp}{11pt}\selectfont\interlinepenalty=10000\noindent \textbf{Note.} Entries are Treatment minus Control, with reader-bootstrap CIs and Holm-adjusted P values across eight comparisons within each exclusion (sections S30 and S31). Top-10 differences are percentage points. Exclusions remove consumption on political topics while retaining the original article-and-answer counts for reader eligibility. Popularity is ranked again among the remaining topics. Individual Control/Treatment samples are 7,976/6,675, 7,777/6,649 and 7,740/6,641 for the narrow, broad and broadest exclusions, respectively.\par}
\medskip

\Needspace{6\baselineskip}
\subsection*{S11. Do the results hold under a different topic model?}
\phantomsection\label{sec:S11}

The alternative topic model also shows lower concentration and a shift toward less-popular topics at both levels (all eight Holm-adjusted P \ensuremath{<} 0.001, Table S13). Refitting the model tests whether these results depend on one grouping of the news archive.

We refit the topic model while keeping the article data and model settings unchanged, apart from the random seed. The resulting model retains 72 topics rather than 58 (section S22).

\Needspace{12\baselineskip}
\setcounter{table}{12}
\captionof{table}{Results under an alternative topic model}\label{tab:S13}

% SI physical table block 16; authoritative Table S13.
\begingroup
\fontsize{9bp}{11.5pt}\selectfont
\setlength{\tabcolsep}{3pt}
\renewcommand{\arraystretch}{1.13}
\setlength{\LTleft}{0pt}\setlength{\LTright}{0pt}
\setlength{\LTpre}{4pt}\setlength{\LTpost}{5pt}
\renewcommand{\theHtable}{supplement.block16.\arabic{table}}
\begin{longtable}{@{}>{\raggedright\arraybackslash}p{\dimexpr0.2769231\linewidth-2\tabcolsep\relax}>{\raggedright\arraybackslash}p{\dimexpr0.1045940\linewidth-2\tabcolsep\relax}>{\raggedright\arraybackslash}p{\dimexpr0.1107906\linewidth-2\tabcolsep\relax}>{\raggedright\arraybackslash}p{\dimexpr0.1076923\linewidth-2\tabcolsep\relax}>{\raggedright\arraybackslash}p{\dimexpr0.1980769\linewidth-2\tabcolsep\relax}>{\raggedright\arraybackslash}p{\dimexpr0.2019231\linewidth-2\tabcolsep\relax}@{}}
\toprule
\textbf{\textbf{Measure}} & \textbf{\textbf{Control}} & \textbf{\textbf{Treatment}} & \textbf{\textbf{Difference}} & \textbf{\textbf{95\% CI}} & \textbf{\textbf{Holm-adjusted P}} \\
\midrule
\endfirsthead
\multicolumn{6}{@{}l@{}}{\textbf{Table S13 (continued)}}\\
\toprule
\textbf{\textbf{Measure}} & \textbf{\textbf{Control}} & \textbf{\textbf{Treatment}} & \textbf{\textbf{Difference}} & \textbf{\textbf{95\% CI}} & \textbf{\textbf{Holm-adjusted P}} \\
\midrule
\endhead
\midrule
\multicolumn{6}{@{}r@{}}{Continued on next page}\\
\endfoot
\bottomrule
\endlastfoot
\addlinespace[3pt]
\multicolumn{6}{@{}p{\linewidth}@{}}{\textit{Individual level}} \\*
Gini coefficient & 0.9495 & 0.9457 & \ensuremath{-}0.0038 & [\ensuremath{-}0.0052, \ensuremath{-}0.0024] & \ensuremath{<} 0.001 \\
Effective number of topics & 5.74 & 6.06 & +0.32 & [0.19, 0.45] & \ensuremath{<} 0.001 \\
Mean popularity rank & 21.43 & 22.69 & +1.26 & [0.91, 1.62] & \ensuremath{<} 0.001 \\
Top-10 popular share (\%) & 36.2 & 33.1 & \ensuremath{-}3.0 & [\ensuremath{-}4.0, \ensuremath{-}2.1] & \ensuremath{<} 0.001 \\
\addlinespace[3pt]
\multicolumn{6}{@{}p{\linewidth}@{}}{\textit{Aggregate level}} \\*
Gini coefficient & 0.4252 & 0.4049 & \ensuremath{-}0.0203 & [\ensuremath{-}0.0294, \ensuremath{-}0.0102] & \ensuremath{<} 0.001 \\
Effective number of topics & 54.02 & 55.52 & +1.49 & [0.74, 2.20] & \ensuremath{<} 0.001 \\
Mean popularity rank & 21.44 & 22.66 & +1.22 & [0.88, 1.57] & \ensuremath{<} 0.001 \\
Top-10 popular share (\%) & 36.2 & 33.5 & \ensuremath{-}2.7 & [\ensuremath{-}3.6, \ensuremath{-}1.7] & \ensuremath{<} 0.001 \\
\end{longtable}
\endgroup

{\fontsize{8bp}{11pt}\selectfont\interlinepenalty=10000\noindent \textbf{Note.} Individual estimates include 8,057 Control and 6,581 Treatment readers. Aggregate results combine all consumption with topic estimates. Confidence intervals and tests follow sections S30 and S31. Each comparison uses the topics and popularity ranking from its own model.\par}
\medskip

The models do not assign topic estimates to exactly the same articles and answers. We therefore repeat the comparison using only content measured by both models, with the same cited articles representing each answer. Both models still show lower concentration and a shift toward less-popular topics (Table S14).

\Needspace{12\baselineskip}
\setcounter{table}{13}
\captionof{table}{Both topic models applied to the same articles and answers}\label{tab:S14}

% SI physical table block 17; authoritative Table S14.
\begingroup
\fontsize{9bp}{11.5pt}\selectfont
\setlength{\tabcolsep}{3pt}
\renewcommand{\arraystretch}{1.13}
\setlength{\LTleft}{0pt}\setlength{\LTright}{0pt}
\setlength{\LTpre}{4pt}\setlength{\LTpost}{5pt}
\renewcommand{\theHtable}{supplement.block17.\arabic{table}}
\begin{longtable}{@{}>{\raggedright\arraybackslash}p{\dimexpr0.3230769\linewidth-2\tabcolsep\relax}>{\raggedright\arraybackslash}p{\dimexpr0.3384615\linewidth-2\tabcolsep\relax}>{\raggedright\arraybackslash}p{\dimexpr0.3384615\linewidth-2\tabcolsep\relax}@{}}
\toprule
\textbf{\textbf{Measure}} & \textbf{\textbf{58-topic difference [95\% CI]}} & \textbf{\textbf{72-topic difference [95\% CI]}} \\
\midrule
\endfirsthead
\multicolumn{3}{@{}l@{}}{\textbf{Table S14 (continued)}}\\
\toprule
\textbf{\textbf{Measure}} & \textbf{\textbf{58-topic difference [95\% CI]}} & \textbf{\textbf{72-topic difference [95\% CI]}} \\
\midrule
\endhead
\midrule
\multicolumn{3}{@{}r@{}}{Continued on next page}\\
\endfoot
\bottomrule
\endlastfoot
\addlinespace[3pt]
\multicolumn{3}{@{}p{\linewidth}@{}}{\textit{Individual level}} \\*
Gini coefficient & \ensuremath{-}0.0049 [\ensuremath{-}0.0063, \ensuremath{-}0.0034] & \ensuremath{-}0.0053 [\ensuremath{-}0.0068, \ensuremath{-}0.0039] \\
Effective number of topics & +0.36 [0.25, 0.47] & +0.47 [0.33, 0.60] \\
Mean popularity rank & +0.81 [0.51, 1.11] & +0.95 [0.59, 1.31] \\
Top-10 popular share (\%) & \ensuremath{-}3.2 [\ensuremath{-}4.2, \ensuremath{-}2.1] & \ensuremath{-}2.1 [\ensuremath{-}3.1, \ensuremath{-}1.1] \\
\addlinespace[3pt]
\multicolumn{3}{@{}p{\linewidth}@{}}{\textit{Aggregate level}} \\*
Gini coefficient & \ensuremath{-}0.0214 [\ensuremath{-}0.0309, \ensuremath{-}0.0110] & \ensuremath{-}0.0171 [\ensuremath{-}0.0261, \ensuremath{-}0.0069] \\
Effective number of topics & +1.27 [0.63, 1.82] & +1.13 [0.38, 1.82] \\
Mean popularity rank & +0.98 [0.69, 1.27] & +1.12 [0.78, 1.47] \\
Top-10 popular share (\%) & \ensuremath{-}3.8 [\ensuremath{-}4.8, \ensuremath{-}2.7] & \ensuremath{-}2.5 [\ensuremath{-}3.5, \ensuremath{-}1.6] \\
\end{longtable}
\endgroup

{\fontsize{8bp}{11pt}\selectfont\interlinepenalty=10000\noindent \textbf{Note.} Both models use the same 51,594 article opens and 8,840 answers. Individual estimates include 7,728 Control and 5,875 Treatment readers; aggregate results use all retained consumption. For this comparison, each cited article's retained topic shares are scaled to sum to one before averaging into an answer. This keeps the sources and their weights the same across models. Confidence intervals follow section S30.\par}
\medskip

\Needspace{6\baselineskip}
\subsection*{S12. Do the results depend on how answers are measured or weighted?}
\phantomsection\label{sec:S12}

Shared information consumption increases under alternative ways of measuring or weighting answers (Table S15). Total consumption, including opened articles, also remains less concentrated and shifts toward less-popular topics at both levels (Table S16). The size of these differences depends on how answers are measured.

We check whether the results change when answer topics are estimated from the answer text or when greater weight is given to sources cited earlier. We also compare the methods using the same answers, so differences in coverage do not drive the comparison. Finally, we count each answer as half or twice an article to test the equal-weight assumption. Section S24 describes these measures.

\Needspace{12\baselineskip}
\setcounter{table}{14}
\captionof{table}{Shared information under alternative answer measurements}\label{tab:S15}

% SI physical table block 18; authoritative Table S15.
\begingroup
\fontsize{9bp}{11.5pt}\selectfont
\setlength{\tabcolsep}{3pt}
\renewcommand{\arraystretch}{1.13}
\setlength{\LTleft}{0pt}\setlength{\LTright}{0pt}
\setlength{\LTpre}{4pt}\setlength{\LTpost}{5pt}
\renewcommand{\theHtable}{supplement.block18.\arabic{table}}
\begin{longtable}{@{}>{\raggedright\arraybackslash}p{\dimexpr0.2538543\linewidth-2\tabcolsep\relax}>{\raggedright\arraybackslash}p{\dimexpr0.1796199\linewidth-2\tabcolsep\relax}>{\raggedright\arraybackslash}p{\dimexpr0.1995776\linewidth-2\tabcolsep\relax}>{\raggedright\arraybackslash}p{\dimexpr0.2090813\linewidth-2\tabcolsep\relax}>{\raggedright\arraybackslash}p{\dimexpr0.1578669\linewidth-2\tabcolsep\relax}@{}}
\toprule
\textbf{\textbf{Answer representation}} & \textbf{\textbf{Shared consumption}\newline{}\textbf{[95\% CI]}} & \textbf{\textbf{Non-core consumption}\newline{}\textbf{[95\% CI]}} & \textbf{\textbf{Cosine similarity}\newline{}\textbf{[95\% CI]}} & \textbf{\textbf{Shared information per reader pair}} \\
\midrule
\endfirsthead
\multicolumn{5}{@{}l@{}}{\textbf{Table S15 (continued)}}\\
\toprule
\textbf{\textbf{Answer representation}} & \textbf{\textbf{Shared consumption}\newline{}\textbf{[95\% CI]}} & \textbf{\textbf{Non-core consumption}\newline{}\textbf{[95\% CI]}} & \textbf{\textbf{Cosine similarity}\newline{}\textbf{[95\% CI]}} & \textbf{\textbf{Shared information per reader pair}} \\
\midrule
\endhead
\midrule
\multicolumn{5}{@{}r@{}}{Continued on next page}\\
\endfoot
\bottomrule
\endlastfoot
\addlinespace[3pt]
\multicolumn{5}{@{}p{\linewidth}@{}}{\textit{Content measurement \textemdash{} all available answers}} \\*
Cited sources (main) & +0.469 [0.434, 0.503] & +0.652 [0.611, 0.692] & +0.0425 [0.0409, 0.0440] & +0.1082 \\
Citation-order weights & +0.467 [0.433, 0.502] & +0.653 [0.613, 0.693] & +0.0409 [0.0393, 0.0424] & +0.1056 \\
Answer text & +0.468 [0.434, 0.502] & +0.695 [0.655, 0.735] & +0.0765 [0.0739, 0.0791] & +0.1905 \\
\addlinespace[3pt]
\multicolumn{5}{@{}p{\linewidth}@{}}{\textit{Content measurement \textemdash{} the same 10,948 answers}} \\*
Cited sources (main) & +0.462 [0.428, 0.497] & +0.645 [0.605, 0.685] & +0.0417 [0.0402, 0.0432] & +0.1063 \\
Citation-order weights & +0.461 [0.427, 0.495] & +0.646 [0.607, 0.688] & +0.0401 [0.0386, 0.0416] & +0.1036 \\
Answer text & +0.447 [0.411, 0.482] & +0.660 [0.620, 0.701] & +0.0695 [0.0669, 0.0720] & +0.1739 \\
\addlinespace[3pt]
\multicolumn{5}{@{}p{\linewidth}@{}}{\textit{Answer weight \textemdash{} cited-source measure}} \\*
Half answer weight & +0.239 [0.207, 0.271] & +0.361 [0.324, 0.400] & +0.0393 [0.0379, 0.0408] & +0.0609 \\
Double answer weight & +0.928 [0.887, 0.968] & +1.233 [1.186, 1.283] & +0.0442 [0.0427, 0.0458] & +0.1899 \\
\end{longtable}
\endgroup

{\fontsize{8bp}{11pt}\selectfont\interlinepenalty=10000\noindent \textbf{Note.} Cited-source measurements cover 11,092 distinct answers and answer-text measurements cover 11,539. The same-answer comparison uses the 10,948 answers measured by all three methods. Differences are Treatment minus Control. Confidence intervals and tests follow sections S30 and S31. All tested differences have unadjusted \textit{P} \ensuremath{<} 0.001; the last column has no separate test.\par}
\medskip

\Needspace{12\baselineskip}
\setcounter{table}{15}
\captionof{table}{Total information consumption under alternative answer measurements}\label{tab:S16}

\Needspace{10\baselineskip}
\noindent\textbf{\textbf{A. Topic concentration}}\par
\smallskip

% SI physical table block 19; authoritative Table S16.
\begingroup
\fontsize{9bp}{11.5pt}\selectfont
\setlength{\tabcolsep}{3pt}
\renewcommand{\arraystretch}{1.13}
\setlength{\LTleft}{0pt}\setlength{\LTright}{0pt}
\setlength{\LTpre}{4pt}\setlength{\LTpost}{5pt}
\renewcommand{\theHtable}{supplement.block19.\arabic{table}}
\begin{longtable}{@{}>{\raggedright\arraybackslash}p{\dimexpr0.1958817\linewidth-2\tabcolsep\relax}>{\raggedright\arraybackslash}p{\dimexpr0.2280887\linewidth-2\tabcolsep\relax}>{\raggedright\arraybackslash}p{\dimexpr0.1615628\linewidth-2\tabcolsep\relax}>{\raggedright\arraybackslash}p{\dimexpr0.2280887\linewidth-2\tabcolsep\relax}>{\raggedright\arraybackslash}p{\dimexpr0.1863780\linewidth-2\tabcolsep\relax}@{}}
\toprule
\textbf{\textbf{Answer representation}} & \textbf{\textbf{Aggregate}\newline{}\textbf{Gini}} & \textbf{\textbf{Aggregate}\newline{}\textbf{effective topics}} & \textbf{\textbf{Individual}\newline{}\textbf{Gini}} & \textbf{\textbf{Individual}\newline{}\textbf{effective topics}} \\
\midrule
\endfirsthead
\multicolumn{5}{@{}l@{}}{\textbf{Table S16 (continued)}}\\
\toprule
\textbf{\textbf{Answer representation}} & \textbf{\textbf{Aggregate}\newline{}\textbf{Gini}} & \textbf{\textbf{Aggregate}\newline{}\textbf{effective topics}} & \textbf{\textbf{Individual}\newline{}\textbf{Gini}} & \textbf{\textbf{Individual}\newline{}\textbf{effective topics}} \\
\midrule
\endhead
\midrule
\multicolumn{5}{@{}r@{}}{Continued on next page}\\
\endfoot
\bottomrule
\endlastfoot
\addlinespace[3pt]
\multicolumn{5}{@{}p{\linewidth}@{}}{\textit{Content measurement \textemdash{} all available answers}} \\*
Cited sources (main) & \ensuremath{-}0.0341 [\ensuremath{-}0.0434, \ensuremath{-}0.0244]\newline{}\ensuremath{<} 0.001 & +2.05 [1.46, 2.58]\newline{}\ensuremath{<} 0.001 & \ensuremath{-}0.0131 [\ensuremath{-}0.0146, \ensuremath{-}0.0117]\newline{}\ensuremath{<} 0.001 & +0.97 [0.86, 1.08]\newline{}\ensuremath{<} 0.001 \\
Citation-order weights & \ensuremath{-}0.0346 [\ensuremath{-}0.0440, \ensuremath{-}0.0252]\newline{}\ensuremath{<} 0.001 & +2.08 [1.51, 2.62]\newline{}\ensuremath{<} 0.001 & \ensuremath{-}0.0118 [\ensuremath{-}0.0132, \ensuremath{-}0.0103]\newline{}\ensuremath{<} 0.001 & +0.88 [0.77, 0.99]\newline{}\ensuremath{<} 0.001 \\
Answer text & \ensuremath{-}0.0471 [\ensuremath{-}0.0563, \ensuremath{-}0.0377]\newline{}\ensuremath{<} 0.001 & +2.22 [1.66, 2.76]\newline{}\ensuremath{<} 0.001 & \ensuremath{-}0.0942 [\ensuremath{-}0.0977, \ensuremath{-}0.0906]\newline{}\ensuremath{<} 0.001 & +4.22 [4.01, 4.42]\newline{}\ensuremath{<} 0.001 \\
\addlinespace[3pt]
\multicolumn{5}{@{}p{\linewidth}@{}}{\textit{Content measurement \textemdash{} the same 10,948 answers}} \\*
Cited sources (main) & \ensuremath{-}0.0335 [\ensuremath{-}0.0430, \ensuremath{-}0.0236]\newline{}\ensuremath{<} 0.001 & +2.02 [1.43, 2.55]\newline{}\ensuremath{<} 0.001 & \ensuremath{-}0.0130 [\ensuremath{-}0.0144, \ensuremath{-}0.0115]\newline{}\ensuremath{<} 0.001 & +0.96 [0.85, 1.07]\newline{}\ensuremath{<} 0.001 \\
Citation-order weights & \ensuremath{-}0.0341 [\ensuremath{-}0.0437, \ensuremath{-}0.0244]\newline{}\ensuremath{<} 0.001 & +2.06 [1.46, 2.60]\newline{}\ensuremath{<} 0.001 & \ensuremath{-}0.0116 [\ensuremath{-}0.0130, \ensuremath{-}0.0102]\newline{}\ensuremath{<} 0.001 & +0.87 [0.76, 0.98]\newline{}\ensuremath{<} 0.001 \\
Answer text & \ensuremath{-}0.0447 [\ensuremath{-}0.0538, \ensuremath{-}0.0349]\newline{}\ensuremath{<} 0.001 & +2.10 [1.51, 2.63]\newline{}\ensuremath{<} 0.001 & \ensuremath{-}0.0903 [\ensuremath{-}0.0938, \ensuremath{-}0.0868]\newline{}\ensuremath{<} 0.001 & +4.03 [3.83, 4.24]\newline{}\ensuremath{<} 0.001 \\
\addlinespace[3pt]
\multicolumn{5}{@{}p{\linewidth}@{}}{\textit{Answer weight \textemdash{} cited-source measure}} \\*
Half answer weight & \ensuremath{-}0.0278 [\ensuremath{-}0.0379, \ensuremath{-}0.0172]\newline{}\ensuremath{<} 0.001 & +1.68 [1.04, 2.26]\newline{}\ensuremath{<} 0.001 & \ensuremath{-}0.0082 [\ensuremath{-}0.0096, \ensuremath{-}0.0068]\newline{}\ensuremath{<} 0.001 & +0.61 [0.50, 0.71]\newline{}\ensuremath{<} 0.001 \\
Double answer weight & \ensuremath{-}0.0405 [\ensuremath{-}0.0496, \ensuremath{-}0.0312]\newline{}\ensuremath{<} 0.001 & +2.41 [1.85, 2.93]\newline{}\ensuremath{<} 0.001 & \ensuremath{-}0.0161 [\ensuremath{-}0.0175, \ensuremath{-}0.0146]\newline{}\ensuremath{<} 0.001 & +1.19 [1.08, 1.31]\newline{}\ensuremath{<} 0.001 \\
\end{longtable}
\endgroup

\Needspace{10\baselineskip}
\noindent\textbf{\textbf{B. Topic popularity}}\par
\smallskip

% SI physical table block 20; authoritative Table S16.
\begingroup
\fontsize{9bp}{11.5pt}\selectfont
\setlength{\tabcolsep}{3pt}
\renewcommand{\arraystretch}{1.13}
\setlength{\LTleft}{0pt}\setlength{\LTright}{0pt}
\setlength{\LTpre}{4pt}\setlength{\LTpost}{5pt}
\renewcommand{\theHtable}{supplement.block20.\arabic{table}}
\begin{longtable}{@{}>{\raggedright\arraybackslash}p{\dimexpr0.1863780\linewidth-2\tabcolsep\relax}>{\raggedright\arraybackslash}p{\dimexpr0.2444562\linewidth-2\tabcolsep\relax}>{\raggedright\arraybackslash}p{\dimexpr0.2000000\linewidth-2\tabcolsep\relax}>{\raggedright\arraybackslash}p{\dimexpr0.1845829\linewidth-2\tabcolsep\relax}>{\raggedright\arraybackslash}p{\dimexpr0.1845829\linewidth-2\tabcolsep\relax}@{}}
\toprule
\textbf{\textbf{Answer representation}} & \textbf{\textbf{Aggregate}\newline{}\textbf{mean rank}} & \textbf{\textbf{Aggregate}\newline{}\textbf{top-10 share}} & \textbf{\textbf{Individual}\newline{}\textbf{mean rank}} & \textbf{\textbf{Individual}\newline{}\textbf{top-10 share}} \\
\midrule
\endfirsthead
\multicolumn{5}{@{}l@{}}{\textbf{Table S16 (continued)}}\\
\toprule
\textbf{\textbf{Answer representation}} & \textbf{\textbf{Aggregate}\newline{}\textbf{mean rank}} & \textbf{\textbf{Aggregate}\newline{}\textbf{top-10 share}} & \textbf{\textbf{Individual}\newline{}\textbf{mean rank}} & \textbf{\textbf{Individual}\newline{}\textbf{top-10 share}} \\
\midrule
\endhead
\midrule
\multicolumn{5}{@{}r@{}}{Continued on next page}\\
\endfoot
\bottomrule
\endlastfoot
\addlinespace[3pt]
\multicolumn{5}{@{}p{\linewidth}@{}}{\textit{Content measurement \textemdash{} all available answers}} \\*
Cited sources (main) & +1.31 [1.03, 1.58]\newline{}\ensuremath{<} 0.001 & \ensuremath{-}4.4 [\ensuremath{-}5.4, \ensuremath{-}3.4]\newline{}\ensuremath{<} 0.001 & +1.16 [0.88, 1.45]\newline{}\ensuremath{<} 0.001 & \ensuremath{-}4.0 [\ensuremath{-}5.0, \ensuremath{-}3.0]\newline{}\ensuremath{<} 0.001 \\
Citation-order weights & +1.33 [1.05, 1.60]\newline{}\ensuremath{<} 0.001 & \ensuremath{-}4.5 [\ensuremath{-}5.4, \ensuremath{-}3.5]\newline{}\ensuremath{<} 0.001 & +1.18 [0.89, 1.46]\newline{}\ensuremath{<} 0.001 & \ensuremath{-}4.0 [\ensuremath{-}5.0, \ensuremath{-}3.0]\newline{}\ensuremath{<} 0.001 \\
Answer text & +1.88 [1.61, 2.15]\newline{}\ensuremath{<} 0.001 & \ensuremath{-}5.7 [\ensuremath{-}6.7, \ensuremath{-}4.8]\newline{}\ensuremath{<} 0.001 & +1.74 [1.46, 2.02]\newline{}\ensuremath{<} 0.001 & \ensuremath{-}5.3 [\ensuremath{-}6.3, \ensuremath{-}4.3]\newline{}\ensuremath{<} 0.001 \\
\addlinespace[3pt]
\multicolumn{5}{@{}p{\linewidth}@{}}{\textit{Content measurement \textemdash{} the same 10,948 answers}} \\*
Cited sources (main) & +1.30 [1.02, 1.57]\newline{}\ensuremath{<} 0.001 & \ensuremath{-}4.4 [\ensuremath{-}5.4, \ensuremath{-}3.4]\newline{}\ensuremath{<} 0.001 & +1.15 [0.87, 1.44]\newline{}\ensuremath{<} 0.001 & \ensuremath{-}3.9 [\ensuremath{-}5.0, \ensuremath{-}3.0]\newline{}\ensuremath{<} 0.001 \\
Citation-order weights & +1.31 [1.03, 1.60]\newline{}\ensuremath{<} 0.001 & \ensuremath{-}4.4 [\ensuremath{-}5.4, \ensuremath{-}3.5]\newline{}\ensuremath{<} 0.001 & +1.17 [0.89, 1.45]\newline{}\ensuremath{<} 0.001 & \ensuremath{-}4.0 [\ensuremath{-}5.0, \ensuremath{-}3.0]\newline{}\ensuremath{<} 0.001 \\
Answer text & +1.80 [1.52, 2.07]\newline{}\ensuremath{<} 0.001 & \ensuremath{-}5.6 [\ensuremath{-}6.5, \ensuremath{-}4.6]\newline{}\ensuremath{<} 0.001 & +1.65 [1.37, 1.94]\newline{}\ensuremath{<} 0.001 & \ensuremath{-}5.0 [\ensuremath{-}6.1, \ensuremath{-}4.0]\newline{}\ensuremath{<} 0.001 \\
\addlinespace[3pt]
\multicolumn{5}{@{}p{\linewidth}@{}}{\textit{Answer weight \textemdash{} cited-source measure}} \\*
Half answer weight & +1.08 [0.79, 1.39]\newline{}\ensuremath{<} 0.001 & \ensuremath{-}3.9 [\ensuremath{-}4.9, \ensuremath{-}2.8]\newline{}\ensuremath{<} 0.001 & +1.07 [0.77, 1.35]\newline{}\ensuremath{<} 0.001 & \ensuremath{-}3.7 [\ensuremath{-}4.7, \ensuremath{-}2.7]\newline{}\ensuremath{<} 0.001 \\
Double answer weight & +1.56 [1.29, 1.83]\newline{}\ensuremath{<} 0.001 & \ensuremath{-}5.0 [\ensuremath{-}6.0, \ensuremath{-}4.1]\newline{}\ensuremath{<} 0.001 & +1.28 [0.99, 1.57]\newline{}\ensuremath{<} 0.001 & \ensuremath{-}4.2 [\ensuremath{-}5.2, \ensuremath{-}3.2]\newline{}\ensuremath{<} 0.001 \\
\end{longtable}
\endgroup

{\fontsize{8bp}{11pt}\selectfont\interlinepenalty=10000\noindent \textbf{Note.} Cells report Treatment-Control differences in concentration and popularity of total information consumption, with 95\% intervals and P values without adjustment for multiple comparisons (sections S30 and S31). Top-10 differences are percentage points. Individual Treatment estimates include 6,676 readers with cited-source measures, 6,764 with answer-text measures, and 6,624 in the same-answer comparison.\par}
\medskip

The findings therefore persist when answers are measured differently or counted as less or more than one article. Section S8 examines consumption using text volume and time spent.

\Needspace{6\baselineskip}
\subsection*{S13. What changes when only opened articles are counted?}
\phantomsection\label{sec:S13}

Article consumption alone shifts toward less-popular topics at both the individual and aggregate levels. Aggregate article consumption also becomes less concentrated, while the concentration differences within individual readers are small and not statistically significant (Table S17).

\Needspace{12\baselineskip}
\setcounter{table}{16}
\captionof{table}{Concentration and popularity of article consumption}\label{tab:S17}

% SI physical table block 21; authoritative Table S17.
\begingroup
\fontsize{9bp}{11.5pt}\selectfont
\setlength{\tabcolsep}{3pt}
\renewcommand{\arraystretch}{1.13}
\setlength{\LTleft}{0pt}\setlength{\LTright}{0pt}
\setlength{\LTpre}{4pt}\setlength{\LTpost}{5pt}
\renewcommand{\theHtable}{supplement.block21.\arabic{table}}
\begin{longtable}{@{}>{\raggedright\arraybackslash}p{\dimexpr0.3077086\linewidth-2\tabcolsep\relax}>{\raggedright\arraybackslash}p{\dimexpr0.1122492\linewidth-2\tabcolsep\relax}>{\raggedright\arraybackslash}p{\dimexpr0.1184794\linewidth-2\tabcolsep\relax}>{\raggedright\arraybackslash}p{\dimexpr0.3615628\linewidth-2\tabcolsep\relax}>{\raggedright\arraybackslash}p{\dimexpr0.1000000\linewidth-2\tabcolsep\relax}@{}}
\toprule
\textbf{\textbf{Measure}} & \textbf{\textbf{Control}} & \textbf{\textbf{Treatment}} & \textbf{\textbf{Difference [95\% CI]}} & \textbf{\textbf{Holm P}} \\
\midrule
\endfirsthead
\multicolumn{5}{@{}l@{}}{\textbf{Table S17 (continued)}}\\
\toprule
\textbf{\textbf{Measure}} & \textbf{\textbf{Control}} & \textbf{\textbf{Treatment}} & \textbf{\textbf{Difference [95\% CI]}} & \textbf{\textbf{Holm P}} \\
\midrule
\endhead
\midrule
\multicolumn{5}{@{}r@{}}{Continued on next page}\\
\endfoot
\bottomrule
\endlastfoot
Individual Gini concentration & 0.953 & 0.953 & \ensuremath{-}0.0003 [\ensuremath{-}0.0021, 0.0014] & 1.000 \\
Individual Effective number of topics & 4.58 & 4.61 & +0.026 [\ensuremath{-}0.104, 0.157] & 1.000 \\
Individual Mean popularity rank & 17.81 & 18.31 & +0.503 [0.143, 0.860] & 0.029 \\
Individual Top-10 popular share (\%) & 40.7 & 38.1 & \ensuremath{-}2.58 [\ensuremath{-}3.86, \ensuremath{-}1.31] & 0.002 \\
Aggregate Gini concentration & 0.414 & 0.399 & \ensuremath{-}0.0153 [\ensuremath{-}0.0261, \ensuremath{-}0.0029] & 0.029 \\
Aggregate Effective number of topics & 44.37 & 45.31 & +0.938 [0.166, 1.559] & 0.029 \\
Aggregate Mean popularity rank & 17.75 & 18.44 & +0.694 [0.351, 1.036] & \ensuremath{<} 0.001 \\
Aggregate Top-10 popular share (\%) & 41.1 & 38.2 & \ensuremath{-}2.89 [\ensuremath{-}4.07, \ensuremath{-}1.70] & \ensuremath{<} 0.001 \\
\end{longtable}
\endgroup

{\fontsize{8bp}{11pt}\selectfont\interlinepenalty=10000\noindent \textbf{Note.} Individual estimates use 7,992 Control and 3,423 Treatment readers with at least two article opens and some measured topic consumption (section S28). Aggregate results combine all article consumption with topic estimates. Confidence intervals and eight-comparison Holm-adjusted P values follow sections S30 and S31.\par}
\medskip

\Needspace{6\baselineskip}
\subsection*{S14. How do AI answers broaden consumption and add less-popular topics?}
\phantomsection\label{sec:S14}

\Needspace{3\baselineskip}
\paragraph*{What answers add for the same readers}

Among Treatment readers who opened articles, adding AI answers to their article consumption spreads consumption across more topics and shifts it toward less-popular topics (Table S18). Aggregate consumption shows the same pattern.

\Needspace{12\baselineskip}
\setcounter{table}{17}
\captionof{table}{Adding answers to article consumption}\label{tab:S18}

\Needspace{10\baselineskip}
\noindent\textbf{\textbf{A. Within individual readers}}\par
\smallskip

% SI physical table block 22; authoritative Table S18.
\begingroup
\fontsize{9bp}{11.5pt}\selectfont
\setlength{\tabcolsep}{3pt}
\renewcommand{\arraystretch}{1.13}
\setlength{\LTleft}{0pt}\setlength{\LTright}{0pt}
\setlength{\LTpre}{4pt}\setlength{\LTpost}{5pt}
\renewcommand{\theHtable}{supplement.block22.\arabic{table}}
\begin{longtable}{@{}>{\raggedright\arraybackslash}p{\dimexpr0.2434002\linewidth-2\tabcolsep\relax}>{\raggedright\arraybackslash}p{\dimexpr0.2027455\linewidth-2\tabcolsep\relax}>{\raggedright\arraybackslash}p{\dimexpr0.1869060\linewidth-2\tabcolsep\relax}>{\raggedright\arraybackslash}p{\dimexpr0.1710665\linewidth-2\tabcolsep\relax}>{\raggedright\arraybackslash}p{\dimexpr0.1958817\linewidth-2\tabcolsep\relax}@{}}
\toprule
\textbf{\textbf{Measure}} & \textbf{\textbf{Articles only}} & \textbf{\textbf{Articles and answers}} & \textbf{\textbf{Difference}} & \textbf{\textbf{95\% CI}} \\
\midrule
\endfirsthead
\multicolumn{5}{@{}l@{}}{\textbf{Table S18 (continued)}}\\
\toprule
\textbf{\textbf{Measure}} & \textbf{\textbf{Articles only}} & \textbf{\textbf{Articles and answers}} & \textbf{\textbf{Difference}} & \textbf{\textbf{95\% CI}} \\
\midrule
\endhead
\midrule
\multicolumn{5}{@{}r@{}}{Continued on next page}\\
\endfoot
\bottomrule
\endlastfoot
Gini coefficient & 0.9632 & 0.9410 & \ensuremath{-}0.0223 & [\ensuremath{-}0.0229, \ensuremath{-}0.0216] \\
Effective number of topics & 3.81 & 5.48 & +1.67 & [1.62, 1.72] \\
Mean popularity rank & 18.49 & 18.81 & +0.32 & [0.23, 0.41] \\
Top-10 popular share (\%) & 37.85 & 37.12 & \ensuremath{-}0.73 & [\ensuremath{-}1.05, \ensuremath{-}0.42] \\
\end{longtable}
\endgroup

{\fontsize{8bp}{11pt}\selectfont\interlinepenalty=10000\noindent \textbf{Note.} Panel A compares articles alone with articles plus answers for the same 6,025 Treatment readers who opened an article with estimated topic shares. Differences are total minus article consumption, with top-10 differences in percentage points. Paired reader-bootstrap CIs keep each reader's measurements together (section S32). The effective-topic and mean-popularity-rank differences have unadjusted P \ensuremath{<} 0.001 from paired t tests. This is not a Treatment-Control comparison.\par}
\medskip

\Needspace{10\baselineskip}
\noindent\textbf{\textbf{B. Within the Treatment group in aggregate}}\par
\smallskip

% SI physical table block 23; authoritative Table S18.
\begingroup
\fontsize{9bp}{11.5pt}\selectfont
\setlength{\tabcolsep}{3pt}
\renewcommand{\arraystretch}{1.13}
\setlength{\LTleft}{0pt}\setlength{\LTright}{0pt}
\setlength{\LTpre}{4pt}\setlength{\LTpost}{5pt}
\renewcommand{\theHtable}{supplement.block23.\arabic{table}}
\begin{longtable}{@{}>{\raggedright\arraybackslash}p{\dimexpr0.2529039\linewidth-2\tabcolsep\relax}>{\raggedright\arraybackslash}p{\dimexpr0.1805702\linewidth-2\tabcolsep\relax}>{\raggedright\arraybackslash}p{\dimexpr0.2203801\linewidth-2\tabcolsep\relax}>{\raggedright\arraybackslash}p{\dimexpr0.3461457\linewidth-2\tabcolsep\relax}@{}}
\toprule
\textbf{\textbf{Measure}} & \textbf{\textbf{Articles only}} & \textbf{\textbf{Articles and answers}} & \textbf{\textbf{Difference [95\% CI]}} \\
\midrule
\endfirsthead
\multicolumn{4}{@{}l@{}}{\textbf{Table S18 (continued)}}\\
\toprule
\textbf{\textbf{Measure}} & \textbf{\textbf{Articles only}} & \textbf{\textbf{Articles and answers}} & \textbf{\textbf{Difference [95\% CI]}} \\
\midrule
\endhead
\midrule
\multicolumn{4}{@{}r@{}}{Continued on next page}\\
\endfoot
\bottomrule
\endlastfoot
Gini coefficient & 0.3985 & 0.3798 & \ensuremath{-}0.0187 [\ensuremath{-}0.0237, \ensuremath{-}0.0152] \\
Effective number of topics & 45.31 & 46.42 & +1.11 [0.91, 1.42] \\
Mean popularity rank & 18.44 & 19.06 & +0.62 [0.49, 0.74] \\
Top-10 popular share (\%) & 38.19 & 36.66 & \ensuremath{-}1.53 pp [\ensuremath{-}1.93, \ensuremath{-}1.12] \\
\end{longtable}
\endgroup

{\fontsize{8bp}{11pt}\selectfont\interlinepenalty=10000\noindent \textbf{Note.} Panel B combines consumption across all 10,922 Treatment readers. Differences are total minus article consumption, with paired reader-bootstrap CIs (section S32). This describes what answers add to observed consumption without isolating their causal effect.\par}
\medskip

\Needspace{3\baselineskip}
\paragraph*{How answer topics compare with Control articles}

Using cited articles to estimate answer topics, aggregate AI-answer consumption is less concentrated and falls on less-popular topics than Control article consumption (Table S19). The popularity difference also persists when readers receive equal weight.

\Needspace{12\baselineskip}
\setcounter{table}{18}
\captionof{table}{Concentration and popularity of answers and Control articles}\label{tab:S19}

% SI physical table block 24; authoritative Table S19.
\begingroup
\fontsize{9bp}{11.5pt}\selectfont
\setlength{\tabcolsep}{3pt}
\renewcommand{\arraystretch}{1.13}
\setlength{\LTleft}{0pt}\setlength{\LTright}{0pt}
\setlength{\LTpre}{4pt}\setlength{\LTpost}{5pt}
\renewcommand{\theHtable}{supplement.block24.\arabic{table}}
\begin{longtable}{@{}>{\raggedright\arraybackslash}p{\dimexpr0.3669483\linewidth-2\tabcolsep\relax}>{\raggedright\arraybackslash}p{\dimexpr0.1520591\linewidth-2\tabcolsep\relax}>{\raggedright\arraybackslash}p{\dimexpr0.1900739\linewidth-2\tabcolsep\relax}>{\raggedright\arraybackslash}p{\dimexpr0.2909187\linewidth-2\tabcolsep\relax}@{}}
\toprule
\textbf{\textbf{Measure}} & \textbf{\textbf{Control articles}} & \textbf{\textbf{AI answers}} & \textbf{\textbf{Difference [95\% CI]}} \\
\midrule
\endfirsthead
\multicolumn{4}{@{}l@{}}{\textbf{Table S19 (continued)}}\\
\toprule
\textbf{\textbf{Measure}} & \textbf{\textbf{Control articles}} & \textbf{\textbf{AI answers}} & \textbf{\textbf{Difference [95\% CI]}} \\
\midrule
\endhead
\midrule
\multicolumn{4}{@{}r@{}}{Continued on next page}\\
\endfoot
\bottomrule
\endlastfoot
Gini coefficient & 0.4139 & 0.3624 & \ensuremath{-}0.0515 [\ensuremath{-}0.0603, \ensuremath{-}0.0420] \\
Effective number of topics & 44.37 & 47.38 & +3.01 [2.47, 3.52] \\
Mean popularity rank & 17.75 & 19.89 & +2.14 [1.87, 2.40] \\
Top-10 popular share (\%) & 41.08 & 34.64 & \ensuremath{-}6.44 pp [\ensuremath{-}7.34, \ensuremath{-}5.52] \\
Mean popularity rank (equal reader weights) & 18.04 & 19.94 & +1.91 [1.66, 2.16] \\
\end{longtable}
\endgroup

{\fontsize{8bp}{11pt}\selectfont\interlinepenalty=10000\noindent \textbf{Note.} The first four rows pool 11,092 distinct answers from 8,844 Treatment readers and 36,303 article opens by 16,094 Control readers after the first search. Repeated displays of an answer count once. The final row averages each reader's mean popularity rank, giving readers equal weight. Differences are answers minus Control articles, with reader-bootstrap CIs independently within groups (section S32).\par}
\medskip

Table S20 compares two cited-source measures of answer content: equal source weights and citation-order weights. Under both measures, answer consumption is less concentrated and falls on less-popular topics than Control article consumption. These answer-only comparisons are distinct from the total-consumption checks in Table S16, which also include opened articles.

\Needspace{12\baselineskip}
\setcounter{table}{19}
\captionof{table}{Answer concentration and popularity under two cited-source measurements}\label{tab:S20}

\Needspace{10\baselineskip}
\noindent\textbf{\textbf{A. All answers with cited-source topic estimates}}\par
\smallskip

% SI physical table block 25; authoritative Table S20.
\begingroup
\fontsize{9bp}{11.5pt}\selectfont
\setlength{\tabcolsep}{3pt}
\renewcommand{\arraystretch}{1.13}
\setlength{\LTleft}{0pt}\setlength{\LTright}{0pt}
\setlength{\LTpre}{4pt}\setlength{\LTpost}{5pt}
\renewcommand{\theHtable}{supplement.block25.\arabic{table}}
\begin{longtable}{@{}>{\raggedright\arraybackslash}p{\dimexpr0.2077086\linewidth-2\tabcolsep\relax}>{\raggedright\arraybackslash}p{\dimexpr0.2077086\linewidth-2\tabcolsep\relax}>{\raggedright\arraybackslash}p{\dimexpr0.2077086\linewidth-2\tabcolsep\relax}>{\raggedright\arraybackslash}p{\dimexpr0.1922914\linewidth-2\tabcolsep\relax}>{\raggedright\arraybackslash}p{\dimexpr0.1845829\linewidth-2\tabcolsep\relax}@{}}
\toprule
\textbf{\textbf{Answer measurement}} & \textbf{\textbf{Gini coefficient}} & \textbf{\textbf{Effective number of topics}} & \textbf{\textbf{Mean popularity rank}} & \textbf{\textbf{Top-10 share (pp)}} \\
\midrule
\endfirsthead
\multicolumn{5}{@{}l@{}}{\textbf{Table S20 (continued)}}\\
\toprule
\textbf{\textbf{Answer measurement}} & \textbf{\textbf{Gini coefficient}} & \textbf{\textbf{Effective number of topics}} & \textbf{\textbf{Mean popularity rank}} & \textbf{\textbf{Top-10 share (pp)}} \\
\midrule
\endhead
\midrule
\multicolumn{5}{@{}r@{}}{Continued on next page}\\
\endfoot
\bottomrule
\endlastfoot
Cited sources (main) & \ensuremath{-}0.0515 [\ensuremath{-}0.0603, \ensuremath{-}0.0420] & +3.01 [2.47, 3.52] & +2.14 [1.87, 2.40] & \ensuremath{-}6.44 [\ensuremath{-}7.34, \ensuremath{-}5.52] \\
Citation-order weighted & \ensuremath{-}0.0525 [\ensuremath{-}0.0613, \ensuremath{-}0.0431] & +3.07 [2.53, 3.58] & +2.18 [1.91, 2.45] & \ensuremath{-}6.59 [\ensuremath{-}7.49, \ensuremath{-}5.67] \\
\end{longtable}
\endgroup

\Needspace{10\baselineskip}
\noindent\textbf{\textbf{B. Answers measured by all three methods in section S12}}\par
\smallskip

% SI physical table block 26; authoritative Table S20.
\begingroup
\fontsize{9bp}{11.5pt}\selectfont
\setlength{\tabcolsep}{3pt}
\renewcommand{\arraystretch}{1.13}
\setlength{\LTleft}{0pt}\setlength{\LTright}{0pt}
\setlength{\LTpre}{4pt}\setlength{\LTpost}{5pt}
\renewcommand{\theHtable}{supplement.block26.\arabic{table}}
\begin{longtable}{@{}>{\raggedright\arraybackslash}p{\dimexpr0.2077086\linewidth-2\tabcolsep\relax}>{\raggedright\arraybackslash}p{\dimexpr0.2077086\linewidth-2\tabcolsep\relax}>{\raggedright\arraybackslash}p{\dimexpr0.2077086\linewidth-2\tabcolsep\relax}>{\raggedright\arraybackslash}p{\dimexpr0.1922914\linewidth-2\tabcolsep\relax}>{\raggedright\arraybackslash}p{\dimexpr0.1845829\linewidth-2\tabcolsep\relax}@{}}
\toprule
\textbf{\textbf{Answer measurement}} & \textbf{\textbf{Gini coefficient}} & \textbf{\textbf{Effective number of topics}} & \textbf{\textbf{Mean popularity rank}} & \textbf{\textbf{Top-10 share (pp)}} \\
\midrule
\endfirsthead
\multicolumn{5}{@{}l@{}}{\textbf{Table S20 (continued)}}\\
\toprule
\textbf{\textbf{Answer measurement}} & \textbf{\textbf{Gini coefficient}} & \textbf{\textbf{Effective number of topics}} & \textbf{\textbf{Mean popularity rank}} & \textbf{\textbf{Top-10 share (pp)}} \\
\midrule
\endhead
\midrule
\multicolumn{5}{@{}r@{}}{Continued on next page}\\
\endfoot
\bottomrule
\endlastfoot
Cited sources (main) & \ensuremath{-}0.0504 [\ensuremath{-}0.0595, \ensuremath{-}0.0412] & +2.96 [2.42, 3.48] & +2.11 [1.84, 2.39] & \ensuremath{-}6.38 [\ensuremath{-}7.31, \ensuremath{-}5.44] \\
Citation-order weighted & \ensuremath{-}0.0515 [\ensuremath{-}0.0606, \ensuremath{-}0.0422] & +3.02 [2.48, 3.55] & +2.16 [1.88, 2.43] & \ensuremath{-}6.53 [\ensuremath{-}7.46, \ensuremath{-}5.59] \\
\end{longtable}
\endgroup

{\fontsize{8bp}{11pt}\selectfont\interlinepenalty=10000\noindent \textbf{Note.} Entries are differences from Control article consumption, with reader-bootstrap CIs and no adjustment for multiple comparisons (section S32). Panel A uses all answers measured by both cited-source methods. Panel B uses the same-answer subset from section S12. Control values are reported in Table S19. This table compares only the two cited-source measures; Table S16 separately examines total consumption under all three answer measures.\par}
\medskip

\Needspace{6\baselineskip}
\subsection*{S15. How does cited-source article consumption differ from other article consumption?}
\phantomsection\label{sec:S15}

Within Treatment, aggregate consumption of cited-source articles is less concentrated and favors less-popular topics than consumption of other articles (Table S21, panel A). The comparison includes article opens after the first search, including other articles reached through conventional results, browsing, or later article opens.

\Needspace{12\baselineskip}
\setcounter{table}{20}
\captionof{table}{Article consumption across all routes}\label{tab:S21}

\Needspace{10\baselineskip}
\noindent\textbf{\textbf{A. Aggregate consumption through two routes within Treatment}}\par
\smallskip

% SI physical table block 27; authoritative Table S21.
\begingroup
\fontsize{9bp}{11.5pt}\selectfont
\setlength{\tabcolsep}{3pt}
\renewcommand{\arraystretch}{1.13}
\setlength{\LTleft}{0pt}\setlength{\LTright}{0pt}
\setlength{\LTpre}{4pt}\setlength{\LTpost}{5pt}
\renewcommand{\theHtable}{supplement.block27.\arabic{table}}
\begin{longtable}{@{}>{\raggedright\arraybackslash}p{\dimexpr0.3076030\linewidth-2\tabcolsep\relax}>{\raggedright\arraybackslash}p{\dimexpr0.1692714\linewidth-2\tabcolsep\relax}>{\raggedright\arraybackslash}p{\dimexpr0.1692714\linewidth-2\tabcolsep\relax}>{\raggedright\arraybackslash}p{\dimexpr0.3538543\linewidth-2\tabcolsep\relax}@{}}
\toprule
\textbf{\textbf{Measure}} & \textbf{\textbf{Other Treatment articles}} & \textbf{\textbf{Cited-source articles}} & \textbf{\textbf{Cited minus other [95\% CI]}} \\
\midrule
\endfirsthead
\multicolumn{4}{@{}l@{}}{\textbf{Table S21 (continued)}}\\
\toprule
\textbf{\textbf{Measure}} & \textbf{\textbf{Other Treatment articles}} & \textbf{\textbf{Cited-source articles}} & \textbf{\textbf{Cited minus other [95\% CI]}} \\
\midrule
\endhead
\midrule
\multicolumn{4}{@{}r@{}}{Continued on next page}\\
\endfoot
\bottomrule
\endlastfoot
Gini coefficient & 0.408 & 0.377 & \ensuremath{-}0.031 [\ensuremath{-}0.045, \ensuremath{-}0.011] \\
Effective number of topics & 44.65 & 46.61 & +1.96 [0.80, 2.78] \\
Mean popularity rank & 18.11 & 19.63 & +1.52 [1.01, 2.03] \\
Top-10 popular share (\%) & 39.75 & 32.66 & \ensuremath{-}7.09 pp [\ensuremath{-}8.78, \ensuremath{-}5.39] \\
\end{longtable}
\endgroup

{\fontsize{8bp}{11pt}\selectfont\interlinepenalty=10000\noindent \textbf{Note.} Panel A pools 12,342 other Treatment article opens and 3,443 cited-source opens with estimated topic shares. Citation attribution follows section S27. Differences are cited-source minus other article consumption, with reader-bootstrap CIs (section S32).\par}
\medskip

Panel B compares the popularity of cited-source articles consumed by Treatment readers with that of articles consumed by Control readers. Cited-source consumption favors less-popular topics whether the comparison includes readers with at least one or at least two relevant opens.

\Needspace{10\baselineskip}
\noindent\textbf{\textbf{B. Individual popularity of Treatment cited articles and Control articles}}\par
\smallskip

% SI physical table block 28; authoritative Table S21.
\begingroup
\fontsize{9bp}{11.5pt}\selectfont
\setlength{\tabcolsep}{3pt}
\renewcommand{\arraystretch}{1.13}
\setlength{\LTleft}{0pt}\setlength{\LTright}{0pt}
\setlength{\LTpre}{4pt}\setlength{\LTpost}{5pt}
\renewcommand{\theHtable}{supplement.block28.\arabic{table}}
\begin{longtable}{@{}>{\raggedright\arraybackslash}p{\dimexpr0.2615628\linewidth-2\tabcolsep\relax}>{\raggedright\arraybackslash}p{\dimexpr0.1529039\linewidth-2\tabcolsep\relax}>{\raggedright\arraybackslash}p{\dimexpr0.2851109\linewidth-2\tabcolsep\relax}>{\raggedright\arraybackslash}p{\dimexpr0.2185850\linewidth-2\tabcolsep\relax}>{\raggedright\arraybackslash}p{\dimexpr0.0818374\linewidth-2\tabcolsep\relax}@{}}
\toprule
\textbf{\textbf{Measure}} & \textbf{\textbf{Control}\newline{}\textbf{articles}} & \textbf{\textbf{Treatment}\newline{}\textbf{cited-source articles}} & \textbf{\textbf{Difference [95\% CI]}} & \textbf{\textbf{P}} \\
\midrule
\endfirsthead
\multicolumn{5}{@{}l@{}}{\textbf{Table S21 (continued)}}\\
\toprule
\textbf{\textbf{Measure}} & \textbf{\textbf{Control}\newline{}\textbf{articles}} & \textbf{\textbf{Treatment}\newline{}\textbf{cited-source articles}} & \textbf{\textbf{Difference [95\% CI]}} & \textbf{\textbf{P}} \\
\midrule
\endhead
\midrule
\multicolumn{5}{@{}r@{}}{Continued on next page}\\
\endfoot
\bottomrule
\endlastfoot
\addlinespace[3pt]
\multicolumn{5}{@{}p{\linewidth}@{}}{\textit{At least two relevant article opens per reader (Control n = 7,801; Treatment n = 574)}} \\*
Mean popularity rank & 17.86 & 19.88 & +2.02 [1.19, 2.84] & \ensuremath{<} 0.001 \\
Top-10 popular share (\%) & 40.56 & 30.07 & \ensuremath{-}10.49 pp [\ensuremath{-}13.25, \ensuremath{-}7.66] & \ensuremath{<} 0.001 \\
\addlinespace[3pt]
\multicolumn{5}{@{}p{\linewidth}@{}}{\textit{At least one relevant article open per reader (Control n = 16,106; Treatment n = 2,547)}} \\*
Mean popularity rank & 18.04 & 19.53 & +1.49 [1.06, 1.92] & \ensuremath{<} 0.001 \\
Top-10 popular share (\%) & 40.17 & 33.73 & \ensuremath{-}6.44 pp [\ensuremath{-}7.94, \ensuremath{-}4.94] & \ensuremath{<} 0.001 \\
\end{longtable}
\endgroup

{\fontsize{8bp}{11pt}\selectfont\interlinepenalty=10000\noindent \textbf{Note.} Relevant opens are cited-source articles in Treatment and articles from any route in Control, with topic estimates and after the first search. Repeated opens count toward the inclusion criteria. Each reader receives equal weight. Samples differ slightly from section S13 because this comparison uses navigation records. Confidence intervals and two-measure Holm-adjusted Welch tests follow section S32. These selected-reader comparisons do not isolate the causal effect of citations.\par}
\medskip

\Needspace{6\baselineskip}
\subsection*{S16. Do readers ask about, or does search offer, less-popular topics?}
\phantomsection\label{sec:S16}

The shift toward less-popular consumption is not mirrored in readers' queries or in the articles offered for the same query (Tables S22 and S23). Query topics instead lean toward more-popular topics. For the offered lists, differences in mean popularity rank and top-10 share are small and not statistically significant.

\Needspace{3\baselineskip}
\paragraph*{Topics readers request}

We estimate query topics in three ways to check whether the result depends on how short queries are classified. The methods use the fitted topic model, the topic most similar to each query, or several topics weighted by similarity. All use the same Control popularity ranking (Table S22).

\Needspace{12\baselineskip}
\setcounter{table}{21}
\captionof{table}{Popularity of requested topics}\label{tab:S22}

% SI physical table block 29; authoritative Table S22.
\begingroup
\fontsize{9bp}{11.5pt}\selectfont
\setlength{\tabcolsep}{3pt}
\renewcommand{\arraystretch}{1.13}
\setlength{\LTleft}{0pt}\setlength{\LTright}{0pt}
\setlength{\LTpre}{4pt}\setlength{\LTpost}{5pt}
\renewcommand{\theHtable}{supplement.block29.\arabic{table}}
\begin{longtable}{@{}>{\raggedright\arraybackslash}p{\dimexpr0.4049630\linewidth-2\tabcolsep\relax}>{\raggedright\arraybackslash}p{\dimexpr0.1520591\linewidth-2\tabcolsep\relax}>{\raggedright\arraybackslash}p{\dimexpr0.1615628\linewidth-2\tabcolsep\relax}>{\raggedright\arraybackslash}p{\dimexpr0.1615628\linewidth-2\tabcolsep\relax}>{\raggedright\arraybackslash}p{\dimexpr0.1198522\linewidth-2\tabcolsep\relax}@{}}
\toprule
\textbf{\textbf{Query topic measure}} & \textbf{\textbf{Control}} & \textbf{\textbf{Treatment}} & \textbf{\textbf{Difference}} & \textbf{\textbf{P}} \\
\midrule
\endfirsthead
\multicolumn{5}{@{}l@{}}{\textbf{Table S22 (continued)}}\\
\toprule
\textbf{\textbf{Query topic measure}} & \textbf{\textbf{Control}} & \textbf{\textbf{Treatment}} & \textbf{\textbf{Difference}} & \textbf{\textbf{P}} \\
\midrule
\endhead
\midrule
\multicolumn{5}{@{}r@{}}{Continued on next page}\\
\endfoot
\bottomrule
\endlastfoot
Topic model, equal reader weights & 19.69 & 19.12 & \ensuremath{-}0.56 & 0.153 \\
Closest topic, aggregate & \textemdash{} & \textemdash{} & \ensuremath{-}1.24 & \textemdash{} \\
Closest topic, equal reader weights & 27.76 & 26.75 & \ensuremath{-}1.01 & 0.002 \\
Several topics, aggregate & \textemdash{} & \textemdash{} & \ensuremath{-}0.43 & \textemdash{} \\
Several topics, equal reader weights & 28.98 & 28.59 & \ensuremath{-}0.39 & \ensuremath{<} 0.001 \\
Closest topic, reader mean (at least two queries) & 27.63 & 26.90 & \ensuremath{-}0.72 & 0.104 \\
\end{longtable}
\endgroup

{\fontsize{8bp}{11pt}\selectfont\interlinepenalty=10000\noindent \textbf{Note.} Higher rank means less-popular topics. The topic-model comparison uses 3,750 readers with at least two queries. Reader-weighted estimates first average queries within each reader, then average readers equally. Aggregate estimates weight queries equally. The final row restricts the closest-topic comparison to readers with at least two queries. Dashes indicate values not reported.\par}
\medskip

The direction of the query differences is opposite to the shift in consumption.

\Needspace{3\baselineskip}
\paragraph*{Articles offered for the same query}

For each query, we compare cited sources with conventional results, then the combined lists with conventional results alone. Neither comparison shows a statistically significant difference in mean popularity rank or the share on the ten most popular topics (Table S23).

\Needspace{12\baselineskip}
\setcounter{table}{22}
\captionof{table}{Topic popularity within the same query}\label{tab:S23}

% SI physical table block 30; authoritative Table S23.
\begingroup
\fontsize{9bp}{11.5pt}\selectfont
\setlength{\tabcolsep}{3pt}
\renewcommand{\arraystretch}{1.13}
\setlength{\LTleft}{0pt}\setlength{\LTright}{0pt}
\setlength{\LTpre}{4pt}\setlength{\LTpost}{5pt}
\renewcommand{\theHtable}{supplement.block30.\arabic{table}}
\begin{longtable}{@{}>{\raggedright\arraybackslash}p{\dimexpr0.2624076\linewidth-2\tabcolsep\relax}>{\raggedright\arraybackslash}p{\dimexpr0.0855333\linewidth-2\tabcolsep\relax}>{\raggedright\arraybackslash}p{\dimexpr0.1330517\linewidth-2\tabcolsep\relax}>{\raggedright\arraybackslash}p{\dimexpr0.2090813\linewidth-2\tabcolsep\relax}>{\raggedright\arraybackslash}p{\dimexpr0.1140444\linewidth-2\tabcolsep\relax}>{\raggedright\arraybackslash}p{\dimexpr0.1112988\linewidth-2\tabcolsep\relax}>{\raggedright\arraybackslash}p{\dimexpr0.0845829\linewidth-2\tabcolsep\relax}@{}}
\toprule
\textbf{\textbf{Comparison}} & \textbf{\textbf{Queries}} & \textbf{\textbf{Conventional}} & \textbf{\textbf{Cited or combined}} & \textbf{\textbf{Difference}} & \textbf{\textbf{95\% CI}} & \textbf{\textbf{P}} \\
\midrule
\endfirsthead
\multicolumn{7}{@{}l@{}}{\textbf{Table S23 (continued)}}\\
\toprule
\textbf{\textbf{Comparison}} & \textbf{\textbf{Queries}} & \textbf{\textbf{Conventional}} & \textbf{\textbf{Cited or combined}} & \textbf{\textbf{Difference}} & \textbf{\textbf{95\% CI}} & \textbf{\textbf{P}} \\
\midrule
\endhead
\midrule
\multicolumn{7}{@{}r@{}}{Continued on next page}\\
\endfoot
\bottomrule
\endlastfoot
Cited sources: mean rank & 3,253 & 19.37 & 19.37 & +0.002 & [\ensuremath{-}0.206, 0.211] & 0.834 \\
Cited sources: top-10 share (\%) & 3,253 & 35.93 & 36.50 & +0.57 & [\ensuremath{-}0.10, 1.24] & 0.385 \\
Combined lists: mean rank & 4,114 & 19.52 & 19.59 & +0.067 & [\ensuremath{-}0.033, 0.167] & 0.378 \\
Combined lists: top-10 share (\%) & 4,114 & 35.01 & 35.10 & +0.08 & [\ensuremath{-}0.23, 0.41] & 0.368 \\
\end{longtable}
\endgroup

{\fontsize{8bp}{11pt}\selectfont\interlinepenalty=10000\noindent \textbf{Note.} Each list contains at least two articles with estimated topic shares, and queries receive equal weight. Top-10 share uses the Control ranking in section S28; differences and CIs are percentage points. Paired query-bootstrap CIs and unadjusted Wilcoxon tests follow section S32. Available lists cover 10\% of Control and 26\% of Treatment searches for the cited-source comparison, and 13\% and 32\% for the combined-list comparison.\par}
\medskip

\Needspace{6\baselineskip}
\subsection*{S17. What do readers do next after a search?}
\phantomsection\label{sec:S17}

After a search, Treatment readers less often open a conventional-result article or turn to browsing. They more often open a cited-source article or search again. About 21\% of searches end the session in both groups (Table S24). Each search is assigned to one of these five next actions.

\Needspace{12\baselineskip}
\setcounter{table}{23}
\captionof{table}{The next action after a search}\label{tab:S24}

% SI physical table block 31; authoritative Table S24.
\begingroup
\fontsize{9bp}{11.5pt}\selectfont
\setlength{\tabcolsep}{3pt}
\renewcommand{\arraystretch}{1.13}
\setlength{\LTleft}{0pt}\setlength{\LTright}{0pt}
\setlength{\LTpre}{4pt}\setlength{\LTpost}{5pt}
\renewcommand{\theHtable}{supplement.block31.\arabic{table}}
\begin{longtable}{@{}>{\raggedright\arraybackslash}p{\dimexpr0.4000000\linewidth-2\tabcolsep\relax}>{\raggedright\arraybackslash}p{\dimexpr0.1615628\linewidth-2\tabcolsep\relax}>{\raggedright\arraybackslash}p{\dimexpr0.1615628\linewidth-2\tabcolsep\relax}>{\raggedright\arraybackslash}p{\dimexpr0.2768743\linewidth-2\tabcolsep\relax}@{}}
\toprule
\textbf{\textbf{Action}} & \textbf{\textbf{Control (\%)}} & \textbf{\textbf{Treatment (\%)}} & \textbf{\textbf{Difference in pp [95\% CI]}} \\
\midrule
\endfirsthead
\multicolumn{4}{@{}l@{}}{\textbf{Table S24 (continued)}}\\
\toprule
\textbf{\textbf{Action}} & \textbf{\textbf{Control (\%)}} & \textbf{\textbf{Treatment (\%)}} & \textbf{\textbf{Difference in pp [95\% CI]}} \\
\midrule
\endhead
\midrule
\multicolumn{4}{@{}r@{}}{Continued on next page}\\
\endfoot
\bottomrule
\endlastfoot
Conventional-result article & 52.9 & 32.1 & \ensuremath{-}20.8 [\ensuremath{-}21.82, \ensuremath{-}19.75] \\
Cited-source article & 0.0 & 14.3 & +14.3 [13.73, 14.86] \\
Follow-up search & 10.4 & 21.6 & +11.1 [9.94, 12.41] \\
Browsing & 15.5 & 10.8 & \ensuremath{-}4.8 [\ensuremath{-}5.41, \ensuremath{-}4.17] \\
Session end & 21.1 & 21.3 & +0.1 [\ensuremath{-}0.58, 0.89] \\
\end{longtable}
\endgroup

{\fontsize{8bp}{11pt}\selectfont\interlinepenalty=10000\noindent \textbf{Note.} Shares use all 46,508 Control and 24,683 Treatment searches, with reader-bootstrap CIs (section S30). Figure 3 rounds shares and changes separately. Section S27 defines the actions and citation attribution.\par}
\medskip

\Needspace{6\baselineskip}
\subsection*{S18. How do more searches add up to more consumption?}
\phantomsection\label{sec:S18}

More frequent searching offsets lower article consumption per search. Searches per reader rise by 29.4\%, article consumption per search falls by 18.2\%, and article consumption per reader rises by 5.9\% (Table S25). Including answers raises total information consumption by 82.2\% per reader and 40.8\% per search.

\Needspace{12\baselineskip}
\setcounter{table}{24}
\captionof{table}{Search volume and consumption per reader and per search}\label{tab:S25}

% SI physical table block 32; authoritative Table S25.
\begingroup
\fontsize{9bp}{11.5pt}\selectfont
\setlength{\tabcolsep}{3pt}
\renewcommand{\arraystretch}{1.13}
\setlength{\LTleft}{0pt}\setlength{\LTright}{0pt}
\setlength{\LTpre}{4pt}\setlength{\LTpost}{5pt}
\renewcommand{\theHtable}{supplement.block32.\arabic{table}}
\begin{longtable}{@{}>{\raggedright\arraybackslash}p{\dimexpr0.2154171\linewidth-2\tabcolsep\relax}>{\raggedright\arraybackslash}p{\dimexpr0.1000000\linewidth-2\tabcolsep\relax}>{\raggedright\arraybackslash}p{\dimexpr0.1307286\linewidth-2\tabcolsep\relax}>{\raggedright\arraybackslash}p{\dimexpr0.1231257\linewidth-2\tabcolsep\relax}>{\raggedright\arraybackslash}p{\dimexpr0.1683210\linewidth-2\tabcolsep\relax}>{\raggedright\arraybackslash}p{\dimexpr0.1330517\linewidth-2\tabcolsep\relax}>{\raggedright\arraybackslash}p{\dimexpr0.1293559\linewidth-2\tabcolsep\relax}@{}}
\toprule
\textbf{\textbf{Measure}} & \textbf{\textbf{Control}} & \textbf{\textbf{Treatment}} & \textbf{\textbf{Difference}} & \textbf{\textbf{95\% CI}} & \textbf{\textbf{Change (\%)}} & \textbf{\textbf{95\% CI}} \\
\midrule
\endfirsthead
\multicolumn{7}{@{}l@{}}{\textbf{Table S25 (continued)}}\\
\toprule
\textbf{\textbf{Measure}} & \textbf{\textbf{Control}} & \textbf{\textbf{Treatment}} & \textbf{\textbf{Difference}} & \textbf{\textbf{95\% CI}} & \textbf{\textbf{Change (\%)}} & \textbf{\textbf{95\% CI}} \\
\midrule
\endhead
\midrule
\multicolumn{7}{@{}r@{}}{Continued on next page}\\
\endfoot
\bottomrule
\endlastfoot
Searches per assigned reader & 1.7459 & 2.2599 & +0.5141 & [0.4642, 0.5658] & +29.4 & [26.5, 32.6] \\
Article opens per assigned reader & 1.7521 & 1.8386 & +0.0865 & [0.0227, 0.1521] & +4.9 & [1.3, 8.7] \\
Article consumption per assigned reader & 1.3628 & 1.4426 & +0.0798 & [0.0239, 0.1362] & +5.9 & [1.7, 10.1] \\
Total information consumption per assigned reader & 1.3628 & 2.4832 & +1.1205 & [1.0580, 1.1840] & +82.2 & [76.7, 88.0] \\
Article opens per search & 1.0036 & 0.8136 & \ensuremath{-}0.1900 & [\ensuremath{-}0.2206, \ensuremath{-}0.1603] & \ensuremath{-}18.9 & [\ensuremath{-}21.8, \ensuremath{-}16.1] \\
Article consumption per search & 0.7806 & 0.6383 & \ensuremath{-}0.1422 & [\ensuremath{-}0.1696, \ensuremath{-}0.1145] & \ensuremath{-}18.2 & [\ensuremath{-}21.5, \ensuremath{-}14.9] \\
Total information consumption per search & 0.7806 & 1.0988 & +0.3182 & [0.2898, 0.3481] & +40.8 & [36.6, 45.2] \\
\end{longtable}
\endgroup

{\fontsize{8bp}{11pt}\selectfont\interlinepenalty=10000\noindent \textbf{Note.} Article opens count all recorded article-page visits. Article consumption includes only opens with estimated topic shares, so its total is smaller (sections S23 and S24). Total information consumption adds AI answers. Per-reader means include all assigned readers. Per-search measures divide group consumption by all searches. CIs resample readers (section S30). These ratios describe the observed totals without isolating the causal contribution of extra searches.\par}
\medskip

\section*{Experimental Design, Measurement, and Inference}

\Needspace{6\baselineskip}
\subsection*{S19. How were readers assigned, and who is in the analytic sample?}
\phantomsection\label{sec:S19}

The publisher reports assigning readers by browser cookie at their first on-site search, before displaying results. Here, a reader is an anonymous cookie-based ID. The assignment code, allocation logs and intended ratio were not disclosed. The Treatment share varies across entry days (\textit{P} \ensuremath{<} 0.001). Our main tests therefore compare assignments within each day while preserving that day's group sizes. Materials and Methods states the assumptions needed to interpret these comparisons.

Table S26 reports reader counts through sample preparation. We screened for automated traffic by examining patterns such as the same search queries recurring in the same order and searches occurring at nearly fixed time intervals. Readers whose first-search query could not be reconstructed were then excluded. At each stage, we use the same twelve first-event characteristics to check whether differences between assigned groups arose during sample preparation. These cover device, subscription, referral, entry page and timing, and differ from the pre-search measures in section S20.

\Needspace{12\baselineskip}
\setcounter{table}{25}
\captionof{table}{Reader counts and assignment predictability across sample preparation}\label{tab:S26}

% SI physical table block 33; authoritative Table S26.
\begingroup
\fontsize{9bp}{11.5pt}\selectfont
\setlength{\tabcolsep}{3pt}
\renewcommand{\arraystretch}{1.13}
\setlength{\LTleft}{0pt}\setlength{\LTright}{0pt}
\setlength{\LTpre}{4pt}\setlength{\LTpost}{5pt}
\renewcommand{\theHtable}{supplement.block33.\arabic{table}}
\begin{longtable}{@{}>{\raggedright\arraybackslash}p{\dimexpr0.5078141\linewidth-2\tabcolsep\relax}>{\raggedright\arraybackslash}p{\dimexpr0.1270327\linewidth-2\tabcolsep\relax}>{\raggedright\arraybackslash}p{\dimexpr0.1270327\linewidth-2\tabcolsep\relax}>{\raggedright\arraybackslash}p{\dimexpr0.1349525\linewidth-2\tabcolsep\relax}>{\raggedright\arraybackslash}p{\dimexpr0.1031679\linewidth-2\tabcolsep\relax}@{}}
\toprule
\textbf{\textbf{Stage}} & \textbf{\textbf{Control}} & \textbf{\textbf{Treatment}} & \textbf{\textbf{Joint F}} & \textbf{\textbf{R\textsuperscript{2}}} \\
\midrule
\endfirsthead
\multicolumn{5}{@{}l@{}}{\textbf{Table S26 (continued)}}\\
\toprule
\textbf{\textbf{Stage}} & \textbf{\textbf{Control}} & \textbf{\textbf{Treatment}} & \textbf{\textbf{Joint F}} & \textbf{\textbf{R\textsuperscript{2}}} \\
\midrule
\endhead
\midrule
\multicolumn{5}{@{}r@{}}{Continued on next page}\\
\endfoot
\bottomrule
\endlastfoot
Baseline clickstream & 35,141 & 13,180 & 14.89 & 0.0037 \\
Automated-traffic screening & 27,181 & 11,000 & 11.19 & 0.0035 \\
Final sample after first-query exclusions & 26,639 & 10,922 & 10.92 & 0.0035 \\
\end{longtable}
\endgroup

{\fontsize{8bp}{11pt}\selectfont\interlinepenalty=10000\noindent \textbf{Note.} The recorded characteristics explain less than 1\% of assignment variation at every stage, although their joint association with assignment is statistically significant (P \ensuremath{<} 0.001). Their explanatory power changes little during sample preparation; the association is already present before cleaning.\par}
\medskip

A reader can return with a new cookie, and the records do not link the same person across devices.

\Needspace{6\baselineskip}
\subsection*{S20. Were the two groups alike before the first search?}
\phantomsection\label{sec:S20}

Individual pre-search differences are small (Table S28), although the characteristics jointly predict assignment. Table S27 defines the measures of prior activity, device, arrival route and entry timing.

\Needspace{12\baselineskip}
\setcounter{table}{26}
\captionof{table}{Definitions of the reader characteristics used to check balance}\label{tab:S27}

% SI physical table block 34; authoritative Table S27.
\begingroup
\fontsize{9bp}{11.5pt}\selectfont
\setlength{\tabcolsep}{3pt}
\renewcommand{\arraystretch}{1.13}
\setlength{\LTleft}{0pt}\setlength{\LTright}{0pt}
\setlength{\LTpre}{4pt}\setlength{\LTpost}{5pt}
\renewcommand{\theHtable}{supplement.block34.\arabic{table}}
\begin{longtable}{@{}>{\raggedright\arraybackslash}p{\dimexpr0.3077086\linewidth-2\tabcolsep\relax}>{\raggedright\arraybackslash}p{\dimexpr0.6922914\linewidth-2\tabcolsep\relax}@{}}
\toprule
\textbf{\textbf{Reader characteristic}} & \textbf{\textbf{Definition}} \\
\midrule
\endfirsthead
\multicolumn{2}{@{}l@{}}{\textbf{Table S27 (continued)}}\\
\toprule
\textbf{\textbf{Reader characteristic}} & \textbf{\textbf{Definition}} \\
\midrule
\endhead
\midrule
\multicolumn{2}{@{}r@{}}{Continued on next page}\\
\endfoot
\bottomrule
\endlastfoot
Any prior activity & Whether any event is recorded before the first search. \\
Prior events & Number of recorded page-view events before the first search, after deduplication. \\
Prior sessions & Number of visits with recorded activity before the first search. This can include the visit in which the first search occurs. \\
Prior article views & Number of article-page views before the first search. \\
Prior home-page views & Number of home-page views before the first search. \\
Prior non-article views & Number of section or topic landing-page views before the first search. \\
Time observed before first search (hours) & Time from the earliest recorded pre-search event to the first search, with zero assigned when no pre-search event is present. \\
On desktop & Whether the device at the first recorded event is desktop. \\
On mobile & Whether the device at the first recorded event is mobile. \\
Arrived from an external link & Whether an external referral is recorded before or at the first search. Arrival at that search precedes display of its results. \\
Entered on the home page & Whether the first recorded event is a home-page view. \\
Entered on an article & Whether the first recorded event is an article-page view. \\
First visit on a weekend & Whether the recorded date of the first event falls on Saturday or Sunday. \\
\end{longtable}
\endgroup

\Needspace{12\baselineskip}
\setcounter{table}{27}
\captionof{table}{Reader characteristics before the first search}\label{tab:S28}

% SI physical table block 35; authoritative Table S28.
\begingroup
\fontsize{9bp}{11.5pt}\selectfont
\setlength{\tabcolsep}{3pt}
\renewcommand{\arraystretch}{1.13}
\setlength{\LTleft}{0pt}\setlength{\LTright}{0pt}
\setlength{\LTpre}{4pt}\setlength{\LTpost}{5pt}
\renewcommand{\theHtable}{supplement.block35.\arabic{table}}
\begin{longtable}{@{}>{\raggedright\arraybackslash}p{\dimexpr0.4691658\linewidth-2\tabcolsep\relax}>{\raggedright\arraybackslash}p{\dimexpr0.1538543\linewidth-2\tabcolsep\relax}>{\raggedright\arraybackslash}p{\dimexpr0.1538543\linewidth-2\tabcolsep\relax}>{\raggedright\arraybackslash}p{\dimexpr0.2231257\linewidth-2\tabcolsep\relax}@{}}
\toprule
\textbf{\textbf{Reader characteristic}} & \textbf{\textbf{Control}} & \textbf{\textbf{Treatment}} & \textbf{\textbf{Standardized difference}} \\
\midrule
\endfirsthead
\multicolumn{4}{@{}l@{}}{\textbf{Table S28 (continued)}}\\
\toprule
\textbf{\textbf{Reader characteristic}} & \textbf{\textbf{Control}} & \textbf{\textbf{Treatment}} & \textbf{\textbf{Standardized difference}} \\
\midrule
\endhead
\midrule
\multicolumn{4}{@{}r@{}}{Continued on next page}\\
\endfoot
\bottomrule
\endlastfoot
Any prior activity & 0.957 & 0.948 & \ensuremath{-}0.042 \\
Prior events & 2.185 & 2.049 & \ensuremath{-}0.044 \\
Prior sessions & 1.143 & 1.088 & \ensuremath{-}0.085 \\
Prior article views & 0.754 & 0.724 & \ensuremath{-}0.016 \\
Prior home-page views & 0.979 & 0.892 & \ensuremath{-}0.097 \\
Prior non-article views & 0.451 & 0.434 & \ensuremath{-}0.013 \\
Time observed before first search (hours) & 13.919 & 11.956 & \ensuremath{-}0.029 \\
On desktop & 0.665 & 0.691 & +0.056 \\
On mobile & 0.314 & 0.287 & \ensuremath{-}0.058 \\
Arrived from an external link & 0.232 & 0.251 & +0.045 \\
Entered on the home page & 0.643 & 0.601 & \ensuremath{-}0.088 \\
Entered on an article & 0.212 & 0.238 & +0.062 \\
First visit on a weekend & 0.176 & 0.186 & +0.027 \\
\end{longtable}
\endgroup

{\fontsize{8bp}{11pt}\selectfont\interlinepenalty=10000\noindent \textbf{Note.} Means include 26,639 Control and 10,922 Treatment readers. For yes-or-no characteristics, means are proportions. Each standardized difference divides the group mean difference by the square root of the average within-group variance. All absolute standardized differences are below 0.10.\par}
\medskip

Together, the characteristics explain little variation in assignment (R\textsuperscript{2} = 0.005), although the joint test is statistically significant (P \ensuremath{<} 0.001). The test omits total prior events because that variable sums the three types of page views. First-search day is addressed separately (section S19). Outcome estimates remain similar after adjustment for these characteristics and entry week (section S21, Tables S29 and S30B).

\Needspace{6\baselineskip}
\subsection*{S21. Do the results hold after adjusting for reader characteristics and entry timing?}
\phantomsection\label{sec:S21}

The results remain similar after accounting for readers' pre-search characteristics and entry week (Table S29 and panel B of Table S30). This checks whether observed differences between the readers in the two groups account for the findings.

The main results also persist after excluding readers who first searched on the first or last study day (Table S30, panel A). This checks whether the findings depend on readers entering at the boundaries of the experiment.

\Needspace{12\baselineskip}
\setcounter{table}{28}
\captionof{table}{Covariate-adjusted shared information and engagement outcomes}\label{tab:S29}

% SI physical table block 36; authoritative Table S29.
\begingroup
\fontsize{9bp}{11.5pt}\selectfont
\setlength{\tabcolsep}{3pt}
\renewcommand{\arraystretch}{1.13}
\setlength{\LTleft}{0pt}\setlength{\LTright}{0pt}
\setlength{\LTpre}{4pt}\setlength{\LTpost}{5pt}
\renewcommand{\theHtable}{supplement.block36.\arabic{table}}
\begin{longtable}{@{}>{\raggedright\arraybackslash}p{\dimexpr0.4429778\linewidth-2\tabcolsep\relax}>{\raggedright\arraybackslash}p{\dimexpr0.1615628\linewidth-2\tabcolsep\relax}>{\raggedright\arraybackslash}p{\dimexpr0.1520591\linewidth-2\tabcolsep\relax}>{\raggedright\arraybackslash}p{\dimexpr0.1356917\linewidth-2\tabcolsep\relax}>{\raggedright\arraybackslash}p{\dimexpr0.1077086\linewidth-2\tabcolsep\relax}@{}}
\toprule
\textbf{\textbf{Per-reader outcome}} & \textbf{\textbf{Original difference}} & \textbf{\textbf{Adjusted difference}} & \textbf{\textbf{HC2 SE}} & \textbf{\textbf{P}} \\
\midrule
\endfirsthead
\multicolumn{5}{@{}l@{}}{\textbf{Table S29 (continued)}}\\
\toprule
\textbf{\textbf{Per-reader outcome}} & \textbf{\textbf{Original difference}} & \textbf{\textbf{Adjusted difference}} & \textbf{\textbf{HC2 SE}} & \textbf{\textbf{P}} \\
\midrule
\endhead
\midrule
\multicolumn{5}{@{}r@{}}{Continued on next page}\\
\endfoot
\bottomrule
\endlastfoot
Shared information consumption per reader & +0.469 & +0.477 & 0.017 & \ensuremath{<} 0.001 \\
Non-core information consumption per reader & +0.652 & +0.653 & 0.020 & \ensuremath{<} 0.001 \\
Total information consumption per reader & +1.120 & +1.130 & 0.032 & \ensuremath{<} 0.001 \\
Searches per reader & +0.514 & +0.510 & 0.026 & \ensuremath{<} 0.001 \\
Time on site per reader & +62.7 & +65.8 & 9.9 & \ensuremath{<} 0.001 \\
\end{longtable}
\endgroup

{\fontsize{8bp}{11pt}\selectfont\interlinepenalty=10000\noindent \textbf{Note.} Regressions include all 37,561 assigned readers and adjust for the twelve pre-search characteristics and entry week. Total prior events is omitted because it sums three other measures. We use ordinary least squares with HC2 standard errors, allowing error variance to differ across readers. P values are unadjusted. Per-search and per-minute ratios are not regression-adjusted.\par}
\medskip

\Needspace{12\baselineskip}
\setcounter{table}{29}
\captionof{table}{Entry-day and covariate checks for concentration and popularity}\label{tab:S30}

\Needspace{10\baselineskip}
\noindent\textbf{\textbf{A. First and last entry days excluded}}\par
\smallskip

% SI physical table block 37; authoritative Table S30.
\begingroup
\fontsize{9bp}{11.5pt}\selectfont
\setlength{\tabcolsep}{3pt}
\renewcommand{\arraystretch}{1.13}
\setlength{\LTleft}{0pt}\setlength{\LTright}{0pt}
\setlength{\LTpre}{4pt}\setlength{\LTpost}{5pt}
\renewcommand{\theHtable}{supplement.block37.\arabic{table}}
\begin{longtable}{@{}>{\raggedright\arraybackslash}p{\dimexpr0.5384615\linewidth-2\tabcolsep\relax}>{\raggedright\arraybackslash}p{\dimexpr0.2307692\linewidth-2\tabcolsep\relax}>{\raggedright\arraybackslash}p{\dimexpr0.2307692\linewidth-2\tabcolsep\relax}@{}}
\toprule
\textbf{\textbf{Measure}} & \textbf{\textbf{Difference}} & \textbf{\textbf{Holm-adjusted P}} \\
\midrule
\endfirsthead
\multicolumn{3}{@{}l@{}}{\textbf{Table S30 (continued)}}\\
\toprule
\textbf{\textbf{Measure}} & \textbf{\textbf{Difference}} & \textbf{\textbf{Holm-adjusted P}} \\
\midrule
\endhead
\midrule
\multicolumn{3}{@{}r@{}}{Continued on next page}\\
\endfoot
\bottomrule
\endlastfoot
Aggregate Gini & \ensuremath{-}0.0353 & \ensuremath{<} 0.001 \\
Aggregate Effective topics & +2.12 & \ensuremath{<} 0.001 \\
Aggregate Mean rank & +1.36 & \ensuremath{<} 0.001 \\
Aggregate Top-10 share (\%) & \ensuremath{-}4.5 & \ensuremath{<} 0.001 \\
Individual: Gini & \ensuremath{-}0.0128 & \ensuremath{<} 0.001 \\
Individual: Effective topics & +0.95 & \ensuremath{<} 0.001 \\
Individual: Mean rank & +1.20 & \ensuremath{<} 0.001 \\
Individual: Top-10 share (\%) & \ensuremath{-}4.0 & \ensuremath{<} 0.001 \\
\end{longtable}
\endgroup

\Needspace{10\baselineskip}
\noindent\textbf{\textbf{B. Adjustment for pre-search characteristics}}\par
\smallskip

% SI physical table block 38; authoritative Table S30.
\begingroup
\fontsize{9bp}{11.5pt}\selectfont
\setlength{\tabcolsep}{3pt}
\renewcommand{\arraystretch}{1.13}
\setlength{\LTleft}{0pt}\setlength{\LTright}{0pt}
\setlength{\LTpre}{4pt}\setlength{\LTpost}{5pt}
\renewcommand{\theHtable}{supplement.block38.\arabic{table}}
\begin{longtable}{@{}>{\raggedright\arraybackslash}p{\dimexpr0.3692308\linewidth-2\tabcolsep\relax}>{\raggedright\arraybackslash}p{\dimexpr0.2153846\linewidth-2\tabcolsep\relax}>{\raggedright\arraybackslash}p{\dimexpr0.2153846\linewidth-2\tabcolsep\relax}>{\raggedright\arraybackslash}p{\dimexpr0.2000000\linewidth-2\tabcolsep\relax}@{}}
\toprule
\textbf{\textbf{Individual measure}} & \textbf{\textbf{Adjusted difference}} & \textbf{\textbf{Original difference}} & \textbf{\textbf{Unadjusted P}} \\
\midrule
\endfirsthead
\multicolumn{4}{@{}l@{}}{\textbf{Table S30 (continued)}}\\
\toprule
\textbf{\textbf{Individual measure}} & \textbf{\textbf{Adjusted difference}} & \textbf{\textbf{Original difference}} & \textbf{\textbf{Unadjusted P}} \\
\midrule
\endhead
\midrule
\multicolumn{4}{@{}r@{}}{Continued on next page}\\
\endfoot
\bottomrule
\endlastfoot
Gini & \ensuremath{-}0.0143 & \ensuremath{-}0.0131 & \ensuremath{<} 0.001 \\
Effective topics & +1.07 & +0.97 & \ensuremath{<} 0.001 \\
Mean rank & +1.06 & +1.16 & \ensuremath{<} 0.001 \\
Top-10 share (\%) & \ensuremath{-}3.5 & \ensuremath{-}4.0 & \ensuremath{<} 0.001 \\
\end{longtable}
\endgroup

{\fontsize{8bp}{11pt}\selectfont\interlinepenalty=10000\noindent \textbf{Note.} Panel A includes 35,885 readers and uses Holm adjustment across eight comparisons. The corresponding shared-information and engagement check retains ten comparisons, omitting shared information per reader pair; all ten remain significant after Holm adjustment. Panel B uses the main individual sample (section S28) and the regression adjustment described in Table S29. Its P values are unadjusted.\par}
\medskip

\Needspace{6\baselineskip}
\subsection*{S22. How was the topic model chosen and curated?}
\phantomsection\label{sec:S22}

We used BERTopic to group articles into recognizable news subjects. We compared candidate models using topic coherence, keyword diversity, and the proportion of articles assigned to a topic. A minimum cluster size of 700 performed best on all three criteria and produced 62 topics (Table S31).

\Needspace{12\baselineskip}
\setcounter{table}{30}
\captionof{table}{Selected candidate topic model settings}\label{tab:S31}

% SI physical table block 39; authoritative Table S31.
\begingroup
\fontsize{9bp}{11.5pt}\selectfont
\setlength{\tabcolsep}{3pt}
\renewcommand{\arraystretch}{1.13}
\setlength{\LTleft}{0pt}\setlength{\LTright}{0pt}
\setlength{\LTpre}{4pt}\setlength{\LTpost}{5pt}
\renewcommand{\theHtable}{supplement.block39.\arabic{table}}
\begin{longtable}{@{}>{\raggedright\arraybackslash}p{\dimexpr0.2384615\linewidth-2\tabcolsep\relax}>{\raggedright\arraybackslash}p{\dimexpr0.1000000\linewidth-2\tabcolsep\relax}>{\raggedright\arraybackslash}p{\dimexpr0.2076923\linewidth-2\tabcolsep\relax}>{\raggedright\arraybackslash}p{\dimexpr0.2153846\linewidth-2\tabcolsep\relax}>{\raggedright\arraybackslash}p{\dimexpr0.2384615\linewidth-2\tabcolsep\relax}@{}}
\toprule
\textbf{\textbf{Minimum cluster size}} & \textbf{\textbf{Topics}} & \textbf{\textbf{Outlier share (\%)}} & \textbf{\textbf{Coherence}} & \textbf{\textbf{Keyword diversity}} \\
\midrule
\endfirsthead
\multicolumn{5}{@{}l@{}}{\textbf{Table S31 (continued)}}\\
\toprule
\textbf{\textbf{Minimum cluster size}} & \textbf{\textbf{Topics}} & \textbf{\textbf{Outlier share (\%)}} & \textbf{\textbf{Coherence}} & \textbf{\textbf{Keyword diversity}} \\
\midrule
\endhead
\midrule
\multicolumn{5}{@{}r@{}}{Continued on next page}\\
\endfoot
\bottomrule
\endlastfoot
300 & 126 & 36.46 & 0.8059 & 0.8444 \\
400 & 105 & 36.13 & 0.8068 & 0.8743 \\
500 & 82 & 31.31 & 0.8060 & 0.9305 \\
600 & 74 & 32.33 & 0.8102 & 0.9108 \\
700 & 62 & 29.61 & 0.8143 & 0.9403 \\
800 & 58 & 31.01 & 0.8098 & 0.9259 \\
\end{longtable}
\endgroup

{\fontsize{8bp}{11pt}\selectfont\interlinepenalty=10000\noindent \textbf{Note.} Six of the eight tested settings are shown. Coherence measures how often leading topic terms occur together. Keyword diversity is the proportion of distinct words among each topic's ten leading words. Outlier share is the percentage of articles assigned to no cluster. These describe the fitted model. Section S23 reports coverage of observed consumption.\par}
\medskip

BERTopic uses all-MiniLM-L6-v2 embeddings, UMAP for dimension reduction, and HDBSCAN for clustering. We tested eight minimum cluster sizes from 300 to 800. The alternative model in section S11 changes the UMAP random seed from 42 to 45, keeping the embeddings and minimum cluster size of 700 unchanged. We apply the same content exclusions using the refitted model's keywords, retaining 72 of 76 topics.

We excluded four categories of publishing-related and routine service content to define the study's scope (Table S32). The table gives the rationale for each choice. The remaining 58 topics define our news-consumption measures.

\Needspace{12\baselineskip}
\setcounter{table}{31}
\captionof{table}{Topics excluded from the analysis}\label{tab:S32}

% SI physical table block 40; authoritative Table S32.
\begingroup
\fontsize{9bp}{11.5pt}\selectfont
\setlength{\tabcolsep}{3pt}
\renewcommand{\arraystretch}{1.13}
\setlength{\LTleft}{0pt}\setlength{\LTright}{0pt}
\setlength{\LTpre}{4pt}\setlength{\LTpost}{5pt}
\renewcommand{\theHtable}{supplement.block40.\arabic{table}}
\begin{longtable}{@{}>{\raggedright\arraybackslash}p{\dimexpr0.0691658\linewidth-2\tabcolsep\relax}>{\raggedright\arraybackslash}p{\dimexpr0.3077086\linewidth-2\tabcolsep\relax}>{\raggedright\arraybackslash}p{\dimexpr0.6231257\linewidth-2\tabcolsep\relax}@{}}
\toprule
\textbf{\textbf{Topic ID}} & \textbf{\textbf{Excluded category}} & \textbf{\textbf{Reason for exclusion}} \\
\midrule
\endfirsthead
\multicolumn{3}{@{}l@{}}{\textbf{Table S32 (continued)}}\\
\toprule
\textbf{\textbf{Topic ID}} & \textbf{\textbf{Excluded category}} & \textbf{\textbf{Reason for exclusion}} \\
\midrule
\endhead
\midrule
\multicolumn{3}{@{}r@{}}{Continued on next page}\\
\endfoot
\bottomrule
\endlastfoot
17 & Journalism and media commentary & Coverage of news production and media practice, including podcasts. This category concerns the production of news itself. \\
33 & Daily weather forecasts & Routine forecast information, including humidity and sky conditions, rather than reporting on weather-related events. \\
35 & Non-article boilerplate & Social-media prompts, newsletter delivery text and other platform material captured in article text. \\
54 & Sports broadcast listings & Broadcast networks and programming information rather than reporting on sporting events. \\
\end{longtable}
\endgroup

{\fontsize{8bp}{11pt}\selectfont\interlinepenalty=10000\noindent \textbf{Note.} IDs refer to the original 62 topics. After exclusions, each article's remaining topic shares are scaled to sum to one. Section S24 describes how cited articles are combined to represent an answer.\par}
\medskip

\Needspace{6\baselineskip}
\subsection*{S23. How much of the observed consumption has topic information?}
\phantomsection\label{sec:S23}

Topic estimates cover similar shares of article opens in Control and Treatment (Table S33). Opens and answers without topic estimates remain in engagement records but do not enter topic-based consumption. Comparing the two models on the same articles and answers preserves the concentration and popularity findings (section S11).

\Needspace{12\baselineskip}
\setcounter{table}{32}
\captionof{table}{Coverage of observed consumption}\label{tab:S33}

% SI physical table block 41; authoritative Table S33.
\begingroup
\fontsize{9bp}{11.5pt}\selectfont
\setlength{\tabcolsep}{3pt}
\renewcommand{\arraystretch}{1.13}
\setlength{\LTleft}{0pt}\setlength{\LTright}{0pt}
\setlength{\LTpre}{4pt}\setlength{\LTpost}{5pt}
\renewcommand{\theHtable}{supplement.block41.\arabic{table}}
\begin{longtable}{@{}>{\raggedright\arraybackslash}p{\dimexpr0.4462513\linewidth-2\tabcolsep\relax}>{\raggedright\arraybackslash}p{\dimexpr0.1768743\linewidth-2\tabcolsep\relax}>{\raggedright\arraybackslash}p{\dimexpr0.1922914\linewidth-2\tabcolsep\relax}>{\raggedright\arraybackslash}p{\dimexpr0.1845829\linewidth-2\tabcolsep\relax}@{}}
\toprule
\textbf{\textbf{Observed consumption}} & \textbf{\textbf{Events or answers}} & \textbf{\textbf{Topic shares estimated}\textbf{\newline{}58-topic model}} & \textbf{\textbf{Topic shares estimated}\textbf{\newline{}Alternative 72-topic model (S11)}} \\
\midrule
\endfirsthead
\multicolumn{4}{@{}l@{}}{\textbf{Table S33 (continued)}}\\
\toprule
\textbf{\textbf{Observed consumption}} & \textbf{\textbf{Events or answers}} & \textbf{\textbf{Topic shares estimated}\textbf{\newline{}58-topic model}} & \textbf{\textbf{Topic shares estimated}\textbf{\newline{}Alternative 72-topic model (S11)}} \\
\midrule
\endhead
\midrule
\multicolumn{4}{@{}r@{}}{Continued on next page}\\
\endfoot
\bottomrule
\endlastfoot
Control article opens & 46,675 & 77.9\% & 79.9\% \\
Treatment article opens & 20,081 & 78.6\% & 80.3\% \\
Treatment AI answers & 14,413 & 77.0\% & 71.8\% \\
\end{longtable}
\endgroup

{\fontsize{8bp}{11pt}\selectfont\interlinepenalty=10000\noindent \textbf{Note.} Denominators include recorded article opens after the first search and distinct substantive answers. Some content is outside the modeled archive or lacks topic estimates after excluded categories are removed. An answer requires topic estimates for at least one cited source and for the resulting answer distribution (section S24).\par}
\medskip

\Needspace{6\baselineskip}
\subsection*{S24. How are article consumption and answer topics measured?}
\phantomsection\label{sec:S24}

Article consumption sums the estimated topic shares from article opens, and total information consumption also includes those from displayed AI answers. Each article open or substantive answer display contributes one consumption count, divided across its topics; empty-answer messages are excluded. For example, an article that is 60\% about politics and 40\% about economics contributes 0.6 and 0.4 to those topics. All measures start at each reader's first search.

Article topics come from the fitted model. For an AI answer, we average the topic shares of its cited articles, giving each source equal weight. We then remove the four excluded topics and rescale the remaining shares to sum to one. The answer's topic distribution therefore reflects the content of the articles it cites.

Article opens and answer displays without topic estimates remain in engagement records but do not contribute to topic-based consumption (section S23). Answer estimates use cited articles with available topic shares. We exclude estimates that are missing, sum to zero, or assign equal shares to every topic, applying the rule before averaging sources and after retaining the 58 topics.

For the sensitivity checks in section S12, citation-order weighting gives sources cited earlier more weight, using weights proportional to 1, 1/2, 1/3, and so on. The answer-text measure applies the fitted article model to the first 2,000 characters after removing HTML and excess spacing, then retains and normalizes the same 58 topics. It can cover answers whose cited sources do not provide topic estimates. The half- and double-weight checks multiply each answer's contribution by 0.5 or 2, leaving article contributions unchanged.

\Needspace{6\baselineskip}
\subsection*{S25. How are time on site and efficiency measured?}
\phantomsection\label{sec:S25}

Consumption per search divides each group's total consumption by its number of searches. Consumption per minute divides total consumption by observed time on site. Shared information consumption per minute uses only consumption on shared topics. These measures use group totals rather than averaging readers' individual ratios.

A session ends after 30 minutes without activity. We measure time on site from the first recorded activity in a session to its recorded departure, or to the last recorded activity when a departure time is unavailable. Departure times are available for about 97\% of sessions in each group. All assigned readers remain in per-reader averages, including those with no recorded time.

\Needspace{6\baselineskip}
\subsection*{S26. How are shared information and reader-pair comparisons defined?}
\phantomsection\label{sec:S26}

We identify broadly encountered topics by the number of distinct readers they reach. Repeated opens by one reader do not increase reach. An article's main topic is the one with its largest estimated share (section S24). We rank topics by the number of Control readers who opened an article mainly about them during the study, including before the first search. The top quartile, 14 of 58 topics, defines shared information in both groups (Table S35). Using Control keeps Treatment answers from determining the benchmark. Section S4 tests other cutoffs and a ranking based only on pre-search Control reading.

Shared information consumption adds the portions of articles and answers assigned to these 14 topics. Non-core information covers the remaining 44 topics. Reach, as an outcome, is the percentage of assigned readers who consume at least one article or answer mainly about a shared topic. Per-reader consumption and reach include all assigned readers, including those with no measured topic consumption.

For political exclusions, we subtract the consumption assigned to the excluded topics and leave the amounts on other topics unchanged.

The shared core identifies topics widely read across the audience; it does not measure what particular reader pairs have in common. We therefore compare reader pairs within each group across all topics. Cosine similarity measures how similarly readers distribute consumption across topics. Shared information per reader pair measures their overlapping consumption on each topic and adds these amounts across topics. For example, if two readers consume 0.6 and 0.4 on one topic, the shared amount is 0.4. Each measure is zero if either reader has no measured topic consumption. Averages include all assigned pairs and thus reflect participation as well as overlap (section S6). Topic overlap does not establish that readers encountered the same articles, facts or viewpoints.

\Needspace{6\baselineskip}
\subsection*{S27. How are search windows and cited-source article opens defined?}
\phantomsection\label{sec:S27}

\Needspace{3\baselineskip}
\paragraph*{Figure 1A search windows}

In Fig. 1A, node areas show topic consumption per assigned reader. The center combines shared topics, while surrounding nodes show other topics. The central and surrounding nodes use separate area scales. The inner outline on a Treatment node shows its Control size.

To identify topics consumed after the same search in Fig. 1A, we group successive article opens until the reader enters a different query, visits a home or section page, starts another visit, or ends the session. Returning to the same query keeps the group together. We count each distinct article and the first substantive AI answer once, provided their topics can be estimated.

An edge connects the main topics of two articles or of an answer and an article. Each pair is counted once, and line width shows how often such pairs occur per assigned reader. An answer alone cannot form an edge. Pairs within one topic or entirely within shared information are counted but not drawn.

\Needspace{3\baselineskip}
\paragraph*{Next actions and citation attribution}

Figure 3 records what happens immediately after each search. An article open is classified as cited-source consumption if that article had appeared in an AI-answer citation at or before the open within the same session. Other article opens recorded immediately after a search are classified as conventional-result opens.

The other next actions are another search, browsing elsewhere on the site, or session end. Browsing means a non-article page visit, including home and section pages. A follow-up search has no intervening article open or other page view. We include all searches, whether or not an answer page was recorded. Sessions are defined in section S25.

\Needspace{6\baselineskip}
\subsection*{S28. How are concentration and popularity measured?}
\phantomsection\label{sec:S28}

Individual measures describe each reader's topic consumption. We calculate them for readers who consume at least two articles or AI answers and have some topic estimates, then average across readers in each group. Aggregate measures combine all measured topic consumption within each group, including consumption by readers with only one article or answer. Sections S9 and S21 examine how the individual results depend on consumption volume, reader inclusion, and pre-search characteristics.

The effective number of topics describes how widely consumption is spread across topics. Consuming four topics equally gives an effective number of four. Concentrating on one of those topics gives a lower value. Formally, it is the exponential of Shannon entropy \citep{hill1973,jost2006}. The normalized Gini coefficient ranges from zero for equal topic shares to one for consumption on a single topic. Materials and Methods gives its definition and supporting references.

Popularity measures address the distinction between consumption of popular and less-popular content in recommender research \citep{fleder2009}. Here, we rank topics by total Control article consumption during the study, including reading before the first search, and use this ranking for both groups. Shared information instead ranks topics by how many readers they reach. Mean popularity rank weights the ranks by consumption, so a higher value means more consumption on less-popular topics. The top-10 share is the percentage consumed on the ten most popular topics. The Control benchmark and the top-10 cutoff are choices for this study.

\Needspace{3\baselineskip}
\paragraph*{Checking popularity ranks set before the first search}

Because the main popularity ranking includes Control reading during the experiment, we also rank topics using only Control reading before each reader's first search. The two rankings are close (Spearman correlation 0.981), and the shift toward less-popular topics persists at both levels (Table S34).

\Needspace{12\baselineskip}
\setcounter{table}{33}
\captionof{table}{Popularity measured with the pre-search ranking}\label{tab:S34}

% SI physical table block 42; authoritative Table S34.
\begingroup
\fontsize{9bp}{11.5pt}\selectfont
\setlength{\tabcolsep}{3pt}
\renewcommand{\arraystretch}{1.13}
\setlength{\LTleft}{0pt}\setlength{\LTright}{0pt}
\setlength{\LTpre}{4pt}\setlength{\LTpost}{5pt}
\renewcommand{\theHtable}{supplement.block42.\arabic{table}}
\begin{longtable}{@{}>{\raggedright\arraybackslash}p{\dimexpr0.2767687\linewidth-2\tabcolsep\relax}>{\raggedright\arraybackslash}p{\dimexpr0.1231257\linewidth-2\tabcolsep\relax}>{\raggedright\arraybackslash}p{\dimexpr0.1231257\linewidth-2\tabcolsep\relax}>{\raggedright\arraybackslash}p{\dimexpr0.1231257\linewidth-2\tabcolsep\relax}>{\raggedright\arraybackslash}p{\dimexpr0.2615628\linewidth-2\tabcolsep\relax}>{\raggedright\arraybackslash}p{\dimexpr0.0922914\linewidth-2\tabcolsep\relax}@{}}
\toprule
\textbf{\textbf{Measure}} & \textbf{\textbf{Control}} & \textbf{\textbf{Treatment}} & \textbf{\textbf{Difference}} & \textbf{\textbf{95\% CI}} & \textbf{\textbf{P}} \\
\midrule
\endfirsthead
\multicolumn{6}{@{}l@{}}{\textbf{Table S34 (continued)}}\\
\toprule
\textbf{\textbf{Measure}} & \textbf{\textbf{Control}} & \textbf{\textbf{Treatment}} & \textbf{\textbf{Difference}} & \textbf{\textbf{95\% CI}} & \textbf{\textbf{P}} \\
\midrule
\endhead
\midrule
\multicolumn{6}{@{}r@{}}{Continued on next page}\\
\endfoot
\bottomrule
\endlastfoot
Aggregate Mean rank & 18.08 & 19.45 & +1.37 & [1.10, 1.65] & \ensuremath{<} 0.001 \\
Aggregate Top-10 share (\%) & 38.6 & 35.2 & \ensuremath{-}3.4 & [\ensuremath{-}4.4, \ensuremath{-}2.4] & \ensuremath{<} 0.001 \\
Individual Mean rank & 18.16 & 19.40 & +1.23 & [0.95, 1.53] & \ensuremath{<} 0.001 \\
Individual Top-10 share (\%) & 38.3 & 35.3 & \ensuremath{-}3.0 & [\ensuremath{-}4.0, \ensuremath{-}2.1] & \ensuremath{<} 0.001 \\
\end{longtable}
\endgroup

{\fontsize{8bp}{11pt}\selectfont\interlinepenalty=10000\noindent \textbf{Note.} The pre-search ranking is rebuilt in each permutation. Confidence intervals and unadjusted P values follow sections S30 and S31. Top-10 differences are percentage points.\par}
\medskip

\Needspace{6\baselineskip}
\subsection*{S29. How are the shared-information tests adjusted for multiple comparisons?}
\phantomsection\label{sec:S29}

Holm adjustment controls the chance of one or more false positives across the 11 shared-information and engagement comparisons. These were chosen before adjustment but were not preregistered. One comparison restricts cosine similarity to pairs in which both readers have measured topic consumption. It is not tabulated because its direction depends on topic representation: positive with topic distributions and negative with a single topic per article. We retain it in the adjustment; the other ten comparisons remain at \textit{P} \ensuremath{<} 0.001 if it is omitted.

The 11 comparisons are shared consumption, reach, non-core consumption, cosine similarity across all reader pairs, shared information per reader pair, conditional cosine similarity, total consumption, searches, consumption per search, time per reader, and consumption per minute. We use 100,000 permutations, except for shared information per reader pair (10,000). The main tests keep the shared-topic definition unchanged. Rebuilding it in each permutation, using either the main or pre-search ranking, also yields P \ensuremath{<} 0.001 for shared consumption, reach, and non-core consumption.

\Needspace{6\baselineskip}
\subsection*{S30. How are intervals and permutation tests computed?}
\phantomsection\label{sec:S30}

Reader-bootstrap confidence intervals resample readers within each group, keeping each reader's articles, answers and searches together. We use 10,000 resamples. Each 95\% interval describes uncertainty for one estimate and is not adjusted for multiple comparisons. The descriptive text and time comparisons in section S8 use 5,000 reader resamples.

Permutation tests repeatedly reassign Treatment labels within each of the 38 first-search days, preserving the two group sizes on each day. This compares Treatment and Control readers who entered on the same day, as described in section S19 and Materials and Methods.

\Needspace{6\baselineskip}
\subsection*{S31. How are the concentration and popularity tests adjusted?}
\phantomsection\label{sec:S31}

Table 1 uses 100,000 permutations and Holm adjustment across its eight concentration and popularity comparisons. Popularity ranks are rebuilt from the reassigned Control readers in every permutation, while bootstrap CIs hold the observed Control ranking fixed. The eight comparisons were chosen before adjustment and were not preregistered.

Each political exclusion, the alternative topic model and the article-only analysis uses its own eight-comparison Holm adjustment. Other supplementary permutation tests use 10,000 permutations and unadjusted P values unless specified.

\Needspace{6\baselineskip}
\subsection*{S32. How are the supplementary comparisons tested?}
\phantomsection\label{sec:S32}

Within-reader comparisons resample paired measurements together; Table S18's effective-topic and mean-popularity-rank comparisons also use two-sided paired t tests. Article-route comparisons keep each reader's contributions through both routes together. Answer-versus-Control-article comparisons resample readers independently within groups. Table S21, panel B, uses Welch tests with Holm adjustment across two measures within each inclusion rule.

For the equal-consumption check in Table S11, panel A, we draw two articles or AI answers without replacement for each eligible reader and repeat this 100 times. We average each reader's concentration across these draws, then use 10,000 reader-bootstrap resamples for confidence intervals. The article-and-answer draws are not repeated within the bootstrap.

Table S23 uses 10,000 paired query-bootstrap resamples for confidence intervals and two-sided Wilcoxon signed-rank tests for P values. The intervals describe mean differences, while the tests use signed ranks, so their conclusions need not coincide. Table notes identify the samples and any adjustment for multiple comparisons.

\Needspace{6\baselineskip}
\subsection*{S33. What are the 58 topics and their labels?}
\phantomsection\label{sec:S33}

Table S35 lists the 58 topics, their five leading model terms and descriptive labels. Labels draw on model keywords and representative article content reviewed with generative AI assistance. They name content categories and do not change the analysis. Bold labels appear in Fig. 1A. Shaded rows identify shared information.

Control reach is the percentage of Control article readers who opened an article mainly about that topic. Consumption share is the percentage of all Control article consumption assigned to the topic, allowing an article to cover several topics. Both use the whole study window, including reading before the first search.

\Needspace{12\baselineskip}
\setcounter{table}{34}
\captionof{table}{The 58 topics and their descriptive labels}\label{tab:S35}

% SI physical table block 43; authoritative Table S35.
\begingroup
\fontsize{7pt}{10pt}\selectfont
\setlength{\tabcolsep}{3pt}
\renewcommand{\arraystretch}{1.13}
\setlength{\LTleft}{0pt}\setlength{\LTright}{0pt}
\setlength{\LTpre}{4pt}\setlength{\LTpost}{5pt}
\renewcommand{\theHtable}{supplement.block43.\arabic{table}}
\begin{longtable}{@{}>{\raggedright\arraybackslash}p{\dimexpr0.0692714\linewidth-2\tabcolsep\relax}>{\raggedright\arraybackslash}p{\dimexpr0.2231257\linewidth-2\tabcolsep\relax}>{\raggedright\arraybackslash}p{\dimexpr0.1220697\linewidth-2\tabcolsep\relax}>{\raggedright\arraybackslash}p{\dimexpr0.1235480\linewidth-2\tabcolsep\relax}>{\raggedright\arraybackslash}p{\dimexpr0.1312566\linewidth-2\tabcolsep\relax}>{\raggedright\arraybackslash}p{\dimexpr0.3307286\linewidth-2\tabcolsep\relax}@{}}
\toprule
\textbf{\textbf{Topic ID}} & \textbf{\textbf{Top five model terms}} & \textbf{\textbf{Control reach (\%)}} & \textbf{\textbf{Control consumption share (\%)}} & \textbf{\textbf{Shared}\textbf{ Information}} & \textbf{\textbf{Topic label}} \\
\midrule
\endfirsthead
\multicolumn{6}{@{}l@{}}{\textbf{Table S35 (continued)}}\\
\toprule
\textbf{\textbf{Topic ID}} & \textbf{\textbf{Top five model terms}} & \textbf{\textbf{Control reach (\%)}} & \textbf{\textbf{Control consumption share (\%)}} & \textbf{\textbf{Shared}\textbf{ Information}} & \textbf{\textbf{Topic label}} \\
\midrule
\endhead
\midrule
\multicolumn{6}{@{}r@{}}{Continued on next page}\\
\endfoot
\bottomrule
\endlastfoot
\rowcolor{black!7}
0 & biden, democrat, senate, presidential, gop & 8.78 & 5.77 & Yes & U.S. National Politics \\
1 & nfl, quarterback, touchdown, redskin, receiver & 3.03 & 1.61 & No & \textbf{American Football} \\
\rowcolor{black!7}
2 & artist, character, album, actor, musical & 5.96 & 3.01 & Yes & Arts \& Entertainment \\
3 & vaccine, virus, vaccination, infection, vaccinate & 2.97 & 1.66 & No & \textbf{Public Health \& Vaccines} \\
\rowcolor{black!7}
4 & dish, sauce, meat, cheese, ingredient & 5.73 & 3.18 & Yes & Food \& Cooking \\
5 & charter, classroom, school district, charter school, academic & 2.57 & 1.67 & No & \textbf{Education} \\
6 & inning, martinez, pitcher, mlb, scherzer & 1.05 & 0.57 & No & Baseball \\
7 & firearm, gun violence, shooter, police department, prosecutor & 2.51 & 1.48 & No & \textbf{Gun Violence} \\
\rowcolor{black!7}
8 & ukraine, ukrainian, putin, kyiv, moscow & 6.37 & 3.13 & Yes & Russia \& Ukraine \\
\rowcolor{black!7}
9 & immigration, migrant, immigrant, asylum, mexican & 9.72 & 5.08 & Yes & Immigration \& Borders \\
10 & nba, wizard, beal, lakers, warrior & 1.10 & 0.51 & No & Basketball \\
11 & hong, hong kong, kong, beijing, taiwan & 3.53 & 1.78 & No & \textbf{China \& East Asia} \\
12 & county police, prince george, george county, gunshot, fatally & 1.12 & 0.60 & No & Local Crime \& Policing \\
\rowcolor{black!7}
13 & israel, israeli, gaza, palestinian, hamas & 12.38 & 6.03 & Yes & Israel \& Palestine \\
14 & transit, tesla, amtrak, scooter, ridership & 1.27 & 0.68 & No & Transit \& Transportation \\
15 & emission, carbon, fossil fuel, fossil, coal & 4.25 & 2.32 & No & \textbf{Climate \& Energy} \\
\rowcolor{black!7}
16 & privacy, zuckerberg, antitrust, apps, iphone & 6.66 & 3.24 & Yes & Technology \& Digital Platforms \\
18 & midfielder, defender, goalkeeper, national team, season complete & 0.60 & 0.47 & No & Soccer \\
\rowcolor{black!7}
19 & kavanaugh, supreme court, supreme, judge, ford & 5.93 & 3.36 & Yes & Supreme Court \& Judicial Appointments \\
20 & iran, iranian, syria, islamic, iraq & 3.13 & 1.82 & No & \textbf{Iran \& Middle East} \\
\rowcolor{black!7}
21 & abortion, roe, pregnancy, ban, supreme court & 10.93 & 5.29 & Yes & Abortion \& Reproductive Rights \\
22 & boeing, faa, aviation, aircraft, flight attendant & 1.67 & 0.86 & No & Aviation \& Air Travel \\
23 & olympic, olympics, tokyo, meter, bile & 1.18 & 0.68 & No & Olympic Sports \\
\rowcolor{black!7}
24 & buyer, bedroom, real estate, condo, bedroom bathroom & 5.09 & 2.52 & Yes & Housing \& Real Estate \\
25 & ovechkin, nhl, hockey, stanley cup, puck & 0.25 & 0.26 & No & Ice Hockey \\
26 & wildlife, zoo, fish, whale, shark & 3.78 & 1.87 & No & \textbf{Wildlife \& Conservation} \\
\rowcolor{black!7}
27 & tornado, tropical, mph, weather service, thunderstorm & 4.84 & 2.74 & Yes & Severe Storms \\
28 & slavery, confederate, statue, slave, african american & 3.32 & 1.74 & No & \textbf{Race \& History} \\
29 & carolyn, adapt online, discussion carolyn, online discussion, adapt & 4.11 & 3.36 & No & \textbf{Relationship \& Advice} \\
\rowcolor{black!7}
30 & amy, daughter, tribune content, content agency, wedding & 4.96 & 2.61 & Yes & Family \& Relationship Advice \\
31 & youngkin, northam, mcauliffe, democrat, fairfax & 0.94 & 0.84 & No & Virginia Politics \\
32 & rude, question manner, manner column, follow realmissmanners, realmissmanners judith & 2.07 & 1.37 & No & Etiquette \& Social Advice \\
34 & taliban, afghan, afghanistan, kabul, troop & 0.58 & 0.40 & No & Afghanistan \\
36 & sudan, ethiopia, coup, kenya, mali & 0.93 & 0.74 & No & African Politics \& Conflict \\
37 & nasa, astronaut, spacecraft, spacex, orbit & 1.95 & 1.27 & No & Space Exploration \\
38 & heloise, puppy, canine, veterinary, veterinarian & 0.49 & 0.31 & No & Pets \& Household Advice \\
39 & saudi, khashoggi, arabia, saudi arabia, crown prince & 0.91 & 0.45 & No & Saudi Arabia \& Yemen \\
40 & north korea, korean, kim, north korean, south korea & 3.07 & 1.49 & No & \textbf{Korean Peninsula} \\
41 & esports, video game, nintendo, twitch, playstation & 0.70 & 0.82 & No & Video Games \& Esports \\
42 & soldier, vietnam, cemetery, pentagon, sailor & 2.18 & 1.50 & No & \textbf{Military \& Veterans} \\
43 & cybersecurity, cyber, hacker, hack, ransomware & 1.85 & 1.05 & No & Cybersecurity \\
44 & brexit, minister, prime minister, labour, parliament & 1.57 & 1.00 & No & \textbf{Parliamentary Politics} \\
45 & medicaid, health care, medicare, aca, obamacare & 4.11 & 2.40 & No & \textbf{Health Care and Policy} \\
\rowcolor{black!7}
46 & breast, muscle, breast cancer, therapy, workout & 5.31 & 3.01 & Yes & Health \& Fitness \\
47 & pga, pga tour, liv, golfer, mickelson & 0.31 & 0.22 & No & Golf \\
48 & county police, pedestrian, prince george, george county, pronounce dead & 0.24 & 0.42 & No & Road Safety \& Collisions \\
49 & opioid, fentanyl, overdose, opioids, addiction & 2.77 & 1.33 & No & Opioids \& Addiction \\
50 & hogan, moore, jealous, cox, alsobrooks & 2.47 & 1.53 & No & Maryland Politics \\
51 & maduro, venezuela, venezuelan, guaid, colombia & 0.80 & 0.56 & No & Venezuela \& Latin America \\
52 & pastor, religious, baptist, evangelical, evangelicals & 2.68 & 1.36 & No & Protestant Christianity \\
53 & tennis, djokovic, grand slam, wimbledon, nadal & 0.92 & 0.41 & No & Tennis \\
55 & india, modi, hindu, kashmir, bjp & 3.38 & 1.71 & No & \textbf{India \& South Asia} \\
56 & harry, meghan, palace, royal family, princess & 1.22 & 0.70 & No & British Royal Family \\
\rowcolor{black!7}
57 & pope, priest, francis, bishop, vatican & 4.46 & 1.75 & Yes & Catholic Church \\
58 & wildfire, drought, forest, air quality, evacuation & 2.14 & 1.15 & No & Wildfires \& Drought \\
59 & bowser, evans, ward, mendelson, silverman & 2.49 & 1.32 & No & \textbf{D.C. Public Affairs} \\
60 & bolsonaro, brazil, brazilian, lula, peru & 1.08 & 0.51 & No & Brazil \& South America \\
61 & macron, pen, minister, parliament, emmanuel & 0.58 & 0.46 & No & French Politics \\
\end{longtable}
\endgroup

{\fontsize{8bp}{11pt}\selectfont\interlinepenalty=10000\noindent \textbf{Note.} Reach uses 18,571 Control article readers. Consumption shares sum to 100\% before rounding. Reach percentages can overlap. Topic IDs retain the model's original numbering.\par}
\medskip

\subsection*{References}

\begingroup\fontsize{8bp}{10bp}\selectfont
\noindent Hill, M. O. (1973). Diversity and evenness: A unifying notation and its consequences. \textit{Ecology}, \textbf{54}, 427--432.\par\smallskip
\noindent Jost, L. (2006). Entropy and diversity. \textit{Oikos}, \textbf{113}, 363--375.\par\smallskip
\noindent Fleder, D., and Hosanagar, K. (2009). Blockbuster culture's next rise or fall: The impact of recommender systems on sales diversity. \textit{Manage. Sci.}, \textbf{55}, 697--712.\par
\endgroup
\setcounter{table}{35}

\end{document}